\documentclass[twocolumn]{aa} 

\usepackage{graphicx}
\usepackage{hyperref}
\usepackage[dvipsnames]{xcolor}
\usepackage{txfonts}
\usepackage{comment}
\usepackage[export]{adjustbox}
\usepackage{float}
\usepackage{placeins}
\usepackage{threeparttable}
\begin{document}

   \title{The hydrostatic-to-lensing mass bias from resolved X-ray and optical-IR data}

    \author{
          M.~Mu\~noz-Echeverr\'ia\inst{\ref{LPSC}}\fnmsep\thanks{miren.munoz@lpsc.in2p3.fr}
          \and
          J. F. Mac\'ias-P\'erez \inst{\ref{LPSC}}
          \and
          G.~W.~Pratt \inst{\ref{CEA}}
          \and
          E.~Pointecouteau\inst{\ref{Toulouse}}
          \and
          I.~Bartalucci \inst{\ref{Milan}}
          \and
          M.~De~Petris \inst{\ref{Roma}}
          \and
          A.~Ferragamo \inst{\ref{IAC}, \ref{Roma}}
          \and
          C.~Hanser \inst{\ref{LPSC}}
          \and
          F.~K\'eruzor\'e \inst{\ref{Argonne}}
          \and
          F.~Mayet\inst{\ref{LPSC}}
          \and
          A.~Moyer-Anin \inst{\ref{LPSC}}
          \and
          A.~Paliwal \inst{\ref{Roma}}
          \and
          L.~Perotto \inst{\ref{LPSC}}
          \and
          G.~Yepes\inst{\ref{Madrid},  \ref{CIAFF}} }

   \institute{
     Univ. Grenoble Alpes, CNRS, LPSC-IN2P3, 53, avenue des Martyrs, 38000 Grenoble, France
     \label{LPSC}
     \and
    Universit\'e Paris-Saclay, Universit\'e Paris Cit\'e, CEA, CNRS, AIM, 91191, Gif-sur-Yvette, France
    \label{CEA}
     \and
     Univ. de Toulouse, UPS-OMP, CNRS, IRAP, 31028 Toulouse, France
     \label{Toulouse}
    \and
    High Energy Physics Division, Argonne National Laboratory, 9700 South Cass Avenue, Lemont, IL 60439, USA
     \label{Argonne}
     \and
     Instituto de Astrof\'isica de Canarias (IAC), C/V\'ia L\'actea s/n, 38205 La Laguna, Tenerife, Spain 
     \label{IAC}
     \and
     Dipartimento di Fisica, Sapienza Universit\`a di Roma, Piazzale Aldo Moro 5, I-00185 Roma, Italy
     \label{Roma}
     \and
     INAF, IASF-Milano, via A. Corti 12, I-20133 Milano, Italy
     \label{Milan}
      \and
     Departamento de F\'isica Te\'orica and CIAFF, Facultad de Ciencias, Modulo 8, Universidad Aut\'onoma de Madrid, 28049 Madrid, Spain\label{Madrid}
     \and
     Centro de Investigaci\'{o}n Avanzada en F\'{i}sica Fundamental (CIAFF), Universidad Aut\'{o}noma de Madrid, Cantoblanco, 28049 Madrid, Spain\label{CIAFF}
   }

   \date{Received ...; accepted ...}

   \abstract{
     An accurate reconstruction of galaxy cluster masses is key to use this population of objects as a cosmological probe. In this work we present a study on the hydrostatic-to-lensing mass scaling relation for a sample of 53 clusters whose masses were reconstructed homogeneously in a redshift range between $z= 0.05$ and $1.07$. The $M_{500}$ mass for each cluster was indeed inferred from the mass profiles extracted from the X-ray and lensing data, without using a priori observable-mass scaling relations. We assessed the systematic dispersion of the masses estimated with our reference analyses with respect to other published mass estimates. Accounting for this systematic scatter does not change our main results, but enables the propagation of the uncertainties related to the mass reconstruction method or used dataset. Our analysis gives a hydrostatic-to-lensing mass bias of $(1-b) =0.739^{+0.075}_{-0.070}$ and no evidence of evolution with redshift. These results are robust against possible subsample differences.}


   \titlerunning{The hydrostatic-to-lensing mass bias}
   \authorrunning{M.~Mu\~noz-Echeverr\'ia et al.}
   \maketitle
%
   \section{Introduction}
 
The distribution of galaxy clusters in mass and redshift is sensitive to the expansion history and matter content of the Universe, as well as to the initial conditions in the primordial Universe \citep{huterer2015}. Thus, cluster masses are valuable tools for the use of these objects in cosmology \citep{Vikhlinin2009, allen2011,Costanzi2019}. Recent results have shown that the cosmological analyses based on cluster number counts seem to favour more lower matter density ($\Omega_{\mathrm{m}}$) and matter power spectrum normalisation ($\sigma_{8}$) values than the studies based on the cosmic microwave background (CMB) power spectrum \citep{planck2014a, salvati2018,Costanzi2019}. Given that cluster masses are not directly observable quantities and have to be estimated under several hypotheses from observations, the uncertainties and systematic errors of those estimates could be the source of tension with the CMB \citep{pratt2019, salvati2020}.

Some cluster number count studies \citep{garrel2022, planck2016a} have relied on cluster masses obtained from scaling relations (SRs) between the Sunyaev-Zel’dovich (SZ) effect \citep{sunyaev} or the X-ray emission of the cluster and the hydrostatic mass reconstructed from X-ray data \citep{arnaud10}. It has been widely investigated and proved that masses reconstructed under the hydrostatic equilibrium (HSE) hypothesis are biased low \citep{Lau_2013,planck2014a,Biffi_2016,gianfagna}. For the cluster number count analyses based on HSE masses, the so-called HSE mass bias could be one of the possibilities to solve the mentioned $\Omega_{\mathrm{m}}-\sigma_{8}$ cosmological tension \citep{planck2014a, salvati2018}. We define the HSE mass bias as the relative difference of HSE mass estimates to the true cluster masses, $ (1-b)= M^{\mathrm{HSE}}/M^{\mathrm{True}}$. A large value of the bias, that is, smaller $(1-b)$, implies larger values for $\Omega_{\mathrm{m}}$ and $\sigma_{8}$ in cluster number count analyses \citep{planck2016a}. 

In the literature different approaches have been developed to estimate this bias. On the one hand, combined CMB power spectrum and cluster number count analyses fit the bias value that is required to get consistent results between the two probes. According to \citet{planck2014a}, $(1-b) = 0.59 \pm 0.05$ would be needed to reconcile the results from the \textit{Planck} CMB analysis in \citet{planck2013cmb} with the cluster count cosmology from \citet{planck2014a}. The posterior analysis of \textit{Planck} data in \citet{planck2016a} obtained $(1-b)$ values in the range between 0.54 and 0.705 considering different priors for the bias (based both on X-ray and lensing data). The updated analysis in \citet{planck2018} provides $(1-b) = 0.62 \pm 0.03$, compatible with the $(1-b) = 0.62\pm 0.07$ from \citet{salvati2018}. Accounting for the power spectrum of the thermal SZ (tSZ) effect together with the cluster number counts, \citet{salvati2018} conclude that the bias needed to reconcile the CMB should be $(1-b) = 0.63 \pm 0.04$. Considering also the trispectrum in the covariance matrix of the tSZ power spectrum analysis, \citet{Bolliet_2018} estimate $(1-b) = 0.58 \pm 0.06$ (68\% C.L.) to be compatible with CMB data.

On the other hand, studies based on simulations have compared the HSE masses of clusters to their true masses.  A large variety of simulations have been used in different works \citep{planck2014a}. Some of them computed the HSE masses by combining, under the HSE hypothesis, the true thermodynamical quantities (density, temperature, and/or pressure) from the intracluster medium in the simulations \citep{Lau_2013,Biffi_2016,gianfagna,Gianfagna_2022}. Others used simulations to mimic mock observations and reconstruct the HSE masses \citep{rasia2012}. However, they all tend to obtain a bias value of $(1-b) > 0.7$ \citep[see Figure 1 in][for a summary]{gianfagna}.

In an attempt to have a reliable estimate of the bias of observational HSE masses, several works have compared the HSE masses to lensing mass estimates, that is, to masses reconstructed from the lensing effect of the cluster on background sources. Under the assumption that lensing masses are unbiased estimates of the true mass of clusters, such HSE-to-lensing mass biases are good estimators of the HSE bias. Most of the studies in the literature are based on lensing masses obtained from the weak lensing signal on background galaxies. Figure 10 in \citet{salvati2018} shows a compilation of HSE-to-lensing mass biases from different works. Despite the heterogeneity of the data and methods used in the various studies, the presented results prefer values of $ M^{\mathrm{HSE}}/M^{\mathrm{lens}}$ above $0.7$. Lensing mass reconstructions from a combination of weak and strong lensing data have also been used to measure the HSE-to-lensing mass bias on small samples \citep{Penna_Lima_2017,ferragamo, munoz2022}, obtaining $ M^{\mathrm{HSE}}/M^{\mathrm{lens}}$ values that span from $\sim 0.6$ to $\sim 1$. The lensing of the CMB anisotropies due to the presence of clusters can also be used to estimate their mass \citep{Melin_2015}. A comparison of HSE and CMB lensing masses based on \textit{Planck} data gave $1/(1-b) = 0.99 \pm 0.19$, approximately $(1-b) = 1.01^{+0.24}_{-0.16}$ \citep{planck2016a}. The posterior analysis in \citet{zubeldia_cosmological_2019} jointly fitted the cosmological parameters and the HSE mass bias in the scaling relation between the SZ signal from \textit{Planck} and cluster masses, using CMB lensing. They determined that the bias is of $(1-b)=0.71 \pm 0.10$. According to the South Pole Telescope (SPT) data analysis in \citet{baxter2015}, the masses inferred from CMB lensing are consistent with those estimated from the SZ.

   Other than lensing, some works in the literature have used the dynamical mass estimate of clusters, based on the velocity dispersion of member galaxies, to compute the bias corresponding to HSE masses \citep[see][and references therein]{ferragamo2021}. According to the analysis with Sloan Digital Sky Survey (SDSS) archival data in \citet{ferragamo2021}, for the 207 galaxy clusters studied, the HSE-to-dynamical bias of \textit{Planck} masses is $(1-b) = 0.83 \pm 0.07\mathrm{(stat.)}  \pm 0.02 \mathrm{(sys.)}$. Also from optical observations, authors in \citet{aguado2021} measured the HSE-to-dynamical mass bias for a different sample of 297 \textit{Planck} clusters and obtained $(1- b) = 0.80 \pm 0.04 \mathrm{(stat.)} \pm 0.05 \mathrm{(sys.)}$.
   
In \citet{wicker} authors investigated the evolution of the (total) HSE bias with mass and redshift by studying the gas mass fraction in galaxy clusters with XMM-\textit{Newton} mass reconstructions from \citet{lovisari2020apj}. The main result in \citet{wicker} is that the value of the HSE bias and its dependence on mass and redshift varies significantly with the analysed cluster sample, in agreement with the conclusions in \citet{salvati2019}. However, according to \citet{wicker} a value of $(1-b)\sim 0.8$ is preferred. Similarly, from the comparison of the gas fraction measured on 12 nearby clusters to the universal gas fraction value, authors in \citet{eckert2019} concluded that the mass bias for SZ-derived estimates is $(1 - b) = 0.85 \pm 0.05$, therefore, inconsistent with the bias needed to reconcile the CMB power spectrum. A different approach was taken in \citet{hurierlacasa}, where the authors used the \textit{Planck} galaxy cluster number counts, tSZ power spectrum, and bispectrum to constrain $(1-b) = 0.71 \pm 0.07$. This was obtained by fitting the normalisation of the SZ-mass SR, interpreting that the bias must appear in the calibration of the scaling relation. They assumed a generalised Navarro-Frenk-White \citep[gNFW,][]{nagai, zhao} pressure profile for the gas in clusters, using the best-fitting parameter values from \citet{arnaud10}, with the normalisation parameter computed to agree with the scaling relation in \citet{planck2014a}. The choice of this particular pressure profile could affect the resulting bias value. 

   There are, therefore, different issues to be considered. Firstly, as stated in \citet{planck2016a}, the main limitation of cosmological analyses with cluster number counts from SZ data is the large uncertainty on the HSE mass bias. But in spite of this large incertitude, the compilation of many studies shows that the bias values estimated with and without considering the need to reconcile CMB results have different tendencies. Such inconsistency is in line with a more general tension between results from early- and late-Universe probes \citep[see][for a review]{Abdalla_2022}. Hence it is essential to have a deeper understanding of the HSE mass bias and its possible evolution with mass and/or redshift. 
   
   In this work, we aim to estimate the HSE-to-lensing mass bias combining single-cluster HSE and lensing $M_{500}$ mass estimates that have been obtained by evaluating mass profiles at their corresponding radius ($R_{500}$). Given the large number of methods and models that can be employed to reconstruct HSE and lensing masses and the potentially different biases that they could be subjected to, we focus on a sample of clusters for which X-ray based HSE and lensing masses have been homogeneously reconstructed. The HSE masses of the \textit{homogeneous} sample have been reconstructed mainly with XMM-\textit{Newton} data and following the method described in Sect.~\ref{sec:xmm}. The lensing masses belong to the CoMaLit compilation of masses from the literature \citep[Sect.~\ref{sec:comalit},][]{serenocomalit}.

   As indicated by \citet{serenoettoricomalit}, and shown also for the well-observed CL~J1226.9+3332 galaxy cluster in Figure 1 in \citet{munoz2022}, cluster mass estimates can vary up to $\sim 40 \%$ from one work to another. Being aware of these important differences that exist between the masses reconstructed by different studies, we gather together results from several works that have also produced estimates based on mass profiles. We use those estimates to measure the systematic dispersion with respect to our XMM-\textit{Newton} and CoMaLit masses. In this work, we analyse a sample of clusters that spans a large redshift range, select homogeneous mass estimates, and propagate the systematic dispersion, which is one step beyond previous studies \citep{lovisari2020,serenoettoricomalit,sereno2020}.

This paper is structured as follows. In Sect.~\ref{sec:data} we present the data, describing the \textit{homogeneous} and \textit{comparison} cluster samples. The method used to match clusters from different catalogues and the measurement of the systematic dispersion of the reference masses with respect to other estimates is described in Sect.~\ref{sec:systematic}. The \textit{reference} sample is built in the same section. In Sect.~\ref{sec:bias} we present the HSE-to-lensing  mass bias and its evolution with redshift. The scaling relation between HSE and lensing masses is obtained in Sect.~\ref{sec:SR}, with all the related systematic tests in the same section. Finally, results are compared to similar works in the literature in Sect.~\ref{sec:comparison} and conclusions are presented in Sect.~\ref{sec:conclusions}. Throughout the paper `log' corresponds to the logarithm to base 10 and `ln' is the natural logarithm. When needed, we assume a flat $\Lambda\mathrm{CDM}$ cosmology with $H_{0} = 70\; \mathrm{km/s/Mpc}$ and $\Omega_{\mathrm{m}} = 0.3$.

\section{Cluster sample construction}
\label{sec:data}

   
\subsection{\textit{Homogeneous} sample}
\label{sec:homogsample}

This study is built aiming for a clusters sample with resolved HSE and lensing masses that are comparable amongst all the objects (homogeneous reconstruction procedure) and cover the largest possible redshift range. We present in this section the mass reconstruction and regularisation procedures of the XMM-\textit{Newton} and CoMaLit clusters, which constitute our \textit{homogeneous} sample.

\subsubsection{CoMaLit sample}
\label{sec:comalit}
The CoMaLit sample contains the clusters with lensing masses that we used to build the \textit{homogeneous} sample. They correspond to the clusters from the Literature Catalogs of weak Lensing Clusters (LC$^2$) compilation presented in \citet{serenocomalit}. The LC$^2$ contains 806 clusters (in the 3.9 version of the LC$^{2}$-\textit{single} catalogue\footnote{\url{http://pico.oabo.inaf.it/~sereno/CoMaLit/}}) with weak lensing masses obtained from different works in the literature, including the widely used Canadian Cluster Comparison Project \citep[CCCP,][]{hoekstra2012, Hoekstra_2015} and Weighing the Giants \citep[WtG,][]{WtG} clusters.

Although the masses were not derived homogeneously amongst the original works, an effort was made in \citet{serenocomalit} to select the most comparable mass estimates. Only masses reconstructed assuming spherical symmetry were considered, clusters without optical, X-ray or SZ counterpart were excluded and when the same authors or collaborations had published several estimates for the same cluster along a refinement process, only the latest result was considered. In addition, all the masses were standardised to the same cosmology (a flat $\Lambda$CDM cosmology with $\Omega_{\mathrm{m}} = 0.3$ and $H_{0} = 70$ km/s/Mpc) and were given at the overdensities of 2500, 500, 200 as well as at the virial radius. We define $R_{\Delta}$ as the radius at which the mean mass density of the cluster is $\Delta$ times the critical density of the Universe at its redshift, $\rho_{\mathrm{crit}} =3H(z)^{2}/(8\pi G)$, with $H(z)$ the Hubble function. We consider only the masses at an overdensity of $\Delta = 500$. For some cases, the masses given in the original papers had to be extrapolated following the density profile adopted in the original paper or with a Navarro-Frenk-White \citep[NFW,][]{navarro} model.

\subsubsection{XMM-\textit{Newton} sample with the reference X-ray pipeline}
\label{sec:xmm}

Regarding the HSE masses, we built a sample of clusters with masses reconstructed from XMM-\textit{Newton} data and following the same procedure, hereafter XMM-\textit{Newton} or reference X-ray pipeline. Thus, an homogeneous method was applied consistently to the full sample. This pipeline has already been used in previous works \citep{pratt2010, bartalucci2017, ruppin1, keruzore, munoz2022}, proving the reliability of the method.

As described in \citet{Bartalucci_2017}, the X-ray raw data were processed using the standard procedures with the XMM-\textit{Newton} Science Analysis System (SAS) pipeline. The electron density and temperature profiles were reconstructed following the correction and deprojection methods detailed in \citet{pratt2010} and \citet{bartalucci2018}. To obtain the HSE mass profiles, the electron density and temperature were combined in the Monte Carlo procedure described in \citet{democles2010}. The binned HSE mass profiles were interpolated to define the $M_{500}$ masses used in this work. Based on the same XMM-\textit{Newton} data, two differently estimated $M_{500}$ are available per cluster: masses derived from a X-ray calibrated scaling relation \citep{arnaud10} and masses estimated from a forward NFW profile fit to the density and temperature profiles. We do not use these two masses in our main analysis, but they are employed to investigate the consistency of all three estimates in Appendix~\ref{sec:otherxmm}. Amongst the clusters with XMM-\textit{Newton} data, we distinguish three different subsamples along the redshift. 

\subsubsection*{Low-$z$ clusters: ESZ+LoCuSS}
Many of the low redshift ($z<0.5$) clusters detected by \textit{Planck} were also observed by XMM-\textit{Newton}. It is the case of the 62 \textit{Planck} Early Sunyaev-Zel’dovich (ESZ) clusters \citep{planck2011esz}, whose HSE masses were reconstructed with X-ray data in \citet{planck2011eszxmm}. Similarly, based on the Local Cluster Substructure Survey (LoCuSS\footnote{\url{http://www.sr.bham.ac.uk/locuss/home.php}}) sample, \citet{pratt2013} reconstructed the HSE mass of 19 clusters. 

\subsubsection*{Intermediate-$z$ clusters: LPSZ}
The LPSZ stands for the NIKA2 SZ Large Programme \citep{mayet2020, perotto2022}. It is a high angular resolution follow-up of $45$ clusters of galaxies detected with the Atacama Cosmology Telescope \citep[ACT,][]{hilton2018,hilton2020} or the \textit{Planck} satellite \citep{planck2016}. The LPSZ follow-up combines high-resolution SZ data from the NIKA2 instrument \citep{adam1,perotto} with X-ray XMM-\textit{Newton} observations and covers a redshift range between 0.5 and 0.9. Studies on individual clusters from the LPSZ sample have already been published \citep{ruppin1, keruzore, munoz2022}, illustrating the joint SZ and X-ray analysis. Even though this sample was designed to be followed-up in SZ using NIKA2, we emphasise that in this work we do not make use of any SZ data for the mass estimation procedure. Instead, we consider the HSE masses obtained from XMM-\textit{Newton} data only. 

\subsubsection*{High-$z$ clusters: Bartalucci+2018}
\citet{bartalucci2017} and \citet{bartalucci2018} were able to go beyond $z=0.9$ and measure the HSE mass of five individual clusters from resolved mass profiles. Given the difficulties related to the high redshift of the clusters, XMM-\textit{Newton} data were combined with \textit{Chandra} observations. Although supplementary \textit{Chandra} data was added, we consider these masses as homogeneous with respect to the ESZ+LoCuSS and LPSZ samples since the same reconstruction pipeline was employed. However, special care is taken in our analyses when studying the impact of these clusters. Authors in \cite{bartalucci2018} also indicate that the mass estimate for the SPT-CLJ2106-5844 cluster is not reliable, therefore, we exclude it from our analyses.\\

\subsection{\textit{Comparison} sample}
\label{sec:allsamples}
The mass estimate of a cluster often varies from analysis to analysis, because of differences related to raw data or to the mass reconstruction method. In order to try to account for possible systematic biases in the CoMaLit and the reference X-ray pipeline masses, we gathered as many as possible HSE and lensing mass estimates from the literature for the clusters in our \textit{homogeneous} sample. Again, we made sure that the masses in the chosen studies were measured on resolved profiles, excluding masses derived from scaling relations. We only considered HSE masses obtained from X-ray data. Comparing to HSE masses that use SZ data or scaling relations is also of great interest, but it would be an independent analysis in itself and beyond the scope of this paper \citep[see, for example,][]{Hoekstra_2015, sereno2017a, Sereno_2017, schellenberger}. For lensing, in addition to the weak lensing masses, we also compared to masses reconstructed from the combination of strong and weak lensing signal. 

We present in the following a brief description of this \textit{comparison} sample, highlighting the distinctive characteristics of each analysis. We refer the reader to the original works for more details.

\subsubsection{Ettori+2010}
In \citet{ettori2010} \citep[and the \textit{Corrigendum},][]{ettori2010corrigendum}, the authors reconstructed the HSE mass of 44 clusters with redshifts $0.092 < z< 0.307$ using XMM-\textit{Newton} observations. They employed two different methods (M1 and M2) and gave the results in units of $R_{500}$. We converted the $R_{500}$ values into $M_{500}$ masses. The main caveat of these results is that profiles were extrapolated to reach $R_{500}$ assuming an NFW profile. As coordinates of the assumed centres of the clusters are not given in \citet{ettori2010}, we took them from \cite{Yuan_2022}\footnote{\url{http://zmtt.bao.ac.cn/galaxy_clusters/dyXimages/newton.html} } and when missing, from the 4XMM-DR9 source list\footnote{\url{http://xmmssc.irap.omp.eu/Catalogue/4XMM-DR9/4xmmdr9_obslist.html}}. 

\subsubsection{Landry+2013}
In \citet{Landry_2013} the HSE masses of 35 clusters with redshifts between $0.152 < z < 0.3017 $ were obtained using \textit{Chandra} data. Two different mass estimates are given in the paper: either using the Vikhlinin model or the polytropic equation of state. According to the authors, the profiles of seven clusters required `slight' extrapolation to reach $R_{500}$. Again, the coordinates of the assumed centres of the clusters are not given in \citet{Landry_2013}, so most of coordinates were taken from \cite{ebeling1998}. When missing, position coordinates of clusters were found by querying in the Simbad-CDS portal\footnote{\url{http://simbad.u-strasbg.fr/simbad/}} with the cluster name given in Table~1 in \citet{Landry_2013}.

\subsubsection{LoCuSS}
The aforementioned LoCuSS sample contains in all 50 clusters, with $0.152 < z < 0.3$ \citep{Smith_2015}. For our mass comparisons, we used the LoCuSS HSE masses published in \citet{martino2014} and the lensing masses from \citet{okabesmith2015}. The HSE masses were reconstructed with \textit{Chandra} data for 43 clusters and with XMM-\textit{Newton} observations for 39. For some clusters both estimates are available. Central coordinates of clusters were also taken from \citet{martino2014}. The analysis in \citet{zhang} studied 12 out the 50 clusters with XMM-\textit{Newton} and Subaru data. The lensing masses published in \citet{zhang} are equivalent to those in \citet{okabesmith2015}, but the HSE mass profiles were evaluated at the $R_{500}$ corresponding to the lensing analyses. We, therefore, gave preference to the results in \citet{okabesmith2015} and \citet{martino2014} and restricted the LoCuSS masses to the estimates in the latter two studies.

\subsubsection{Mahdavi+2008}
Uniformly estimated masses of 18 clusters were published in \citet{Mahdavi_2008}. Lensing masses were obtained as in \citet{hoekstra2007}, but with the photometric redshift distributions from \citet{Ilbert2006}.  The lensing mass reconstruction was done with a method based on aperture mass estimation, that is, obtaining first projected masses, and subsequently deprojecting by assuming an NFW density model and the concentration-mass scaling relation from \citet{bullock2001}. For the HSE masses, \textit{Chandra} observations were used. As indicated in Table~2 in \citet{Mahdavi_2008}, for 14 out of the 18 clusters the HSE masses at $R_{500}$ were obtained from extrapolation and all of them were measured at the lensing $R_{500}$.

\subsubsection{Mahdavi+2013}
\label{sec:mahdavi}
In \citet{Mahdavi_2013} authors studied a sample of 50 clusters with redshift $0.152 < z<0.55$. The clusters correspond to the CCCP sample. The HSE masses were reconstructed from a combined analysis of XMM-\textit{Newton} and \textit{Chandra} data. For the same sample, lensing estimates were obtained in \citet{hoekstra2012}, using CFH12k and Megacam data from the Canada-France-Hawaii Telescope. HSE masses were measured at the $R_{500}$ obtained from lensing masses.

\subsubsection{Israel+2014}
The analysis in \citet{israel} contains eight clusters with redshift $0.35 < z < 0.80$. The lensing masses were obtained from an NFW fit to the tangential shear profiles of clusters, assuming a mass-concentration relation. To reconstruct the HSE mass, the authors used the electron density profiles of individual clusters, which were estimated from \textit{Chandra} surface brightness maps. The temperature profile of individual clusters being more challenging to obtain, the authors combined the \textit{Chandra} data of all clusters in the sample to reconstruct a single global temperature profile for the whole sample. The HSE masses in \citet{israel} were also evaluated at the $R_{500}$ measured from lensing mass profiles. 

\subsubsection{LPSZ+CLASH}
Within the LPSZ programme, \citet{munozproceedingclash}, \citet{ferragamo}, and \citet{munoz2022} estimated the lensing mass for three clusters in the sample in common with the Cluster Lensing And Supernova survey with Hubble \citep[CLASH,][]{postman, zitrin1, zitrin2, zitrin3, zitrin4}. Masses were reconstructed by fitting a projected NFW mass density profile to the publicly available CLASH convergence maps \citep{zitrin1}. Given that two differently modelled convergence maps were provided, for some clusters two lensing mass estimates are available, named LTM and PIEMD+eNFW following the name of the method used to reconstruct the convergence maps. We also considered the lensing masses published in \citet{umetsu2014} and \citet{merten2015} for the same clusters.

\subsubsection{Bartalucci+2018}
In \citet{bartalucci2018} authors studied the HSE-to-lensing mass bias of five SPT clusters. The weak lensing masses were obtained in \citet{Schrabback_2017} using Hubble Space Telescope (HST) observations. The profiles were centred in the X-ray peak or the SZ peak \citep[indicated in Table~1 in ][]{Schrabback_2017}, giving two different lensing mass estimates per cluster. 


\section{Selection and characterisation of the sample}
\label{sec:systematic}
In this section, we present the comparison of the XMM-\textit{Newton} and CoMaLit mass estimates to the results from other works presented in Sect.~\ref{sec:allsamples}. We briefly describe the procedure used to match and select clusters from different catalogues, and then quantify the scatter based on the comparisons of several mass measurements for each cluster across our sample. Finally, we build the \textit{reference} sample with the XMM-\textit{Newton} and CoMaLit masses that we use for the rest of the analysis.

\subsection{Matching clusters}
\label{sec:matching}

We matched clusters from different catalogues on the basis of their coordinates. We considered that two entries in two distinct catalogues correspond to the same cluster for angular separations smaller than $400^{\prime \prime}$. We further verified every match by checking the redshifts given in the different catalogues. We identified suspicious mismatching between A1606 ($z=0.0963$) and A2029 ($z=0.078$) and excluded it.

At the same time, we discarded clusters that appear as one object in some catalogue and as a combination of multiple substructures in another. For example, the cluster A1758 in \cite{Landry_2013} has four entries in the LC$^{2}$-\textit{single} catalogue: A1758S, A1758NW, A1758N, A1758NE. Similarly, we excluded A222, A223N, and A223S. In addition, we identified and discarded A750 (present in CoMaLit, LoCuSS, Mahdavi+2013, and Mahdavi+2008 catalogues), whose mass estimate can not be reliable since it is superimposed along the line of sight with MS0906+11 \citep{Geller_2014}.

We summarise in Table~\ref{tab:data} the overlap between the \textit{homogeneous} clusters in XMM-\textit{Newton} and CoMaLit samples and those from other works presented in Sect.~\ref{sec:allsamples}. For 36 of the XMM-\textit{Newton} and 82 of the CoMaLit clusters we identified other HSE and lensing mass estimates\footnote{Since the LC$^2$ catalogue is a compilation of masses from many works in the literature, it is not surprising that some CoMaLit masses are directly the estimates published in other works. It is the case for some LoCuSS clusters.}.

\renewcommand{\arraystretch}{1.4}
    \scriptsize
    \begin{table*}[]

      \scriptsize
      \centering
        \caption{Summary of the amount of clusters in the each of the \textit{comparison} samples and their overlap with the \textit{homogeneous} XMM-\textit{Newton} and CoMaLit clusters. }
        \begin{tabular}{c|c|c|c|c|c}

          \hline
          \hline

          Sample &  Redshift & Type of mass & \# of clusters &  \# of clusters in common  & \# of clusters in common \\
          ~ & ~ & ~& ~& with the XMM-\textit{Newton} sample &  with the CoMaLit sample \\\hline
           
           Ettori+2010 & $0.092 < z < 0.307$ & HSE & 44 & 24 & ~\\
           Landry+2013 & $0.152 < z < 0.3017$ & HSE & 35 & 19 & ~ \\
           LoCuSS & $0.152 < z < 0.3$ & HSE and lensing & 50 & 22 & 45 \\
           Mahdavi+2013 & $0.152 < z < 0.55$ & HSE and lensing & 50 & 18 & 44 \\
           Mahdavi+2008 & $0.170 < z< 0.547$ & HSE and lensing & 18 & 11 & 17 \\
           Israel+2014 & $0.35 < z < 0.80$ & HSE and lensing & 8 & 0 & 8 \\
           LPSZ+CLASH & $0.55 < z<0.89$ & lensing & 3 & ~ & 3\\
           Bartalucci+2018 & $0.933 < z<1.066$ & lensing & 4 & ~ & 3 \\\hline
           All & ~ & ~ &~  & 94 & 120 \\
           All without repetition & ~ & ~ & ~ & 36 & 82 \\
           \hline    
        \end{tabular}
        \vspace*{0.2cm}  
        \begin{tablenotes}         
        \small
         \item \textbf{Notes.} We also report the total amount of matches, that is, the data points in Fig.~\ref{fig:initialselection} and the number of different objects.
        \end{tablenotes}        
        \vspace*{0.2cm}     
        \label{tab:data}
    \end{table*}
\normalsize

\subsection{Estimation of systematic dispersion}

\begin{figure*}[h]
        \begin{minipage}[b]{0.5\textwidth}
        \includegraphics[trim={10pt 0pt 0pt 0pt},scale=0.33]{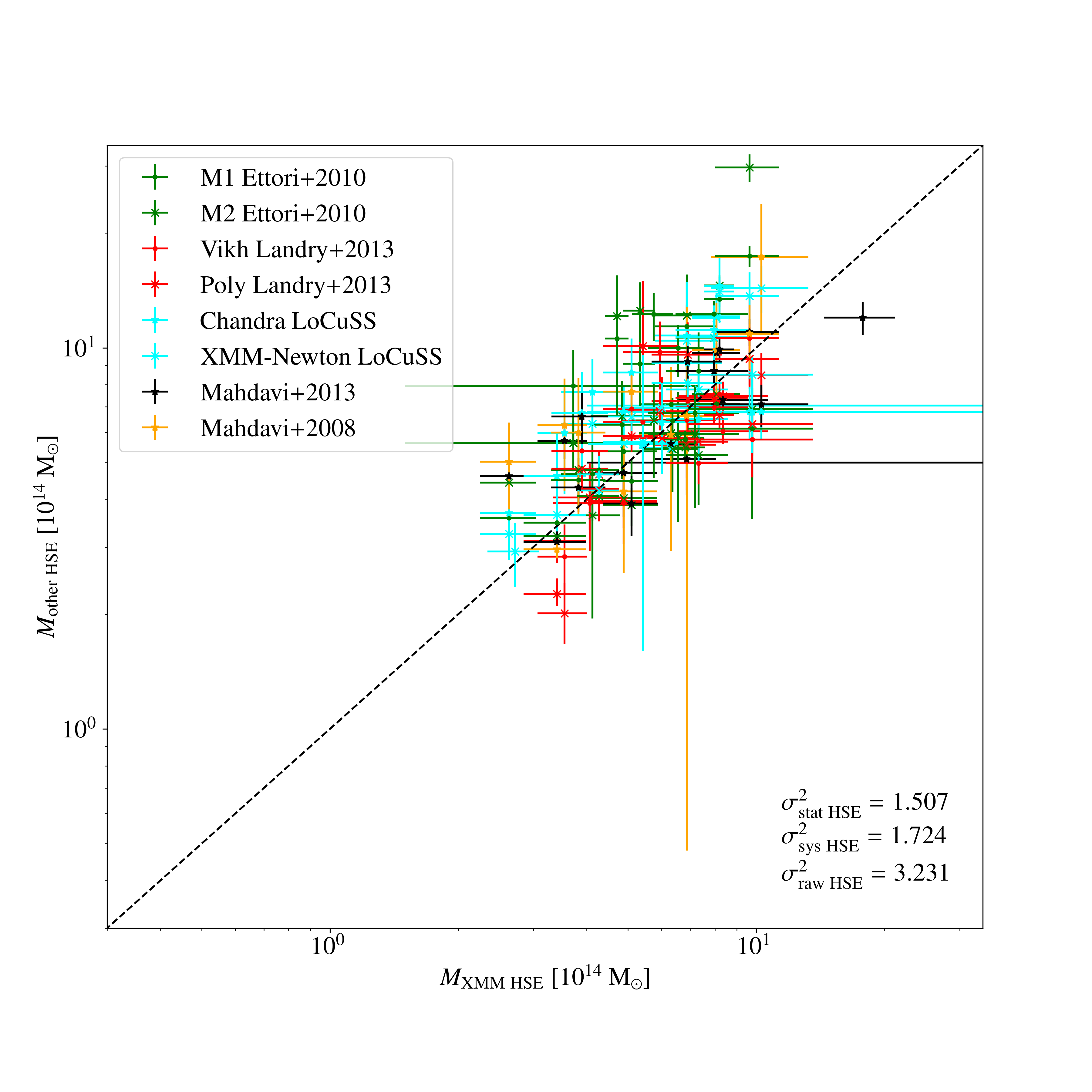}
        \end{minipage}
        \hfill
        \begin{minipage}[b]{0.5\textwidth}
        \includegraphics[trim={10pt 0pt 0pt 0pt},scale=0.33]{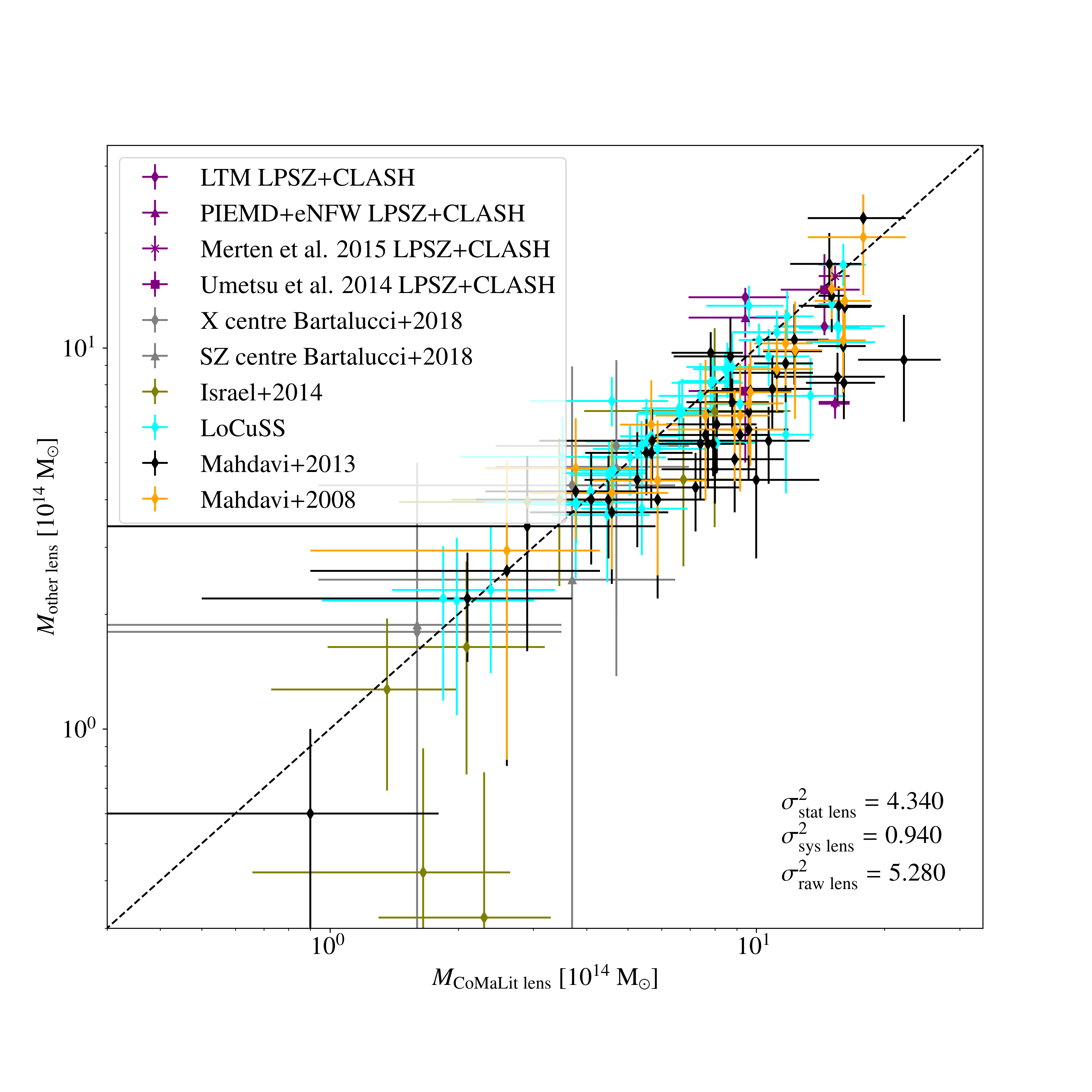}
        \end{minipage}
        \caption{Relation between HSE (left) and lensing (right) masses from the \textit{homogeneous} samples in this work (XMM-\textit{Newton} and CoMaLit) with respect to other estimates from the literature (\textit{comparison} sample). Each colour indicates a different analysis and several results from the same work are differentiated by using different markers. The black dashed lines show the one-to-one relation. We give the statistical, systematic, and raw variances as defined in the text. All the variances are in units of $(10^{14}$ M$_{\odot})^2$. }
    \label{fig:initialselection}
\end{figure*}

We present in the left panel in Fig.~\ref{fig:initialselection} the relation between X-ray HSE masses obtained with the reference X-ray pipeline (\textit{homogeneous} masses) with respect to other X-ray HSE masses from the literature (\textit{comparison} sample). In the right panel, we show the relation between lensing masses from different works with respect to the estimates summarised in CoMaLit. Each colour represents one of the samples described in Sect.~\ref{sec:allsamples} and different estimates of the same work are differentiated with markers. The black dashed line shows the one-to-one relation.

Overall, the agreement between the samples is reasonable, with a significant dispersion around the 1:1 relation. We identify some clusters for which the mass estimates differ significantly. These are Abell521, Abell2390, Abell2163 in X-rays and RXJ1347.5-1145, CL1641, CL1701 in the lensing masses comparison. The cluster shown with the green marker on top of the left panel in Fig.~\ref{fig:initialselection} is Abell2390 and, despite its departure from the 1:1 relation, we do not have strong arguments for excluding it. For lensing masses (in the right panel in Fig.~\ref{fig:initialselection}) there seem also to be a hint of some bias that we do not propagate hereafter. We verified that the bias does not correlate with a \textit{comparison} sample in particular, but rather with high mass clusters. Further investigation would be needed to understand this trend.

To quantify the systematic dispersion with respect to the 1:1 relation, we followed Eq.~3 and 4 in \citet{pratt2009}. For the $N_{\mathrm{lens}} = 120$ matched entries between the CoMaLit catalogue and the other lensing samples (Table~\ref{tab:data}), we defined the raw variance as
\begin{equation}
    \sigma_{\mathrm{raw \; lens}}^2 = \frac{1}{N_{\mathrm{lens}}-2}\sum_{i=1}^{N_{\mathrm{lens}}} w_i(M_{\mathrm{other\; lens}} - M_{\mathrm{CoMaLit\; lens}})^2,
\end{equation}
where $w_{i}$ is the weight of each cluster and $ M_{\mathrm{CoMaLit\; lens}}$ and $M_{\mathrm{other\; lens}}$ are the lensing mass in the CoMaLit catalogue and in a different analysis, respectively. The weight given to each cluster is 
\begin{equation}
  w_i = \frac{1/\sigma_i^2}{1/N\sum_{j=1}^{N}1/\sigma_j^2},
\end{equation}
using $\sigma_i^2 = \delta_{M_{\mathrm{other\; lens}}}^2 +  \delta_{M_{\mathrm{CoMaLit\; lens}}}^2$, the sum of the uncertainties related to each cluster. The $\sigma_{\mathrm{raw \; HSE}}^2 $ was measured in an equivalent way using the HSE masses and uncertainties of each cluster, $M_{\mathrm{XMM\; HSE}}$ and $M_{\mathrm{other\; HSE}}$ and $\delta_{M_{\mathrm{XMM\; HSE}}}^2$ and $\delta_{M_{\mathrm{other\; HSE}}}^2$.

The statistical error associated to the masses was obtained for both lensing and X-ray masses with the above-mentioned weight, $ w_i$, and $ \sigma_i ^2$:
\begin{equation}
    \sigma_{\mathrm{stat}}^2 = \frac{1}{N-2}\sum_{i=1}^{N} w_i \sigma_i ^2 = \frac{1}{N-2}\sum_{i=1}^{N} \frac{N}{\sum_{j=1}^{N}1/\sigma_j^2}.
\end{equation}
This allowed us to define the systematic scatter, that is, the excess of scatter in the raw variance not explained by the statistical uncertainties, as
\begin{equation}
  \sigma_{\mathrm{sys}}^2 = \sigma_{\mathrm{raw}}^2 - \sigma_{\mathrm{stat}}^2.
  \label{eq:sysdef}
\end{equation}

We report in Fig.~\ref{fig:initialselection} (and in Table~\ref{tab:scatters}) the statistical, systematic, and raw scatter for the HSE and lensing masses. The raw dispersion of lensing masses ($ \sigma_{\mathrm{raw \; lens}}^2 = 5.280  \times ( 10^{14}\; \mathrm{M}_{\odot} )^{2}$) is larger than HSE ones ($ \sigma_{\mathrm{raw \; HSE}}^2 = 3.231  \times ( 10^{14}\; \mathrm{M}_{\odot} )^{2}$) and the uncertainties of individual lensing masses being larger, the statistical dispersion is also larger ($ \sigma_{\mathrm{stat \; lens}}^2 = 4.340  \times ( 10^{14}\; \mathrm{M}_{\odot} )^{2}$ and $ \sigma_{\mathrm{stat \; HSE}}^2 = 1.507  \times ( 10^{14}\; \mathrm{M}_{\odot} )^{2}$). Nevertheless, the error bars of HSE masses are not large enough to cover the excess of scatter around the 1:1 relation, making the systematic scatter for HSE masses ($ \sigma_{\mathrm{sys \; HSE}}^2 = 1.724 \times ( 10^{14}\; \mathrm{M}_{\odot} )^{2}$) larger than for lensing ($ \sigma_{\mathrm{sys \; lens}}^2 = 0.940 \times ( 10^{14}\; \mathrm{M}_{\odot} )^{2}$). 

As mentioned in the description of each sample in Sect.~\ref{sec:allsamples}, HSE masses in some works were evaluated at the $R_{500}$ obtained from lensing. We checked the impact of excluding such estimates from the analysis. In the left panel in Fig.~\ref{fig:initialstrangemiscenter}, we present the relation between the XMM-\textit{Newton} reference pipeline masses and X-ray masses from the \textit{comparison} sample without accounting for $M^{\mathrm{HSE}}(<R_{500}^{\mathrm{lens}})$ estimates \citep[that is, without][]{Mahdavi_2008, Mahdavi_2013}. The statistical, raw, and systematic variances change by 0.4, 5, and 10\%, respectively. Hence, taking a conservative approach, in the following sections we consider the largest systematic scatter values obtained.


\subsection{\textit{Reference} sample}
\label{sec:refsample}
\begin{figure*}[h]
    \centering
    \includegraphics[scale=0.28]{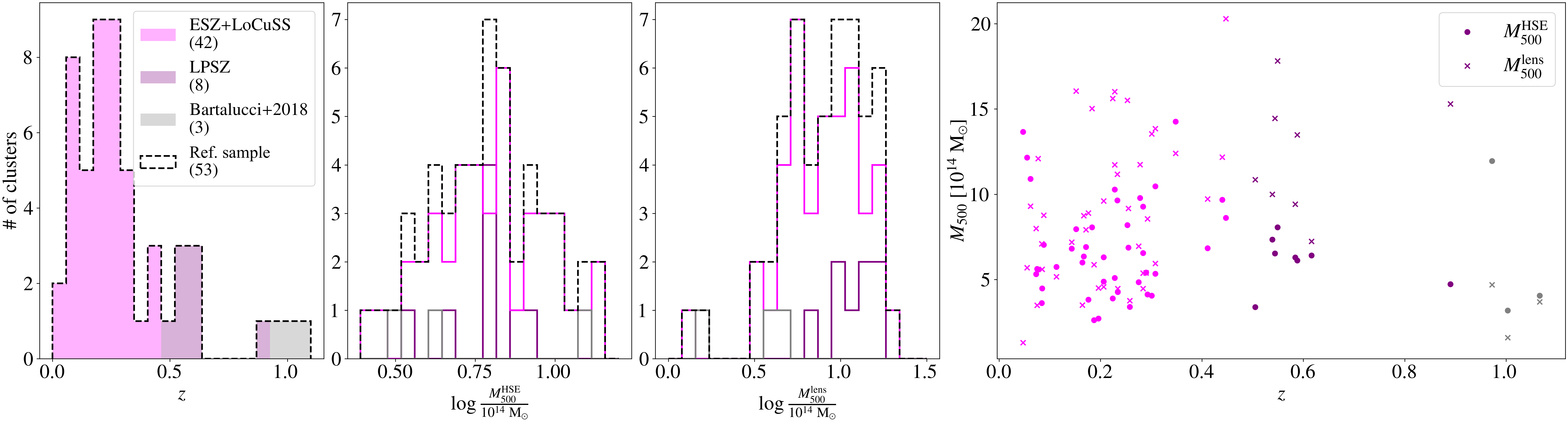}
    \caption{Main characteristics of the 53 clusters in the \textit{reference} sample. Histograms in the left panels show the redshift, HSE mass, and lensing mass distributions. We show in purple, magenta, and grey the distributions for ESZ+LoCuSS, LPSZ, and Bartalucci+2018 clusters, respectively. The black dashed lines represent the distributions of the whole sample. In the right panel we show the HSE and lensing masses as a function of redshift for all the clusters.}
    \label{fig:referencesample}
\end{figure*}

Following the procedure described in Sect.~\ref{sec:matching}, we matched the clusters in the CoMaLit catalogue (Sect.~\ref{sec:comalit}) with the clusters with HSE masses from the XMM-\textit{Newton} reference pipeline (Sect.~\ref{sec:xmm}) and obtained an homogeneous sample composed of 65 clusters. Amongst the 65 clusters, 54 correspond to the ESZ+LoCuSS samples, eight clusters are from the LPSZ, and three from Bartalucci+2018. For these clusters, we performed additional checks and discarded: three clusters with senseless error bars (see Appendix~\ref{sec:uncertainties} for details), and ten clusters (one of them already excluded) for which X-ray and lensing mass reconstruction analyses had assumed very distant centres (see Appendix~\ref{sec:miscentring}).

As a result, our \textit{reference} sample contains 53 clusters with homogeneous HSE and lensing masses that can be used for comparisons (see Table~\ref{tab:all53}). We present in Fig.~\ref{fig:referencesample} a summary of the characteristics of the sample. The histograms in the left show the number of clusters with respect to redshift, HSE mass, and lensing mass. The right panel in Fig.~\ref{fig:referencesample} presents the clusters in the mass-redshift plane. While very few works in the literature go above $z=0.5$, 20\% of the clusters in our sample have redshifts higher than 0.5. However, the distribution in redshift of the sample is dominated by low-$z$ clusters.

After excluding the last 12 clusters (in Appendix~\ref{sec:uncertainties} and \ref{sec:miscentring}) from the XMM-\textit{Newton} and CoMaLit samples, we recalculated the scatter with respect to other HSE and lensing masses. Compared to Fig.~\ref{fig:initialselection}, the raw, statistical, and systematic dispersions remain of the same order, but the impact of individual clusters is again noticeable in the resulting values (less than 10\% of change, see Fig.~\ref{fig:initialstrangemiscenter} and Table~\ref{tab:scatters}). Therefore, we took the most conservative approach and considered that the systematic scatters to be accounted for in the XMM-\textit{Newton} and CoMaLit masses are the largest values we have found: $\sigma_{\mathrm{sys\; lens}}^2 = 1.202 \times ( 10^{14}\; \mathrm{M}_{\odot} )^{2}$ and $\sigma_{\mathrm{sys\; HSE}}^2 = 2.017 \times ( 10^{14}\; \mathrm{M}_{\odot} )^{2}$. We note that the clusters used for these calculations are not necessarily the 53 in our \textit{reference} sample, but the ones in common between XMM-\textit{Newton} and other X-ray samples and between CoMaLit and other lensing works (summarised in Table~\ref{tab:data}). We compare in Fig~\ref{fig:uncertcomparison} the systematic standard deviation values to the individual statistical uncertainties of the masses from the XMM-\textit{Newton} reference pipeline and the CoMaLit catalogue.

In the following sections, we investigate how the HSE-to-lensing mass bias and scaling relation change when accounting for these systematic scatters. In order to propagate the scatters to the final results, we consider that the uncertainties in the mass of each cluster are the quadratic sum of the measurement statistical uncertainties and the systematic scatters derived in this section. Thus, we have
\begin{equation}
    \delta_{\mathrm{lens}}^2 = \delta_{M_{\mathrm{CoMaLit\; lens}}}^2 + \sigma_{\mathrm{sys\; lens}}^2
    \label{eq:intrinlens}
\end{equation}
for the lensing masses, and
\begin{equation}
    \delta_{\mathrm{HSE}}^2 = \delta_{M_{\mathrm{XMM \; HSE}}}^2 + \sigma_{\mathrm{sys\; HSE}}^2
    \label{eq:intrinhse}
\end{equation}
for the hydrostatic ones.

This is a very conservative approach that assumes that the mass estimates from the X-ray reference pipeline and the CoMaLit catalogue may have an additional error (due to, for example, the used dataset or the mass reconstruction method) that can be quantified from the distance to other estimates. Such supplementary error is usually not considered in the literature. For this reason, we also perform the study without accounting for the systematic uncertainties. An alternative approach was considered in \citet{serenoettoricomalit} by separating the analysis in subsamples. A cross-validation of our results by subsamples is presented in Sect.~\ref{sec:caveats}.

\section{Direct HSE-to-lensing mass bias measurement}
\label{sec:bias}
\begin{figure}[!p]  
      \includegraphics[trim={0pt 0pt 0pt 0pt}, scale=0.215]{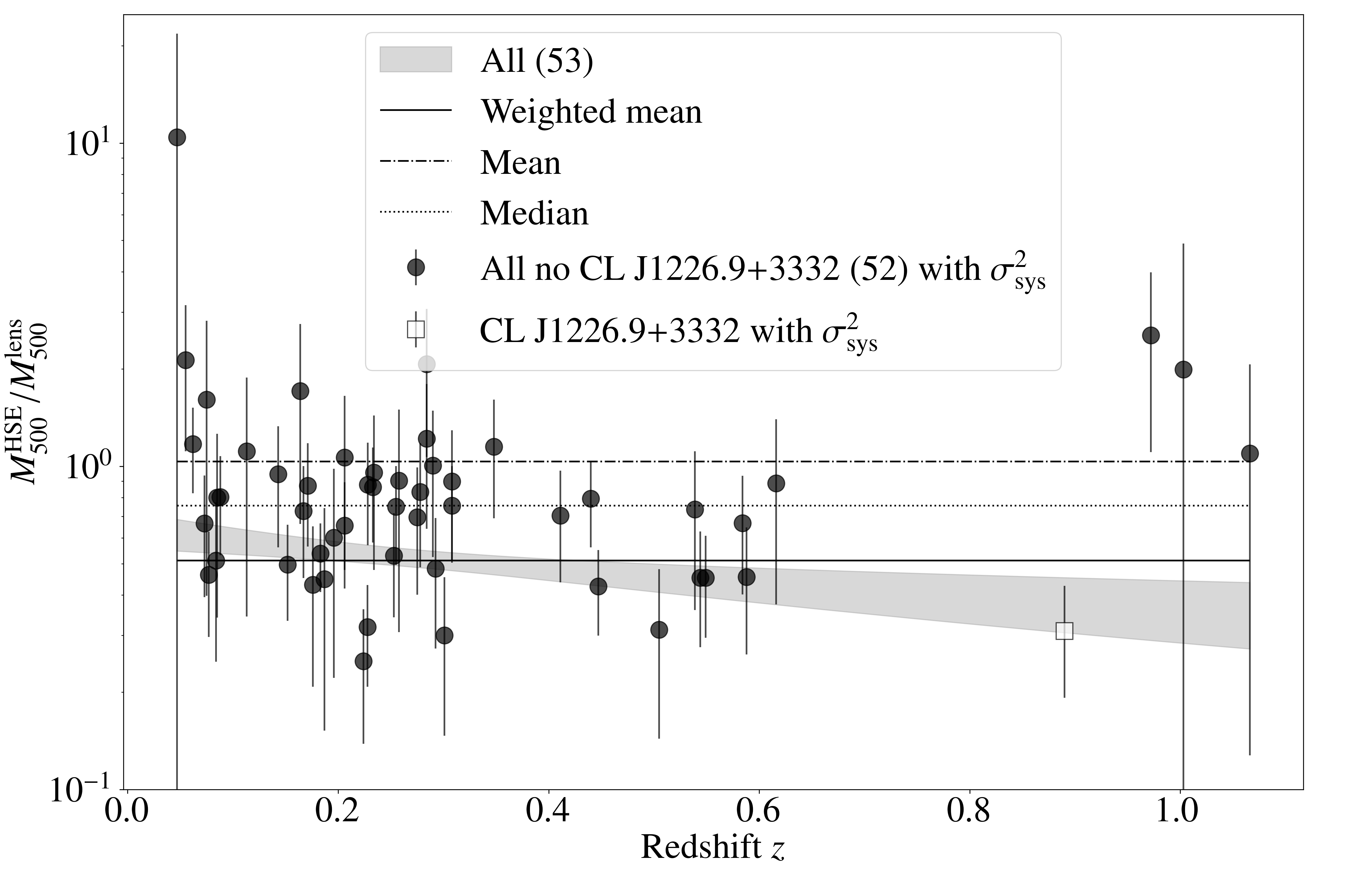}
      \includegraphics[trim={0pt 0pt 0pt 0pt}, scale=0.215]{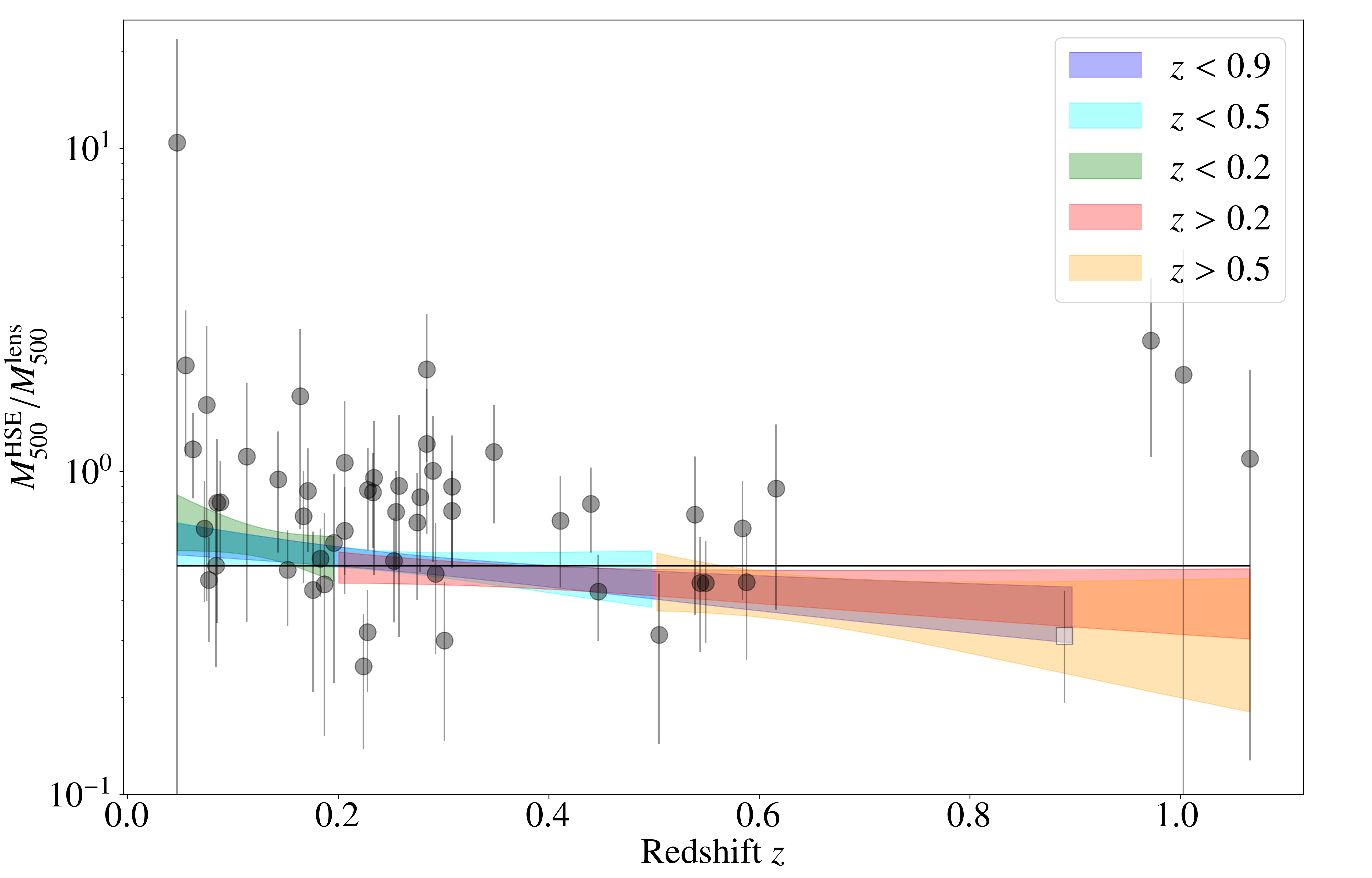}
      \includegraphics[trim={0pt 0pt 0pt 0pt}, scale=0.215]{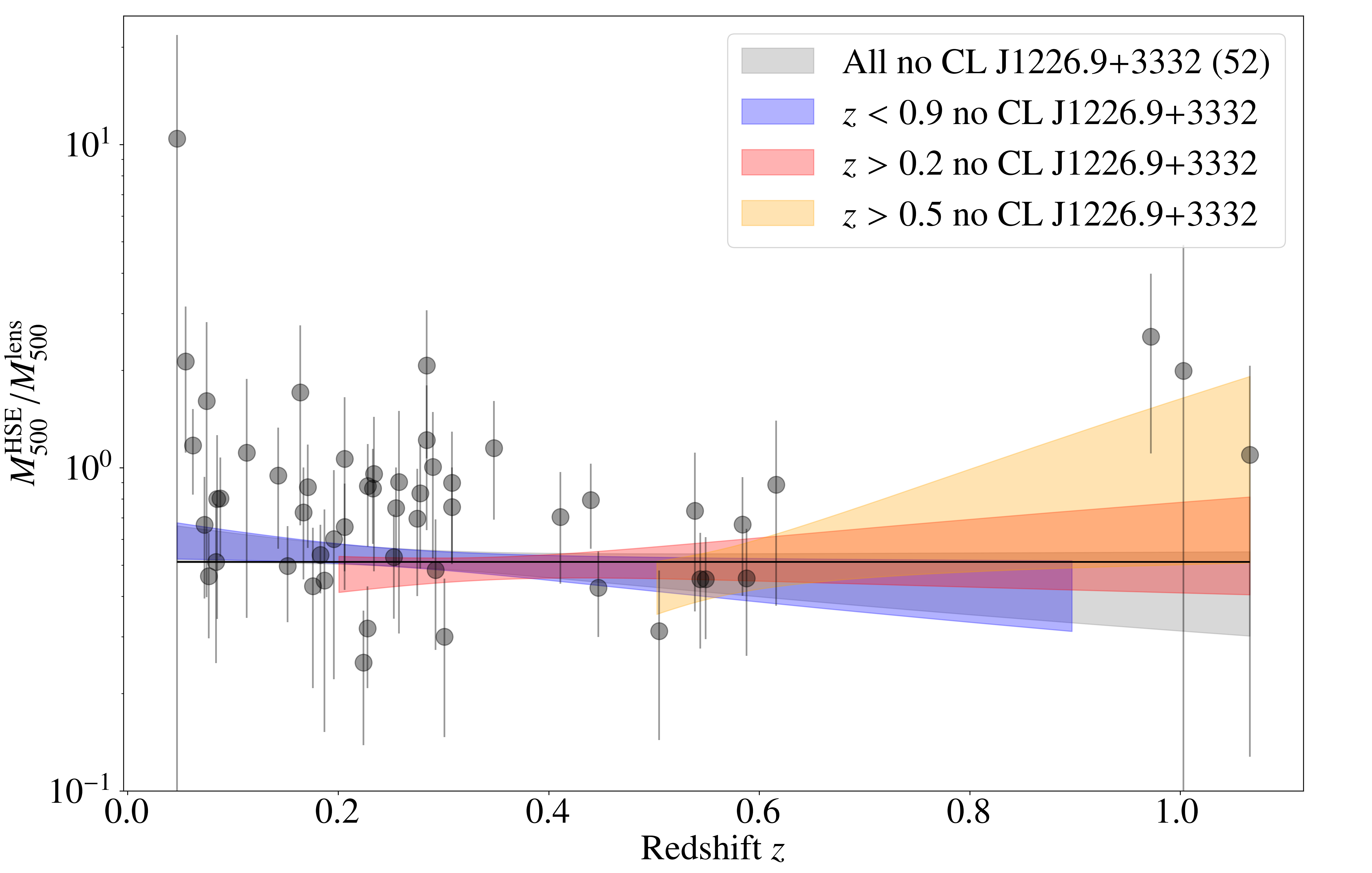}
      \caption{HSE-to-lensing mass ratio with respect to the redshift. Markers with error bars show the ratio of each cluster in the \textit{reference} sample with error bars accounting for the systematic uncertainty. Horizontal solid, dotted, and dash-dotted black lines give respectively the error weighted mean, median, and mean mass ratio for the data points. Shaded areas represent the 16th to 84th percentiles of the bias evolution model obtained by fitting different redshift ranges. Top: the bias evolution model obtained with the 53 clusters in the \textit{reference} sample. Centre: different colours indicate the models fitted to clusters in different redshift ranges. Bottom: grey, blue, red, and orange shaded areas show respectively the bias evolution model fitted to clusters along all the redshift range, at $z<0.9$, at $z>0.2$, and at $z>0.5$, excluding in all the cases the CL~J1226.9+3332 galaxy cluster.}
    \label{fig:biasfitref}
\end{figure}

\begin{figure}[!p]
    \centering
        \centering
        \includegraphics[trim={0pt 0pt 0pt 0pt},scale=0.29]{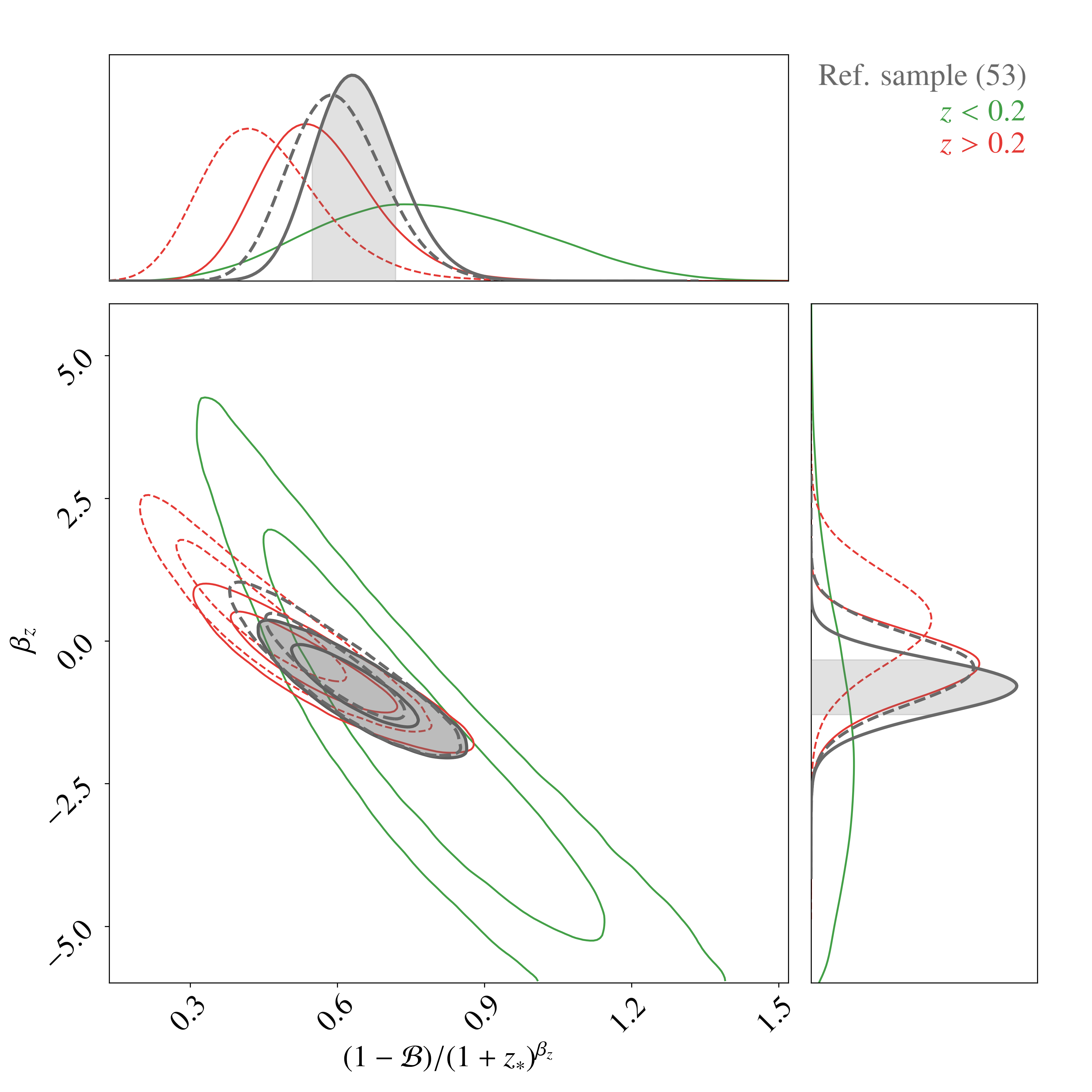}
        \caption{One-dimensional and two-dimensional posterior distributions of the parameters in the redshift dependent mass bias model, accounting for the $\sigma^{2}_{\mathrm{sys}}$ in the error bars. Different colours describe the results for the various samples presented in Table~\ref{tab:bias}. For a good visualisation, we only show in grey, green, and red the results for the whole sample, the $z<0.2$, and the $z>0.2$ ranges, respectively. Dashed distributions have been obtained excluding CL~J1226.9+3332 galaxy cluster.}
    \label{fig:biasfit}
\end{figure}

The bias of HSE masses with respect to lensing estimates is defined from the ratio of the masses,
\begin{equation}
  (1-b_{\mathrm{HSE/lens}}) = M_{500}^{\mathrm{HSE}}/M_{500}^{\mathrm{lens}}.
  \label{eq:lensbias}
\end{equation}  
For simplicity, in the rest of this paper we name the HSE-to-lensing mass bias without subscripts $b = b_{\mathrm{HSE/lens}}$.

As a first approach, and for comparison with other works in the literature, we directly compare the HSE-to-lensing mass ratio among the clusters of the \textit{reference} sample.
Following the parametrisation in \citet{salvati2019} and \citet{wicker}, we describe the redshift evolution of the HSE-to-lensing mass bias as

\begin{equation}
    M_{500}^{\mathrm{HSE}}/M_{500}^{\mathrm{lens}} (z) = (1-b) (z) = (1-\mathcal{B})\left(\frac{1+z}{1+z_{*}}\right) ^{\beta_{z}},
    \label{eq:biasmodel}
\end{equation}
where $ (1- \mathcal{B})$ is the bias normalised at the pivot redshift, $z_{*}$, and $\beta_{z}$ describes the evolution with redshift. As in \citet{salvati2019}, we take $z_{*}$ the median redshift value of the clusters in the analysed sample. In \citet{wicker} the pivot redshift is the mean of the sample. 

With the homogeneous HSE and lensing masses of the 53 clusters in the \textit{reference} sample, we perform a Markov chain Monte Carlo (MCMC) analysis to fit the model (Eq.~\ref{eq:biasmodel}) to data, using the \texttt{emcee} Python package \citep{foreman,goodman}. We consider uniform priors for the parameters, $ (1- \mathcal{B})\sim \mathcal{U}(0, 2)$ and $ \beta_{z}\sim \mathcal{U}(-8, 8)$, and assume a  Gaussian likelihood, uncorrelated between points. 

We show in Fig.~\ref{fig:biasfitref} the HSE-to-lensing mass ratio as a function of redshift for the 53 clusters in the \textit{reference} sample. Here error bars include systematic scatter following Eq.~\ref{eq:intrinlens} and \ref{eq:intrinhse}. The grey shaded area in the top panel indicates the 16th to 84th percentile region of the bias evolution model obtained from the posterior distributions of the fitted parameters. For comparison, the horizontal lines show the mean (dash-dotted line), median (dotted line), and error weighted mean (solid line) HSE-to-lensing mass ratio obtained with the 53 cluster masses. Posterior distributions of the fitted parameters are shown with grey contours in Fig.~\ref{fig:biasfit}. The best-fit values and uncertainties are given in the first row in Table~\ref{tab:bias}. We give $(1-\mathcal{B}) / (1+z_{*})^{\beta_{z}}$, which is the value of the bias at $z=0$. We also report the results without accounting for the systematic scatter of the lensing and HSE masses. As expected, when accounting for $\sigma_{\mathrm{sys}}^2$ the uncertainties of the posterior distributions are enlarged. 

   \renewcommand{\arraystretch}{1.4}    
     \scriptsize
    \begin{table*}[]
      \centering
        \caption{Best-fit values and uncertainties for the normalisation and redshift evolution parameters of the mass bias model in Eq.~\ref{eq:biasmodel} obtained for different subsamples of the \textit{reference} sample.}
        \scriptsize
        \begin{tabular}{c|c|c|c|c|c|c} 
          \hline
          \hline
             Sample & $\#$ of clusters & $z_{*}$&  \multicolumn{2}{|c|}{No $\sigma_{\mathrm{sys}}^2$} &  \multicolumn{2}{|c}{With $\sigma_{\mathrm{sys}}^2$} \\
             ~ & ~ & ~ & $(1-\mathcal{B}) / (1+z_{*})^{\beta_{z}}$ & $\beta_{z}$ & $(1-\mathcal{B}) / (1+z_{*})^{\beta_{z}}$ & $\beta_{z}$ \\ \hline
             
             \textbf{\textit{Reference} sample} &  \textbf{53} &  \textbf{0.253} &  $0.585_{-0.050}^{+0.059}$ &  $-0.797_{-0.373}^{+0.309}$ &  $\mathbf{0.632_{-0.074}^{+0.093}}$  &  $\mathbf{-0.787_{-0.529}^{+0.418}}$ \\
             $z<0.9$ & 50 & 0.234 & $0.591_{-0.050}^{+0.060}$ &  $-0.846_{-0.381}^{+0.306}$ & $0.642_{-0.079}^{+0.094}$ & $-0.860_{-0.528}^{+0.427}$ \\
             $z<0.5$ & 42 & 0.215 & $0.578_{-0.082}^{+0.095}$ & $-0.744_{-0.773}^{+0.724}$ & $0.618_{-0.112}^{+0.130}$ & $-0.661_{-0.948}^{+0.887}$\\
             $z<0.2$ & 19 & 0.113 & $0.716_{-0.144}^{+0.159}$ &   $-1.577_{-1.560}^{+1.589}$ &  $0.802_{-0.235}^{+0.216}$ &$-2.226_{-1.993}^{+2.515}$\\
             $z>0.2$ & 34 & 0.305 & $0.471_{-0.057}^{+0.071}$ & $-0.271_{-0.424}^{+0.344}$ & $0.543_{-0.092}^{+0.131}$ &$-0.392_{-0.668}^{+0.496}$ \\
             $z>0.5$ & 11 & 0.588 & $0.665_{-0.279}^{+0.620}$  & $-1.043_{-1.355}^{+1.020}$ & $0.692_{-0.356}^{+1.267}$ & $-0.987_{-2.249}^{+1.380}$ \\
             
             \textit{Ref.} no CL~J1226.9+3332 & 52 & 0.244 & $0.548_{-0.060}^{+0.069}$ &$-0.467_{-0.505}^{+0.453}$ & $0.594_{-0.086}^{+0.100}$ &$-0.483_{-0.639}^{+0.562}$\\
             $z<0.9$ no CL~J1226.9+3332 & 49 & 0.233 & $0.560_{-0.059}^{+0.071}$  &$-0.578_{-0.517}^{+0.443}$ & $0.610_{-0.085}^{+0.104}$ &$-0.610_{-0.648}^{+0.548}$\\
             $z>0.2$ no CL~J1226.9+3332 & 33 & 0.301 & $0.353_{-0.059}^{+0.076}$ &$0.855_{-0.650}^{+0.552}$ & $0.432_{-0.096}^{+0.135}$ &$0.452_{-0.886}^{+0.732}$ \\
             $z>0.5$ no CL~J1226.9+3332 & 10 & 0.586 & $0.060_{-0.015}^{+0.154}$& $4.593_{-2.867}^{+0.584}$ & $0.068_{-0.018}^{+0.350}$ &$4.398_{-4.097}^{+0.592}$\\\hline
             
        \end{tabular}
        \vspace*{0.2cm}
        \begin{tablenotes}         
        \small
        \item \textbf{Notes.} Columns 1 to 3 present the considered sample, the number of clusters, and the median redshift. Columns 4 to 7 give the best-fit values with 16th and 84th percentiles of the posterior distributions for parametres describing bias evolution, without (columns 4 and 5) and with (columns 6 and 7) the systematic scatters. In bold the values corresponding to the \textit{reference} sample accounting for the systematic scatters. 
        \end{tablenotes}        
        \vspace*{0.2cm}
        \label{tab:bias}
    \end{table*}
    
\normalsize

Due to the significant differences in the mass uncertainties and the non-uniform distribution of the clusters in redshift, certain subsamples might be driving the fit of the model. To check for these effects and investigate any dependence with redshift, we repeat the fit by considering clusters in different redshift ranges.

Considering only the clusters with $z<0.9$ (that is, those in ESZ+LoCuSS and LPSZ samples) and only those with $z<0.5$ (only ESZ+LoCuSS), the results are very close to the ones obtained with the \textit{reference} sample. This means that the grey result in Fig.~\ref{fig:biasfitref} and \ref{fig:biasfit} is most probably dominated by ESZ+LoCuSS clusters. Best-fit values and uncertainties for these two cases are given in Table~\ref{tab:bias}. The corresponding bias evolution models are shown in blue ($z<0.9$) and cyan ($z<0.5$) in the central panel in Fig.~\ref{fig:biasfitref}. 

We find more significant differences when considering only low redshift clusters ($z<0.2$, in green), or, when discarding them ($z>0.2$, in red). For low redshift clusters, the HSE masses at $z = 0$ are less biased with respect to lensing masses ($(1-\mathcal{B}) / (1+z_{*})^{\beta_{z}}$ closer to 1) than for the \textit{reference} sample, but the dependence on redshift is stronger. Exactly the opposite happens when fitting only $z>0.2$ masses: the HSE-to-lensing mass bias is larger at $z=0$ (smaller  $(1-\mathcal{B}) / (1+z_{*})^{\beta_{z}}$), but the redshift evolution is weaker (the absolute value of $\beta_{z}$ smaller). These conclusions agree with the results in \citet{wicker}, where the same cut in redshift is adopted. In \citet{Smith_2015}, the authors also reported a different tendency for \textit{Planck} cluster masses depending on the redshift, with a larger HSE-to-lensing bias value (smaller $1-b$) for \textit{Planck} masses at $z>0.3$, than the bias at $z<0.3$. However, these masses were inferred from the SZ-mass scaling relation and not measured from profiles. Nonetheless, in our analysis $\beta_{z}$ is compatible with no redshift evolution both for $z<0.2$ and $z>0.2$ subsamples (see posterior probability density contours in Fig.~\ref{fig:biasfit}).

As shown in Fig.~\ref{fig:referencesample}, the clusters at high redshift are rare in our sample, with a large gap between $z = 0.62$ and $z=0.89$. Only CL~J1226.9+3332, SPT-CLJ0615-5746, SPT-CLJ0546-5345, and SPT-CLJ2341-5119 are above $z = 0.62$. For CL~J1226.9+3332 the uncertainties on the bias are more than one order of magnitude smaller than the uncertainties of the three SPT clusters. We suspect that this single cluster may be forcing the bias towards lower values at high redshift. To test the impact that CL~J1226.9+3332 has on the fits, we repeat the analyses excluding it. The results without CL~J1226.9+3332 are shown, following the same colour scheme as before, in the bottom panel in Fig.~\ref{fig:biasfitref} and with dashed lines in Fig.~\ref{fig:biasfit}. We observe that $\beta_{z}$ varies significantly when excluding CL~J1226.9+3332 and it tends to be more compatible with no redshift evolution. At the same time, the bias at $z=0$ is slightly shifted towards lower values. All the results are summarised in Table~\ref{tab:bias}. 

The described direct HSE-to-lensing mass bias estimation method neglects the intrinsic scatters of the HSE and lensing mass estimates. As explained in \citet{serenoettoricomalit}, this could influence the resulting bias that relates HSE and lensing masses. For this reason, in the next section we take a different approach to estimate the HSE-to-lensing mass bias.


\section{HSE-to-lensing mass scaling relation}
\label{sec:SR}
\begin{figure*}
        \begin{minipage}[b]{0.5\textwidth}
        \includegraphics[trim={10pt 0pt 10pt 0pt},scale=0.6]{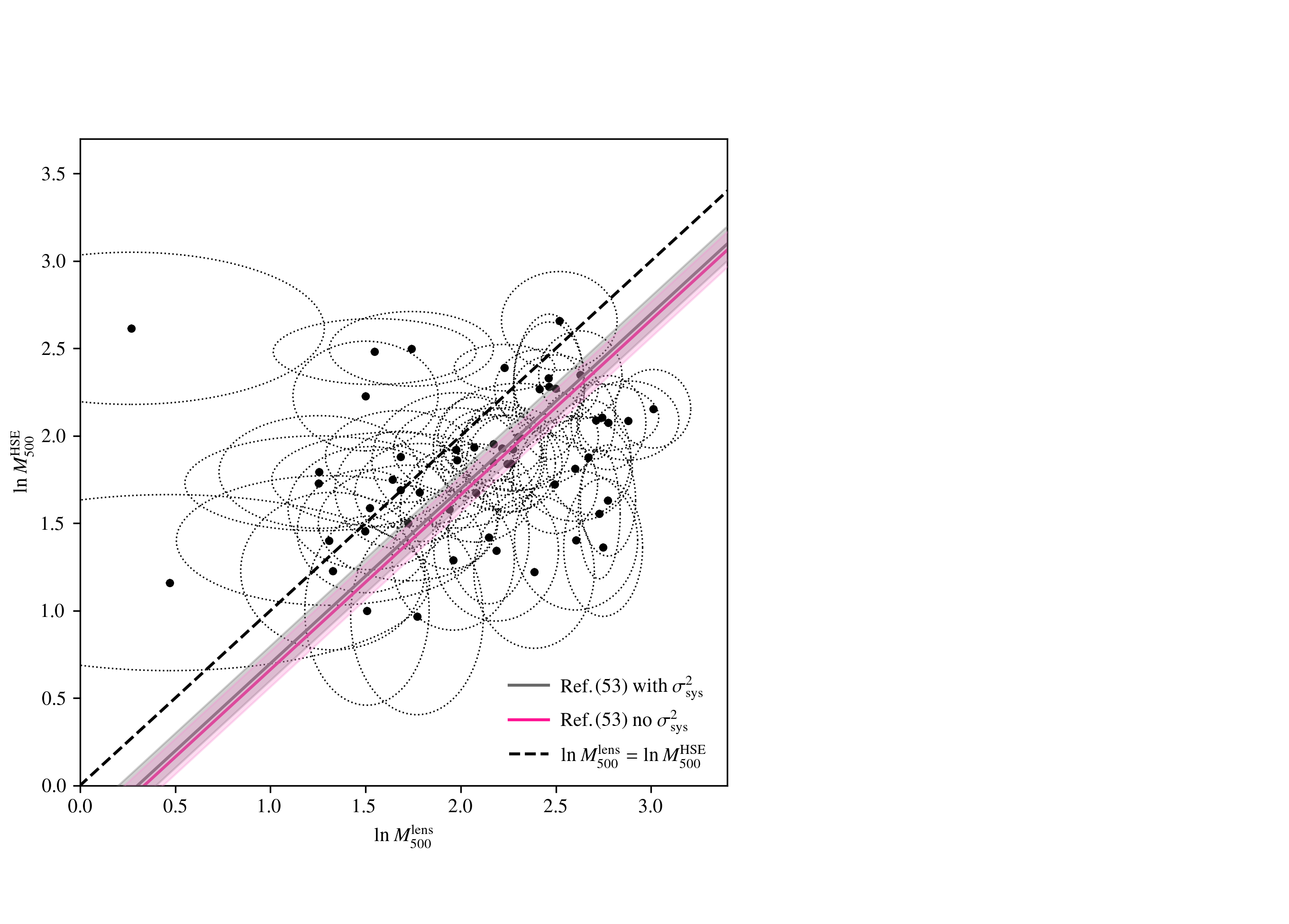}
        \end{minipage}
        \hfill
        \begin{minipage}[b]{0.5\textwidth}
        \includegraphics[trim={0pt 0pt 0pt 0pt},scale=0.3]{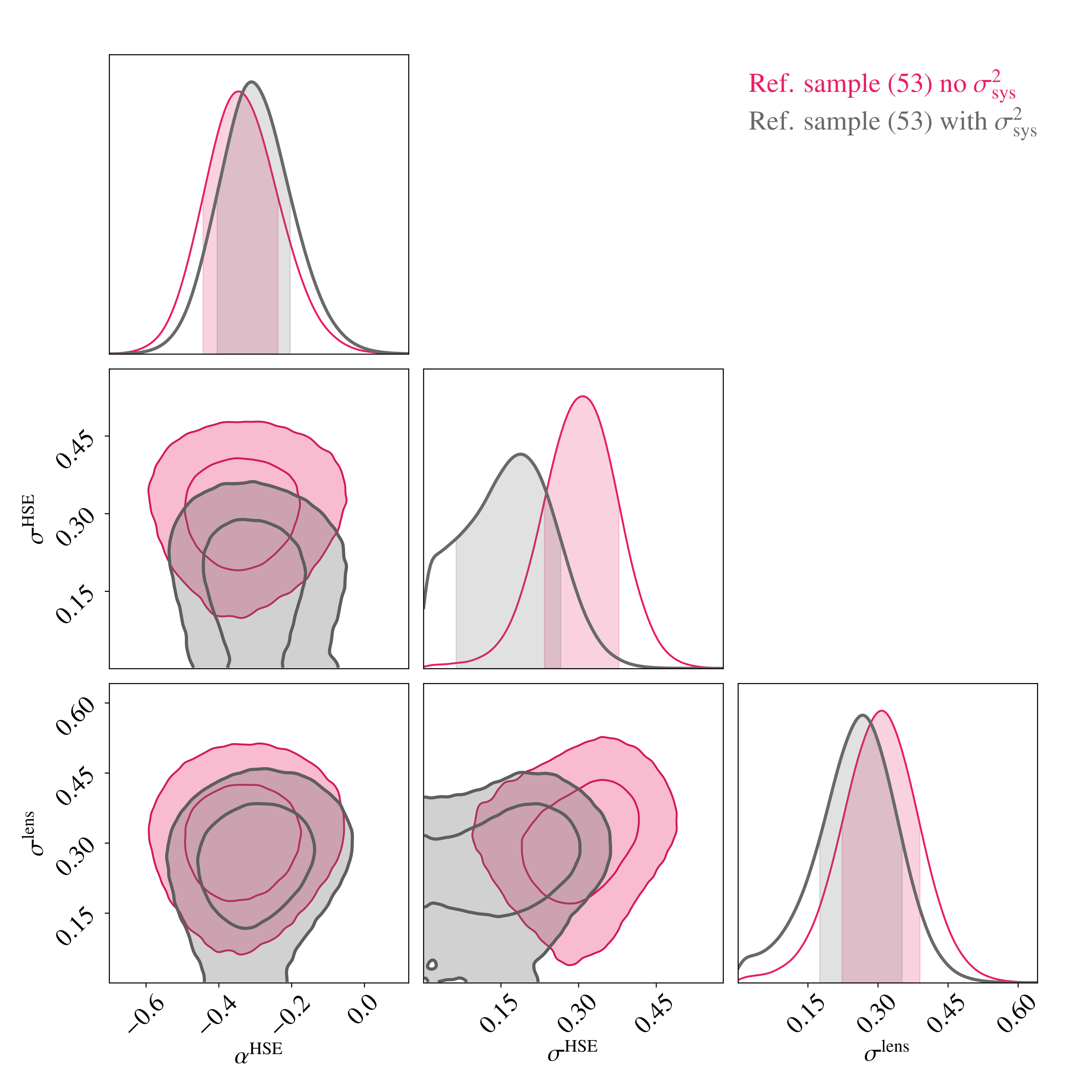}
        \end{minipage}
    \caption{Reference scaling relation ($\beta^{\mathrm{HSE}}= 1$) between HSE and lensing masses in the \textit{reference} sample. Data points with ellipses represent each cluster masses and uncertainties in both axes accounting for the systematic scatter. The pink line corresponds to the SR for the median value of parameters obtained without $\sigma_{\mathrm{sys}}$ and the solid grey line with $\sigma_{\mathrm{sys}}$. The shaded regions show the 16th and 84th percentiles and the black dashed line gives the one-to-one relation. The corner plots in the right panel are the posterior 1D and 2D distributions of the parameters in the SR, including (grey) or not (pink) systematic scatters.}
    \label{fig:refSRnosysandsys}
\end{figure*}

Estimating the scaling relation between HSE and lensing masses is an alternative way for measuring the HSE-to-lensing mass bias (Eq.~\ref{eq:lensbias}), together with the intrinsic scatter associated to HSE and lensing masses. We follow the methodology presented in \citet{serenoettoricomalit} and consider that both the HSE and the lensing masses are scattered and biased estimates of the true mass of clusters, such that
\begin{equation}
  \ln M^{\mathrm{lens}}  \pm \delta_{\mathrm{lens}}  = \alpha^{\mathrm{lens}}+ \beta^{\mathrm{lens}} \ln M^{\mathrm{True}} \pm \sigma^{\mathrm{lens}},
  \label{eq:srlens}
\end{equation}

\begin{equation}
  \ln M^{\mathrm{HSE}} \pm \delta_{\mathrm{HSE}}  = \alpha^{\mathrm{HSE}}+ \beta^{\mathrm{HSE}} \ln M^{\mathrm{True}} \pm \sigma^{\mathrm{HSE}}.
  \label{eq:srhse}
\end{equation}
Here $\delta_{\mathrm{lens}}$ and $\delta_{\mathrm{HSE}}$ are the measurement uncertainties associated with the logarithm of the lensing and HSE mass estimates for each cluster. The natural logarithm of the bias and the deviation from linearity are $\alpha$ and $\beta$, respectively. The intrinsic scatter of the lensing and HSE masses with respect to the true mass are given by $\sigma^{\mathrm{lens}}$ and $\sigma^{\mathrm{HSE}}$. All the masses in the arguments of logarithms are in $10^{14}\; \mathrm{M}_{\odot}$ units.
Authors in \cite{serenoettoricomalit} verified that the scatter and bias results do not vary if $\alpha^{\mathrm{lens}} = 0$ or $\alpha^{\mathrm{HSE}} = 0$ is considered, so following their work, we fix $\alpha^{\mathrm{lens}} = 0$.

We use the LInear Regression in Astronomy \citep[LIRA\footnote{\url{https://cran.r-project.org/web/packages/lira/}},][]{serenolira} R package and the \texttt{pylira}\footnote{\url{https://github.com/fkeruzore/pylira}} Python wrapper to perform the fit of the SR. LIRA performs the Gibbs sampling of a posterior distribution constructed from a MCMC fit based on a Bayesian hierarchical modelling. It can account for heteroscedastic measurement errors, intrinsic scatter, and time evolution of the SR.

\subsection{Reference scaling relation}

The SR of reference in this paper is built using the aforementioned 53 clusters in the \textit{reference} sample, assuming that both the lensing and the HSE masses scale linearly with the true mass, $\beta^{\mathrm{lens}}=1$ and $\beta^{\mathrm{HSE}}=1$, and that there is no evolution of the SR with redshift. The MCMC sampling is performed using 200 chains and $6\times 10^{6}$ steps, with a burn-in of the first half of the steps. Convergence is checked following the $\hat{R}$ test of \citet{gelmanrubin}. We take uniform priors for the free parameters: $\alpha^{\mathrm{HSE}} \sim \mathcal{U}(-4, 4)$, $\sigma^{\mathrm{HSE}}\sim \mathcal{U}(0, 10)$, $\sigma^{\mathrm{lens}}\sim \mathcal{U}(0, 10)$.

We present in the left panel in Fig.~\ref{fig:refSRnosysandsys} the HSE-to-lensing mass scaling relation obtained with the 53 clusters of the \textit{reference} sample. Data points correspond to each one of the clusters in the sample, with the ellipses in the figure indicating the error bars in both axes when considering the systematic scatter (see Eq.~\ref{eq:intrinlens} and \ref{eq:intrinhse}). We assume no correlation between both mass estimates. The grey and pink lines show respectively the scaling relation accounting and not accounting for the systematic scatter in the error bars of each cluster (Eq.~\ref{eq:intrinlens} and \ref{eq:intrinhse}). Shaded areas indicate the $1\sigma$ region. The black dashed line shows the one-to-one relation between HSE and lensing masses.
In the right panel in Fig.~\ref{fig:refSRnosysandsys}, we show the posterior distributions of the fitted scaling relation parameters. The intrinsic scatter related to HSE masses is remarkably shifted towards zero when accounting for the systematic scatter in the error bars of cluster masses. This is expected, since increasing the error bars of clusters reduces the need to have a dispersion around the SR. The median values with the 16th and 84th percentiles of the posterior distributions of $\alpha^{\mathrm{HSE}}$,  $\sigma^{\mathrm{HSE}}$, and $\sigma^{\mathrm{lens}}$ are given in the first row of Table~\ref{tab:fits}. From $\alpha^{\mathrm{HSE}}$ we compute the HSE-to-lensing mass bias at $R_{500}$ (Eq.~\ref{eq:lensbias}), which gives $(1-b) =0.739^{+0.075}_{-0.070}$ considering the systematic scatters.

\renewcommand{\arraystretch}{1.4}
    \scriptsize
    \begin{table*}[]
      \scriptsize
      \centering
       \caption{Summary of the median values and uncertainties at 16th and 84th percentiles of the parameters for the HSE-to-lensing SR assuming linearity ($\beta^{\mathrm{HSE}}=1$). }
        \begin{tabular}{c|c|c|c|c|c}

          \hline
          \hline

             Cluster sample & $\#$ of clusters &  \multicolumn{4}{c}{No $\sigma_{\mathrm{sys}}^2$}  \\
             ~ & ~ & $\alpha^{\mathrm{HSE}}$ &  $e^{\alpha^{\mathrm{HSE}}} = (1-b)$ & $\sigma^{\mathrm{HSE}}$ & $\sigma^{\mathrm{lens}}$ \\ \hline

             \textit{Reference} sample & 53 & $-0.338_{-0.097}^{+0.105}$ &  $0.713_{-0.069}^{+0.075}$ & $0.304_{-0.072}^{+0.069}$ & $0.305_{-0.083}^{+0.080}$ \\
             $z<0.9$ & 50 & $-0.309_{-0.110}^{+0.124}$ & $0.734_{-0.081}^{+0.091}$ & $0.275_{-0.071}^{+0.071}$ & $0.267_{-0.086}^{+0.083}$\\
             $z<0.5$ & 42 & $-0.328_{-0.102}^{+0.111}$ & $0.720_{-0.073}^{+0.080}$ & $0.282_{-0.086}^{+0.080}$ & $0.308_{-0.091}^{+0.090}$\\
             $z<0.2$ & 19 & $-0.215_{-0.166}^{+0.223}$ & $0.806_{-0.133}^{+0.180}$ & $0.332_{-0.128}^{+0.114}$ & $0.368_{-0.152}^{+0.155}$\\
             $z>0.2$ & 34 & $-0.421_{-0.129}^{+0.139}$ & $0.656_{-0.085}^{+0.091}$ & $0.298_{-0.090}^{+0.076}$ & $0.334_{-0.086}^{+0.090}$ \\
             $z>0.5$ & 11 & $-0.668_{-0.320}^{+0.316}$ & $0.513_{-0.164}^{+0.162}$ & $0.403_{-0.116}^{+0.155}$ & $0.307_{-0.157}^{+0.200}$\\
             \textit{Ref.} no CL~J1226.9+3332 & 52 & $-0.350_{-0.092}^{+0.098}$ & $0.705_{-0.065}^{+0.070}$ & $0.294_{-0.075}^{+0.072}$ & $0.295_{-0.087}^{+0.084}$\\
             $z<0.9$ no CL~J1226.9+3332 & 49 & $-0.338_{-0.099}^{+0.114}$ & $0.713_{-0.070}^{+0.081}$ & $0.273_{-0.074}^{+0.073}$ & $0.274_{-0.086}^{+0.083}$\\
             $z>0.2$ no CL~J1226.9+3332 & 33 & $-0.430_{-0.126}^{+0.130}$ & $0.651_{-0.082}^{+0.085}$ & $0.289_{-0.099}^{+0.082}$ & $0.320_{-0.093}^{+0.093}$ \\
             $z>0.5$ no CL~J1226.9+3332 & 10 & $-0.629_{-0.410}^{+0.332}$ & $0.533_{-0.219}^{+0.177}$ & $0.446_{-0.133}^{+0.184}$ & $0.189_{-0.131}^{+0.209}$\\\hline

             ~ & ~ &  \multicolumn{4}{c}{~}   \\ \hline\hline

             ~ & ~ &  \multicolumn{4}{c}{With $\sigma_{\mathrm{sys}}^2$}   \\
             ~ & ~ & $\alpha^{\mathrm{HSE}}$ &  $e^{\alpha^{\mathrm{HSE}}} = (1-b)$ & $\sigma^{\mathrm{HSE}}$ & $\sigma^{\mathrm{lens}}$\\ \hline

             \textbf{\textit{Reference} sample} & \textbf{53} &  $\mathbf{-0.303_{-0.095}^{+0.101}}$& $\mathbf{0.739_{-0.070}^{+0.075}}$ &$\mathbf{0.166_{-0.101}^{+0.086}}$ & $\mathbf{0.257_{-0.092}^{+0.080}}$\\
             $z<0.9$ & 50 & $-0.260_{-0.110}^{+0.124}$ & $0.771_{-0.084}^{+0.095}$ & $0.122_{-0.082}^{+0.090}$ & $0.220_{-0.099}^{+0.084}$\\
             $z<0.5$ & 42 & $-0.280_{-0.104}^{+0.114}$ & $0.756_{-0.079}^{+0.086}$ & $0.136_{-0.090}^{+0.098}$ & $0.262_{-0.100}^{+0.089}$\\
             $z<0.2$ & 19 & $-0.191_{-0.157}^{+0.181}$ & $0.827_{-0.129}^{+0.150}$ & $0.179_{-0.111}^{+0.123}$ & $0.318_{-0.130}^{+0.135}$\\
             $z>0.2$ & 34 & $-0.388_{-0.127}^{+0.136}$ & $0.679_{-0.086}^{+0.092}$ & $0.162_{-0.103}^{+0.099}$ & $0.271_{-0.103}^{+0.095}$\\
             $z>0.5$ & 11 & $-0.621_{-0.308}^{+0.305}$ & $0.538_{-0.166}^{+0.164}$ & $0.293_{-0.150}^{+0.175}$ & $0.252_{-0.156}^{+0.206}$\\
             \textit{Ref.} no CL~J1226.9+3332 & 52 & $-0.312_{-0.091}^{+0.096}$ & $0.732_{-0.067}^{+0.070}$ & $0.154_{-0.097}^{+0.089}$ & $0.249_{-0.096}^{+0.084}$\\
             $z<0.9$ no CL~J1226.9+3332 & 49 & $-0.287_{-0.102}^{+0.116}$ & $0.750_{-0.077}^{+0.087}$ & $0.120_{-0.080}^{+0.089}$ & $0.226_{-0.099}^{+0.085}$\\
             $z>0.2$ no CL~J1226.9+3332 & 33 & $-0.397_{-0.131}^{+0.131}$ & $0.672_{-0.088}^{+0.088}$ & $0.151_{-0.097}^{+0.103}$ & $0.264_{-0.112}^{+0.100}$\\
             $z>0.5$ no CL~J1226.9+3332 & 10 & $-0.587_{-0.427}^{+0.332}$ & $0.556_{-0.238}^{+0.185}$ & $0.339_{-0.170}^{+0.207}$ & $0.190_{-0.132}^{+0.216}$\\\hline 

        \end{tabular}
        \vspace*{0.2cm}  
         \begin{tablenotes}         
        \small
        \item \textbf{Notes.} We present the results for different data subsamples, with and without accounting for the systematic uncertainties in the error bars of the masses. We show in bold the parameters for the scaling relation of reference presented in Sect.~\ref{sec:SR}.
        \end{tablenotes}        
        \vspace*{0.2cm}
        \label{tab:fits}
\end{table*}
\normalsize

\subsubsection*{Impact of particular subsamples in redshift}
\label{sec:subsamples}

As for the bias model in Sect.~\ref{sec:bias}, we also want to check how the SR parameters may vary depending on the chosen redshift range. Therefore, we repeat the analysis for the different redshift subsamples considered in Sect.~\ref{sec:bias}. We present in Fig.~\ref{fig:differentzsys} and \ref{fig:differentz} and in Table~\ref{tab:fits} the different results, with and without $\sigma_{\mathrm{sys}}^2$. Again, we observe that the bias changes for $z<0.2$ and $z>0.2$ clusters, in line with a $(1-b)$ value that decreases with redshift. The scaling relations with $z<0.9$ and $z<0.5$ samples remain almost unchanged with respect to the SR of reference. Not accounting for CL~J1226.9+3332 reduces the lensing scatter for the $z>0.5$ subsample. Overall, we find that the SRs are compatible for the different subsamples.

The posterior distribution of the SR parameters obtained for the $z<0.5$ clusters without $\sigma_{\mathrm{sys}}^2$ (see Fig.~\ref{fig:different0.5}) can be directly compared to Figure 5 in \citet{serenoettoricomalit}. In that work, the 50 CCCP clusters from \citet{Mahdavi_2013} were used to measure the HSE-to-lensing mass scaling relation (even though the HSE masses were evaluated at the $R_{500}$ obtained from lensing). The intrinsic scatters seem to be differently correlated in \citet{serenoettoricomalit} and in this paper. However, in both cases we observe no strong correlation between $\alpha^{\mathrm{HSE}}$ and the intrinsic HSE or lensing scatters.  In our case, for the $z<0.5$ clusters without (with) $\sigma_{\mathrm{sys}}^2$ {we measure $(1-b) = 0.720_{-0.073}^{+0.080}$ ($(1-b) = 0.756^{+0.086}_{-0.079}$). These results (Table~\ref{tab:fits}) are in line with the values reported in Table~6 in \citet{serenoettoricomalit} and Table~2 in \citet{lovisari2020}.

\begin{figure*}
        \begin{minipage}[b]{0.5\textwidth}
        \includegraphics[trim={0pt 0pt 0pt 0pt},scale=0.6]{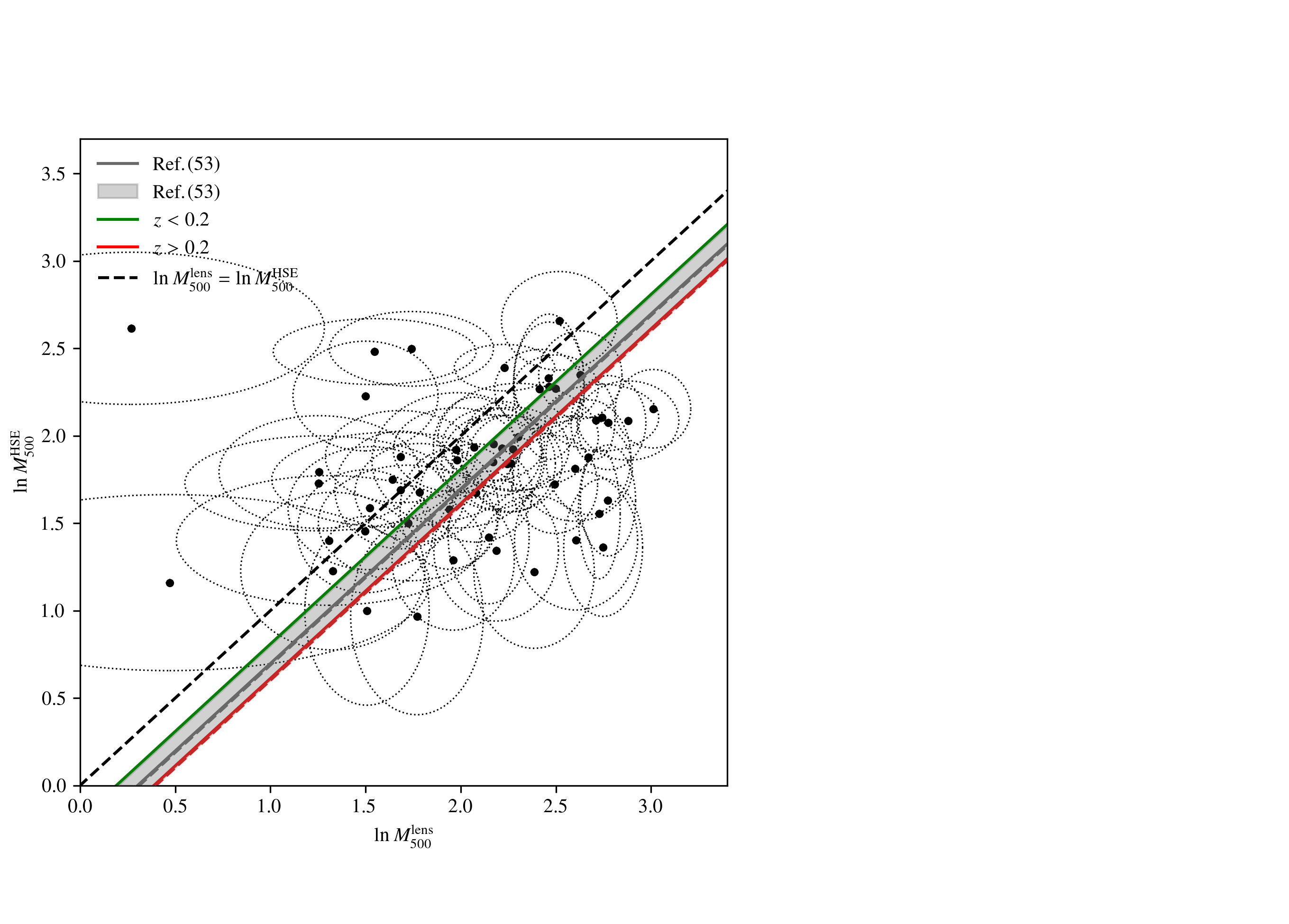}
        \end{minipage}
        \hfill
        \begin{minipage}[b]{0.5\textwidth}
        \includegraphics[trim={0pt 0pt 0pt 0pt},scale=0.3]{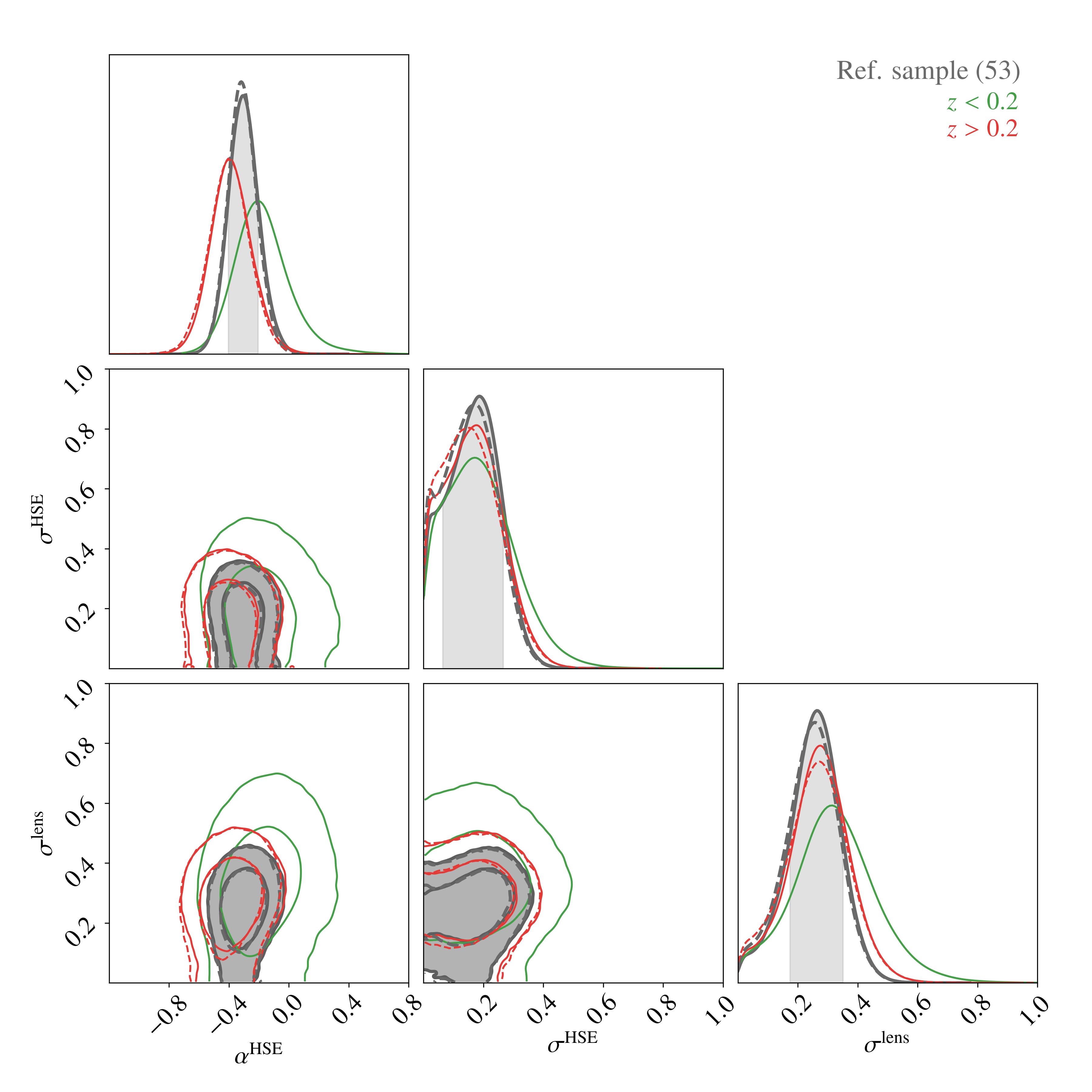}
        \end{minipage}
        \caption{Scaling relation between HSE and lensing masses for the \textit{reference} sample in grey and for different subsamples in colours, all accounting for $\sigma_{\mathrm{sys}}$. Here $\beta^{\mathrm{HSE}}$ is fixed to 1. As in Fig.~\ref{fig:biasfit}, we only show the cases for $z>0.2$ and $z<0.2$. Data points with ellipses represent each cluster masses and uncertainties in both axes accounting for the systematic scatters. The black dashed line shows the equality. The corner plots in the right panel are the posterior 1D and 2D distributions of the parameters in the SR.}
    \label{fig:differentzsys}
\end{figure*}

\subsection{Investigations of possible model extensions}
\label{sec:robustness}

Beyond the reference scaling relation, for which we have assumed no redshift evolution and a linear scaling between the masses, in this section we test if relaxing some of these assumptions improves the description of the data by the scaling relation model. 

\subsubsection*{Deviation from linearity}
\begin{figure*}
        \begin{minipage}[b]{0.5\textwidth}
        \includegraphics[trim={0pt 0pt 0pt 0pt},scale=0.6]{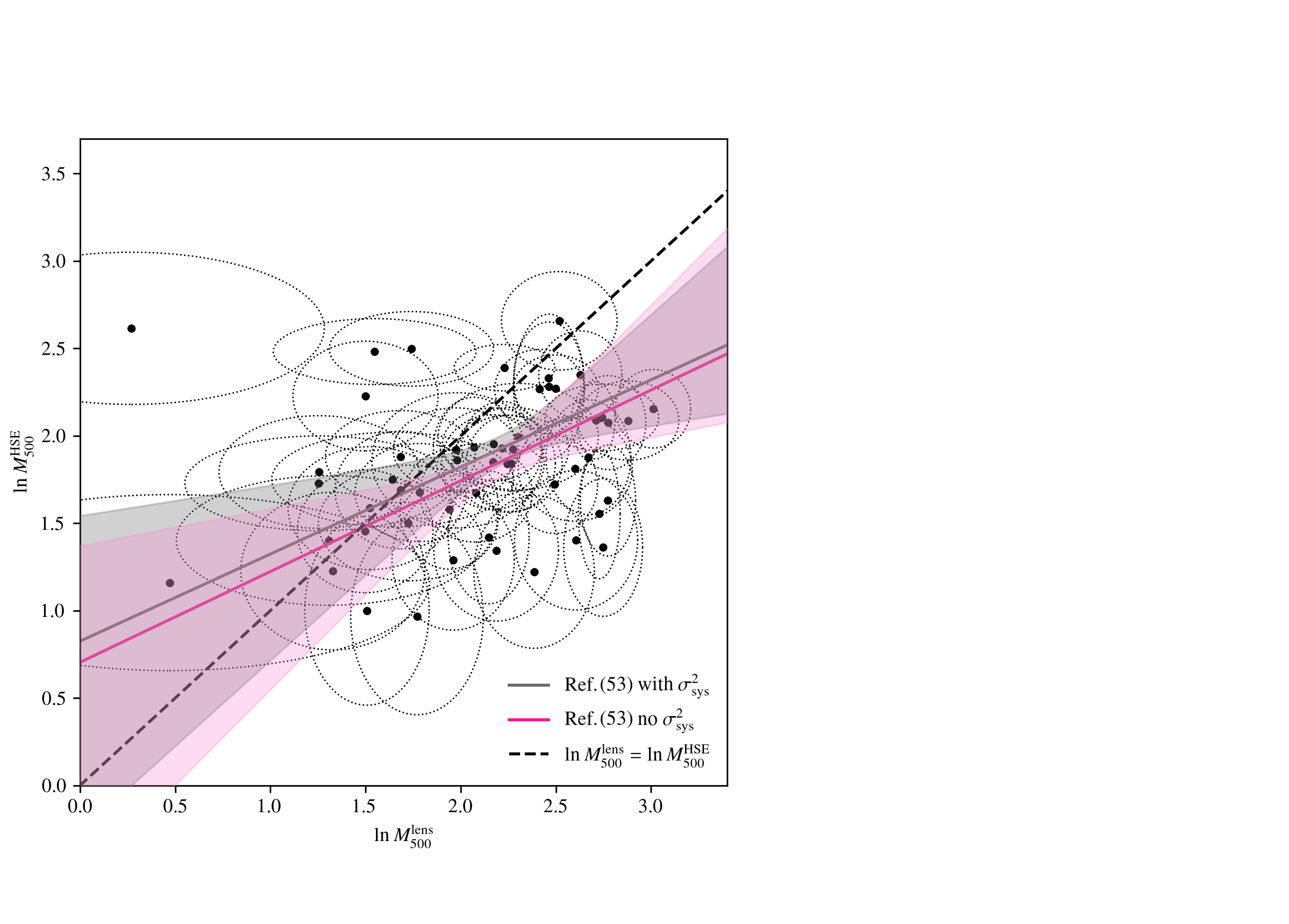}
        \end{minipage}
        \hfill
        \begin{minipage}[b]{0.5\textwidth} 
         \includegraphics[trim={0pt 0pt 0pt 0pt},scale=0.3]{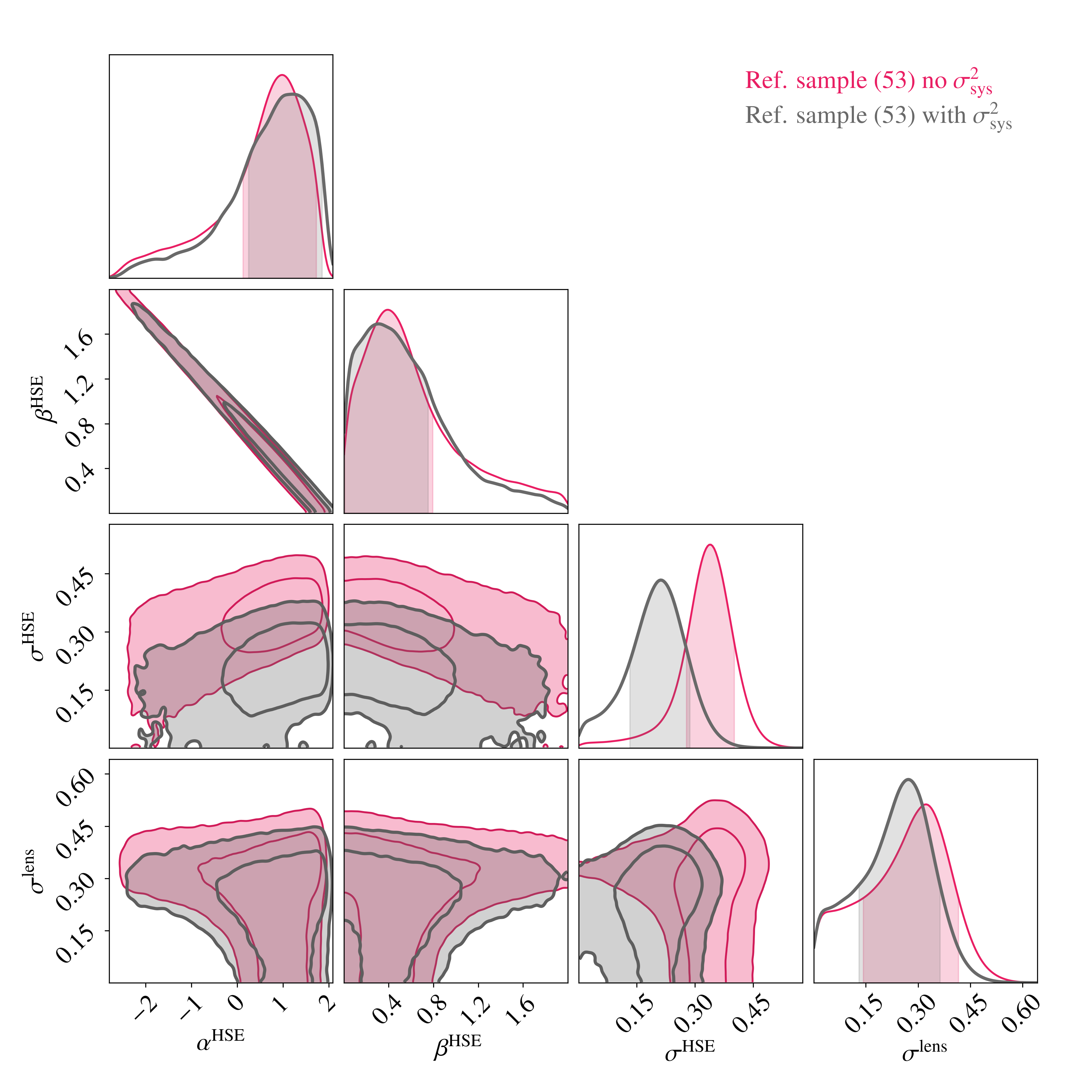}
         \end{minipage}
    \caption{Scaling relation between HSE and lensing masses in the \textit{reference} sample considering a deviation from linearity. Data points with ellipses represent each cluster masses and uncertainties in both axes accounting for the systematic scatters. The pink line corresponds to the SR for the median value of parameters obtained without $\sigma_{\mathrm{sys}}$ and the solid grey line with $\sigma_{\mathrm{sys}}$. The black dashed line shows the equality and shaded regions the 16th and 84th percentiles. The corner plots in the right panel are the posterior 1D and 2D distributions of the parameters in the SR, including (grey) or not (pink) the systematic scatters.}
    \label{fig:refSRnosysandsysbetafree}
\end{figure*}

\renewcommand{\arraystretch}{1.4}
    \scriptsize
    \begin{table*}[]
      \scriptsize
      \centering
       \caption{Summary of the median values and uncertainties at the 16th and 84th percentiles of the parameters in the HSE-to-lensing SR when considering a deviation from linearity, an offset between HSE and lensing masses or an evolution with redshift.}
        \begin{tabular}{c|c|c|c|c|c|c}
          \hline
          \hline
             Cluster sample & $\#$ of clusters &  \multicolumn{5}{c}{No $\sigma_{\mathrm{sys}}^2$}    \\
             ~ & ~ & $\alpha^{\mathrm{HSE}}$ &  $\beta^{\mathrm{HSE}}$ & $\sigma^{\mathrm{HSE}}$ & $\sigma^{\mathrm{lens}}$&  $\gamma^{\mathrm{HSE}}$\\  \hline

             \textit{Reference} sample & 53 & $0.705_{-1.249}^{+0.666}$ & $0.519_{-0.309}^{+0.576}$ & $0.335_{-0.066}^{+0.057}$  & $0.275_{-0.163}^{+0.103}$ & [0] \\
             \textit{Reference} sample (BCES) & 53 & $0.826 \pm 0.886$ &  $0.481 \pm 0.415$ & $0.326^{*}$ & - & [0]  \\
             \textit{Reference} sample & 53 & $-0.219_{-0.134}^{+0.137}$  &  [1] & $0.303_{-0.072}^{+0.069}$ & $0.295_{-0.085}^{+0.082} $ & $-1.742_{-1.083}^{+1.082}$\\
             \textit{Ref.} no CL~J1226.9+3332 & 52 & $-0.263_{-0.136}^{+0.140}$ & [1] & $0.297_{-0.073}^{+0.072}$ & $0.292_{-0.087}^{+0.084}$  & $-1.168_{-1.163}^{+1.156}$   \\\hline

             ~ & ~  & $A^{\mathrm{HSE}}$ [$10^{14}$ M$_{\odot}$] &  $B^{\mathrm{HSE}}$ &  $\sigma^{\mathrm{HSE}}$ [$10^{14}$ M$_{\odot}$] & $\sigma^{\mathrm{lens}}$ [$10^{14}$ M$_{\odot}$] &  $\gamma^{\mathrm{HSE}}$ [$10^{14}$ M$_{\odot}$]\\  \hline
             
              \textit{Reference} sample & 53 & $0.818_{-3.037}^{+2.313}$ &  $0.614_{-0.270}^{+0.351}$ & $1.673_{-0.640}^{+0.413}$  & $3.159_{-0.983}^{+0.701}$ & [0]\\
              \textit{Reference} sample (BCES) & 53 & $4.467 \pm 1.85$ & $0.246 \pm 0.19$ &  $2.109^{*}$ & - & [0]\\
              
              \textit{Reference} sample & 53 & $0.545_{-2.925}^{+2.465}$  & $0.675_{-0.291}^{+0.334}$ & $1.644_{-0.700}^{+0.439}$ & $3.188_{-0.854}^{+0.674}$ & $-2.685_{-4.260}^{+4.654}$ \\ \hline

             ~ & ~ &  \multicolumn{5}{c}{~}\\\hline \hline

             ~ & ~ &  \multicolumn{5}{c}{With $\sigma_{\mathrm{sys}}^2$}\\
             ~ & ~ & $\alpha^{\mathrm{HSE}}$ &   $\beta^{\mathrm{HSE}}$ & $\sigma^{\mathrm{HSE}}$ & $\sigma^{\mathrm{lens}}$ &  $\gamma^{\mathrm{HSE}}$\\\hline

             \textit{Reference} sample & 53  & $0.824_{-1.087}^{+0.719}$  & $0.498_{-0.326}^{+0.481}$ & $0.204_{-0.082}^{+0.068}$ & $0.242_{-0.135}^{+0.091}$ & [0]\\

             \textit{Reference} sample (BCES) & 53 &  $1.000 \pm 0.692$ &  $0.397\pm 0.324$ & $0.191^{*}$ & - & [0]\\
             \textit{Reference} sample & 53 & $-0.193_{-0.134}^{+0.135}$  & [1] & $0.168_{-0.100}^{+0.086}$  & $0.246_{-0.095}^{+0.082}$ & $-1.530_{-1.085}^{+1.071}$ \\
             \textit{Ref.} no CL~J1226.9+3332 & 52 & $-0.242_{-0.137}^{+0.139}$ & [1] & $0.153_{-0.096}^{+0.090}$ & $0.248_{-0.093}^{+0.084}$ & $-0.896_{-1.155}^{+1.154}$\\\hline
             
             ~ & ~  & $A^{\mathrm{HSE}}$ [$10^{14}$ M$_{\odot}$] &$B^{\mathrm{HSE}}$ &   $\sigma^{\mathrm{HSE}}$ [$10^{14}$ M$_{\odot}$] & $\sigma^{\mathrm{lens}}$ [$10^{14}$ M$_{\odot}$] &  $\gamma^{\mathrm{HSE}}$ [$10^{14}$ M$_{\odot}$]\\  \hline
             
             \textit{Reference} sample & 53 & $1.603_{-2.994}^{+1.665}$ & $0.522_{-0.194}^{+0.349}$ & $0.949_{-0.602}^{+0.557}$  &  $2.867_{-1.251}^{+0.806}$ & [0]
\\
             \textit{Reference} sample (BCES) & 53 & $4.644 \pm 1.727$ &   $0.226 \pm 0.176$ & $1.340^{*}$ & - & [0]\\
             \textit{Reference} sample & 53 & $1.434_{-3.005}^{+1.780}$ & $0.570_{-0.210}^{+0.346}$&  $0.950_{-0.612}^{+0.570}$ & $2.927_{-1.131}^{+0.771}$ & $-2.493_{-4.282}^{+4.535}$\\
             \hline
    
        \end{tabular}
        \vspace*{0.2cm}  
         \begin{tablenotes}         
        \small
        \item \textbf{Notes.} We present the results for the \textit{reference} sample, accounting or not for the systematic scatter in the error bars of the masses. For the BCES fit we report the best-fit values and $1\sigma$ uncertainties. $^{(*)}$ We also calculate the scatter with respect to the best BCES scaling relations following Eq.~\ref{eq:sysdef}. 
         \end{tablenotes}        
        \vspace*{0.2cm}
        \label{tab:timeevol-nolinear-offset}
   \end{table*} 
    \normalsize

\begin{figure}[h]
    \centering   
    \includegraphics[trim={0pt 0pt 0pt 0pt}, scale=0.32]{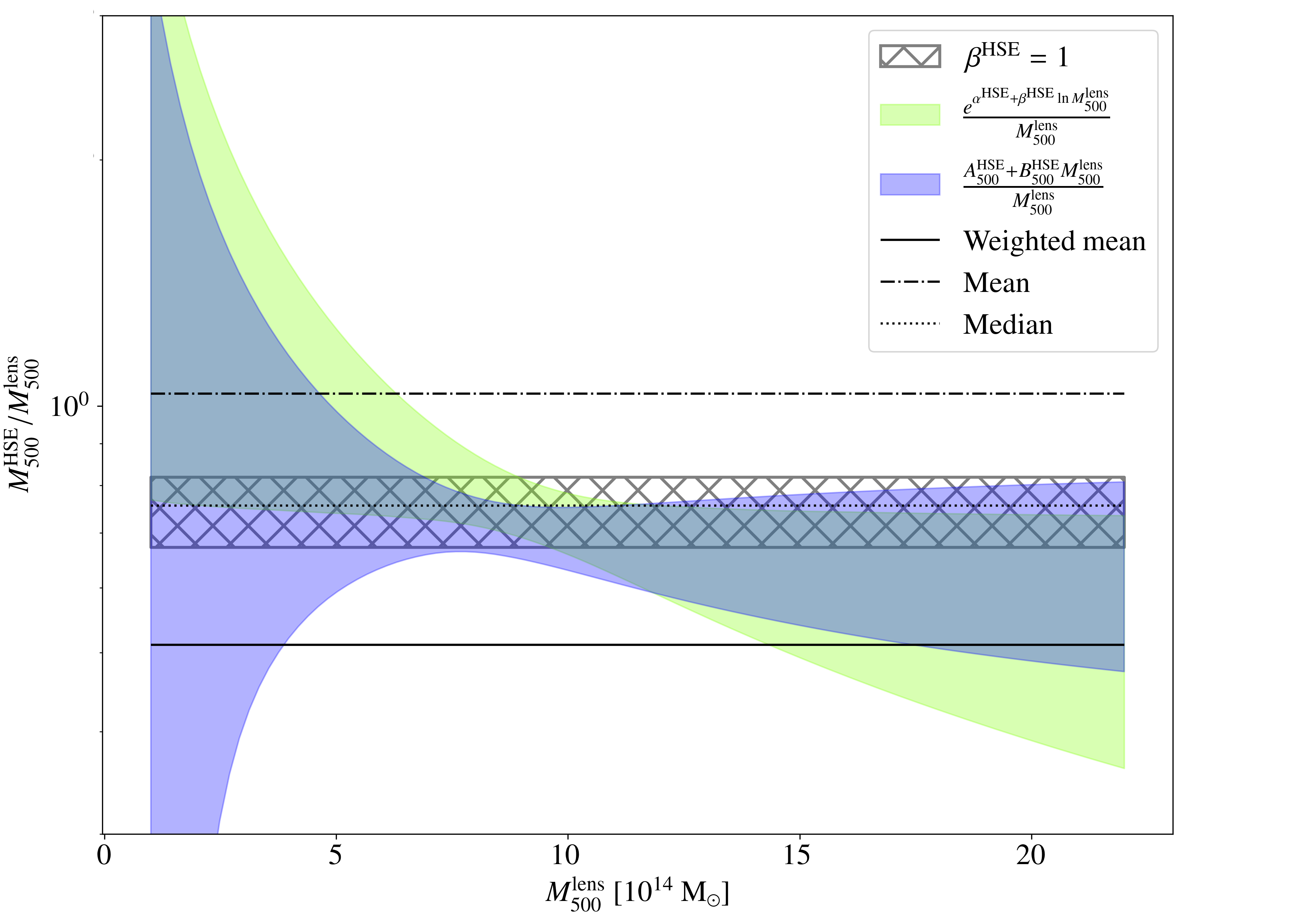}
    \caption{HSE-to-lensing mass ratio with respect to lensing mass. The grey hatched area indicates the 16th to 84th percentiles of the bias without mass dependence, accounting for systematic scatters in the uncertainties of HSE and lensing masses. The green area shows the bias evolution when assuming a deviation from linearity of the HSE and lensing masses. Blue area indicates the bias evolution when considering an offset between HSE and lensing masses. Horizontal solid, dotted, and dash-dotted black lines give respectively the weighted mean, median, and mean mass ratio for the 53 clusters, same as in Fig.~\ref{fig:biasfitref}.}
    \label{fig:evolmass}
\end{figure}

The HSE and/or lensing masses could also scale non-linearly with the true mass, meaning that the HSE-to-lensing bias would depend on the mass of the clusters. In \citet{Hoekstra_2015} and \citet{vonderlinden} authors investigated such dependence on the mass comparing \textit{Planck} results to CCCP and WtG lensing masses, respectively. Both works found modest evidence for a mass-dependence: $M_{Planck} \propto M_{\mathrm{CCCP}}^{0.64\pm 0.17}$ with $\alpha^{\mathrm{HSE}} \sim 0.55$ in \citet{Hoekstra_2015}, and $M_{Planck} \propto M_{\mathrm{WtG}}^{0.68^{+0.15}_{-0.11}}$ with $\alpha^{\mathrm{HSE}} \sim 0.38$ and  $M_{Planck} \propto M_{\mathrm{WtG}}^{0.76^{+0.39}_{-0.20}}$ with $\alpha^{\mathrm{HSE}} \sim 0.19$ in \citet{vonderlinden} for different cluster samples. Physically, this mass dependence could correspond, for example, to an impact of the baryonic physics that would depend on the strength of the clusters potential wells. In this case, low mass clusters having shallower potential wells, we can imagine that baryonic effects are stronger in them \citep{McCarthy2011}. On the contrary, simulations in \cite{rasia2012} also indicate that massive objects are the most disturbed ones and have, probably, more complex temperature structures. 

We also test this hypothesis by fitting the SR in Eq.~\ref{eq:srlens} and \ref{eq:srhse} leaving $\beta^{\mathrm{HSE}}$ as a free parameter. We take a uniform prior for $\beta^{\mathrm{HSE}} \sim \mathcal{U}(0, 2)$ and consider the same priors for $\alpha^{\mathrm{HSE}}$, $\sigma^{\mathrm{HSE}}$, and $\sigma^{\mathrm{lens}}$. The resulting scaling relations are presented in Fig.~\ref{fig:refSRnosysandsysbetafree} and the median values are given in Table~\ref{tab:timeevol-nolinear-offset}. As shown in the corner plot in Fig.~\ref{fig:refSRnosysandsysbetafree}, $\alpha^{\mathrm{HSE}}$ and $\beta^{\mathrm{HSE}}$ are completely degenerated. Nevertheless, our results are in agreement with \citet{Hoekstra_2015} and \citet{vonderlinden}. However, the HSE masses in those works were \textit{Planck} masses from the SZ-mass scaling relation.

For comparison to the results obtained with LIRA, we also perform the fit of the SR using the orthogonal Bivariate Correlated Errors and intrinsic Scatter method \citep[BCES,][]{bces}. BCES favours a larger deviation from linearity, that is, smaller $\beta^{\mathrm{HSE}}$. We also report the results in Table~\ref{tab:timeevol-nolinear-offset}.  Given the large uncertainties on $\alpha^{\mathrm{HSE}}$ and $\beta^{\mathrm{HSE}}$, the scaling relations obtained with LIRA and BCES are compatible. 
    
In Fig.~\ref{fig:evolmass} we present the HSE-to-lensing mass ratio as a function of the lensing mass for the fitted $\alpha^{\mathrm{HSE}}$ and $\beta^{\mathrm{HSE}}$, with the green shaded area showing the 16th to 84th percentiles. The horizontal grey hatched area represents the HSE-to-lensing mass ratio measured in the previous section assuming that HSE and lensing masses scale linearly with the true mass. Given that we obtain $\beta^{\mathrm{HSE}}<1$, on average the difference between HSE and lensing masses is larger for more massive objects. This is in agreement with the mild decreasing tendency for the HSE-to-lensing mass ratio obtained in \citet{Hoekstra_2015}, \citet{vonderlinden}, and \citet{eckert2019}, but different from the trend observed in \citet{salvati2019}. Nevertheless, our results are consistent with no mass dependence of the ratio. The difficulty of disentangling $\alpha^{\mathrm{HSE}}$ and $\beta^{\mathrm{HSE}}$ does not motivate further investigations of the SR model with additional free parameters. Leaving free $\alpha^{\mathrm{lens}}$ would add a free parameter to the model strongly correlated to  $\alpha^{\mathrm{HSE}}$} and $\beta^{\mathrm{HSE}}$. 

\subsubsection*{Considering an offset}

In addition to the HSE-to-lensing mass bias defined in Eq.~\ref{eq:lensbias}, there could be also an offset between the HSE and lensing mass estimates. Thus, the scaling relation could be defined as,
\begin{equation}
  M^{\mathrm{lens}} \pm \delta_{\mathrm{lens}}  =  M^{\mathrm{True}} \pm \sigma^{\mathrm{lens}},
  \label{eq:offset1}
\end{equation}
\begin{equation}
  M^{\mathrm{HSE}} \pm \delta_{\mathrm{HSE}}  = A^{\mathrm{HSE}}+ B^{\mathrm{HSE}}  M^{\mathrm{True}} \pm \sigma^{\mathrm{HSE}},
  \label{eq:offset2}
\end{equation}
where $A^{\mathrm{HSE}}$ and $B^{\mathrm{HSE}}$ are the offset and the multiplicative factor, respectively. Here $\sigma^{\mathrm{HSE}}$ and $\sigma^{\mathrm{lens}}$ are again the scatter of HSE and lensing masses with respect to the SR, but in this case in units of $10^{14}$ M$_{\odot}$.

\begin{figure*}
  \begin{minipage}[b]{0.5\textwidth}
    \centering
        \includegraphics[trim={0pt 0pt 0pt 0pt},scale=0.6]{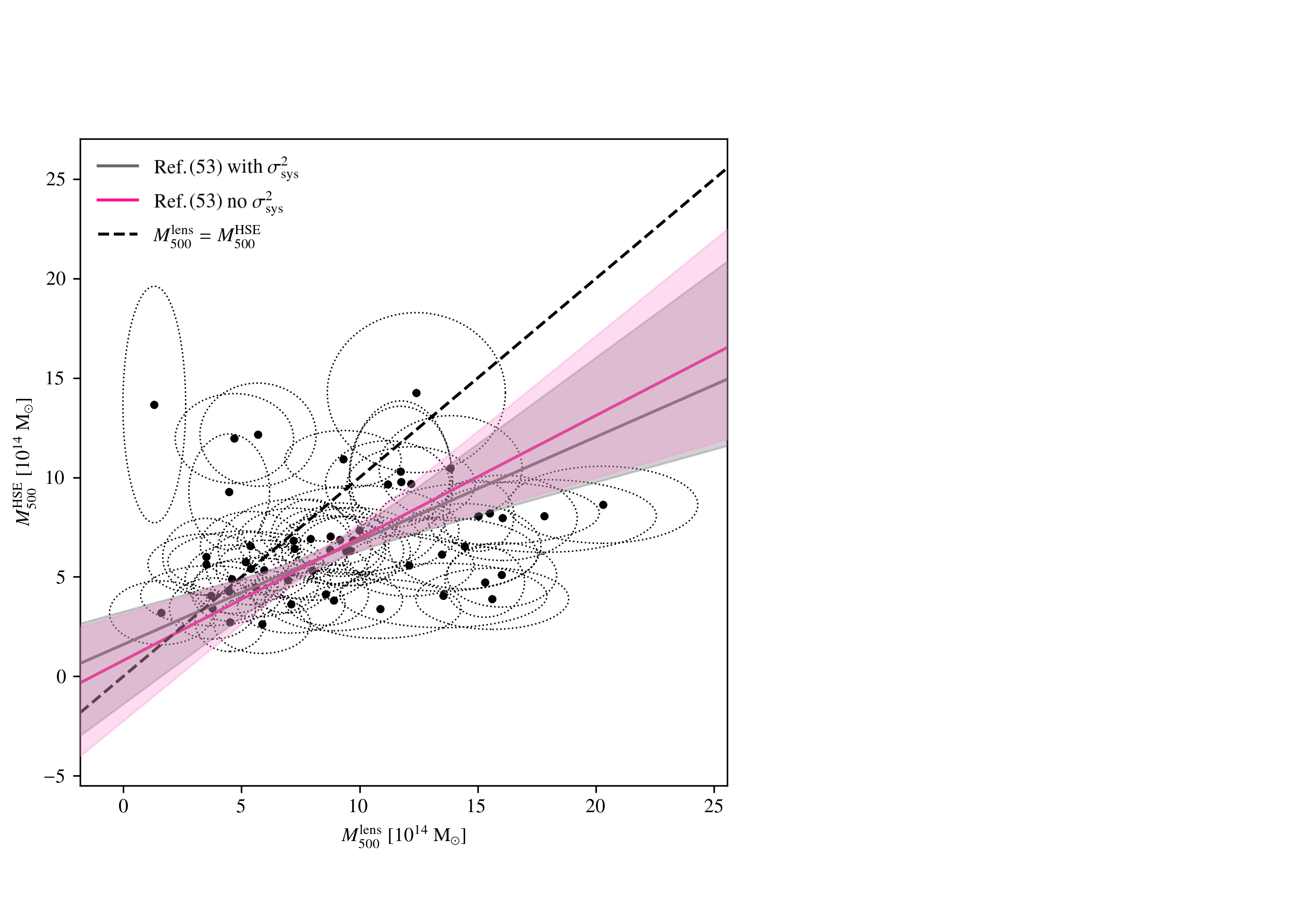}
        \end{minipage}
        \hfill
        \begin{minipage}[b]{0.5\textwidth} 
        \includegraphics[trim={0pt 0pt 0pt 0pt},scale=0.3]{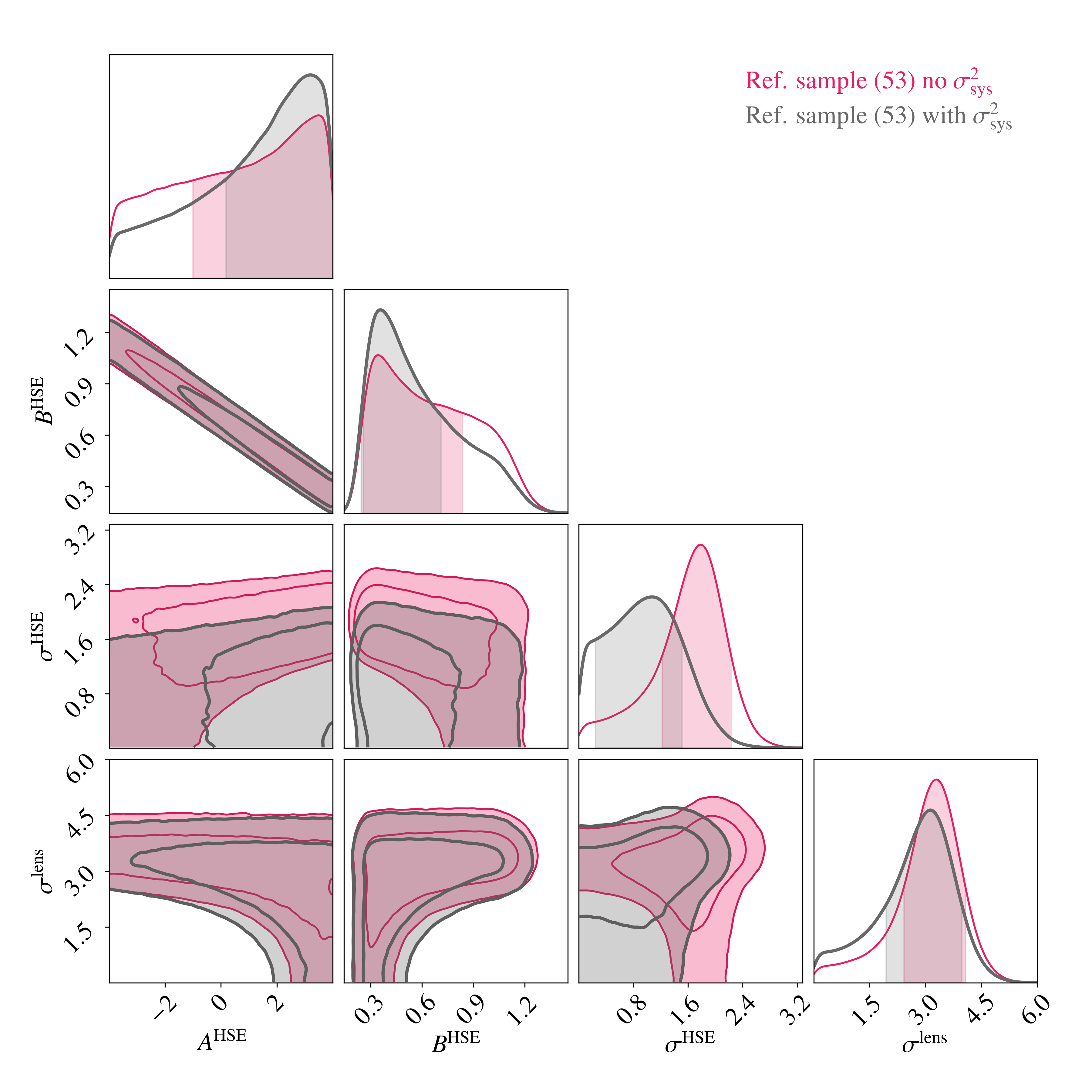}      
        \end{minipage}

        \caption{Scaling relation between HSE and lensing masses in the \textit{reference} sample considering an offset between both mass estimates. Data points with ellipses represent each cluster masses and the uncertainties in both axes accounting for the systematic scatter. The pink line corresponds to the SR obtained without $\sigma_{\mathrm{sys}}$ and the solid grey line with $\sigma_{\mathrm{sys}}$. The black dashed line shows the equality. The corner plots in the right panel are the posterior 1D and 2D distributions of the parameters in the SR, including (grey) or not (pink) the systematic scatters.}
        \label{fig:lin}
\end{figure*}

We perform again the fit of the SR using both the LIRA and BCES methods. We present in Fig.~\ref{fig:lin} and Table~\ref{tab:timeevol-nolinear-offset} the results. As for the non-linear SR fit, $A^{\mathrm{HSE}}$ and $B^{\mathrm{HSE}}$ are completely degenerated. The results obtained with LIRA indicate an offset in mass completely compatible with zero. It is reassuring to verify that the data motivates a scaling relation model for which the HSE mass goes to zero in the limit $M^{\mathrm{True}} \rightarrow 0$. We show in Fig.~\ref{fig:evolmass} the bias evolution in blue, indicating again that there is no significant trend of the HSE-to-lensing mass ratio with cluster mass.

\subsubsection*{Evolution with redshift}
\label{sec:timeevol}
\begin{figure*}
    \centering   
    \includegraphics[trim={0pt 0pt 0pt 0pt}, scale=0.4]{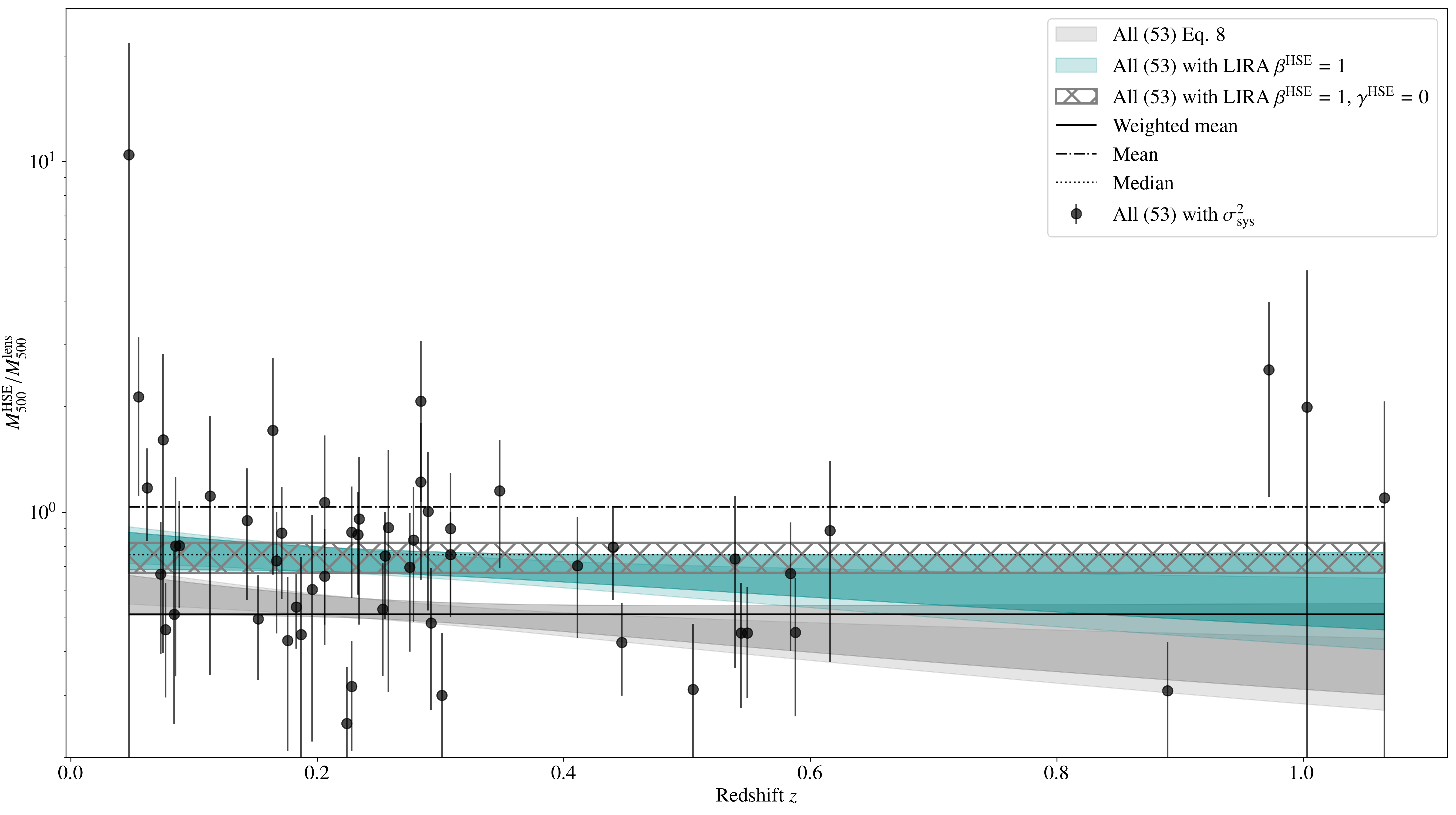}
    \caption{HSE-to-lensing mass ratio with respect to redshift. The grey shaded area shows the evolution from Fig.~\ref{fig:biasfitref} for all the clusters in the sample and in darker excluding CL~J1226.9+3332. The blue area gives the evolution with redshift obtained from the fit of the scaling relation with the \textit{reference} sample and the grey hatched area without considering the redshift evolution. The blue dark area is the evolution obtained for the \textit{reference} sample excluding CL~J1226.9+3332. As in Fig.~\ref{fig:biasfitref}, markers with error bars show the ratio per cluster in the \textit{reference} sample with error bars accounting for the systematic uncertainty. Horizontal solid, dotted, and dash-dotted black lines give respectively the weighted mean, median, and mean mass ratio for the data points.}
    \label{fig:biasfit_lira}
\end{figure*}

LIRA enables fitting a scaling relation that evolves with redshift. Looking for such evolution can be particularly interesting with our \textit{reference} sample, given the large redshift range that it covers ($0.05<z<1.07$). Assuming again that HSE and lensing masses scale linearly with the true mass ($\beta^{\mathrm{lens}} = \beta^{\mathrm{HSE}} =1$), we write
\begin{equation}
    \ln M^{\mathrm{lens}} \pm \delta_{\mathrm{lens}}  =  \ln M^{\mathrm{True}} \pm \sigma^{\mathrm{lens}}, 
\end{equation}
and,
\begin{equation}
  \ln M^{\mathrm{HSE}} \pm \delta_{\mathrm{HSE}}  = \alpha^{\mathrm{HSE}}+  \ln M^{\mathrm{True}} \pm \sigma^{\mathrm{HSE}} + \gamma^{\mathrm{HSE}}T.
  \label{eq:timeevol}
\end{equation}
\noindent We note that $T$ is the time evolution factor, $T = \log \left( \frac{1+z}{1+z_{ref}} \right)$, with $z_{ref} = 0.01$ the normalisation redshift set by default in LIRA. We take flat priors for the parameter describing the evolution with redshift: $\gamma^{\mathrm{HSE}} \sim \mathcal{U}(-10, 10)$. Similarly, we consider the evolution with redshift for the SR defined in Eq.~\ref{eq:offset1} and \ref{eq:offset2}. Given the strong impact of the CL~J1226.9+3332 galaxy cluster on the fits at high redshift (see Sect.~\ref{sec:bias}), we repeat the analysis excluding it. All the results are summarised in Table~\ref{tab:timeevol-nolinear-offset}. 

In Fig.~\ref{fig:biasfit_lira} we present the redshift evolution of the HSE-to-lensing mass ratio for the analyses performed with the \textit{reference} sample and accounting for systematic uncertainties in the HSE and lensing masses. We show in grey the results obtained in Sect.~\ref{sec:bias}, neglecting the intrinsic scatter of HSE and lensing masses with respect to the true masses. In blue we present the bias evolution model resulting from the scaling relation fit in this section. Darker regions show the evolution with redshift obtained when excluding CL~J1226.9+3332 from the analyses.

There seem to be a tendency for a decreasing HSE-to-lensing mass ratio with redshift ($\gamma^{\mathrm{HSE}} = -1.530^{+1.071}_{-1.085}$), but it is not statistically significant when removing CL~J1226.9+3332 from the sample ($\gamma^{\mathrm{HSE}} = -0.896^{+1.154}_{-1.155}$). From the comparison of the grey and blue results we observe directly the impact that accounting for the intrinsic scatters of the SRs has on the bias. Considering the intrinsic scatter reduces the difference between HSE and lensing masses and, therefore, the bias.

We present in Fig.~\ref{nanana} and \ref{jiji} a comparison of the scaling relations and posterior distributions of parameters when accounting for redshift evolution (dashed lines) and not accounting for it (solid lines). The contribution of the redshift evolution factor introduces a change of the order of a few percent (or less) in the intrinsic scatters. Given the correlation of the other parameters with $\gamma^{\mathrm{HSE}}$, the change is of $\sim 30 \%$ for $\alpha^{\mathrm{HSE}}$ and of the order of $10\%$ for $A^{\mathrm{HSE}} $ and $B^{\mathrm{HSE}} $. However, the results are compatible with the ones obtained without considering redshift evolution, so there is no strong evidence of redshift evolution in the data.  

\subsection{Comparison of SR models}
\label{sec:timeevol}

In this section, we compare the tested SR models to assess which is the one preferred by the data. We define the goodness of fit of the scaling relations $\hat{\chi}^2$ following Eq.~3 in \citet{lovisari2020apj}:
\begin{equation}
    \hat{\chi}^2 = \sum_{i=1}^{N_{\mathrm{clusters}}}   \frac{ \left[\ln M^{\mathrm{HSE}}_{i}  - \ln M^{\mathrm{HSE}}\left( \ln M^{\mathrm{lens}}_{i},  z_{i}, \vartheta\right)\right]^2}{\delta_{\mathrm{HSE}, i}^2 +  \left(\sigma^{\mathrm{HSE}} \right)^2+    \left(\beta^{\mathrm{HSE}} \right)^2 \left[ \delta_{\mathrm{lens}, i}^2 +  \left(\sigma^{\mathrm{lens}} \right)^2 \right] },    
   \label{eq:goodnessoffit}
\end{equation}
where the sum is done over the $N_{\mathrm{clusters}} = 53$ clusters in the \textit{reference} sample. In Eq.~\ref{eq:goodnessoffit} $\ln M^{\mathrm{HSE}}\left( \ln M^{\mathrm{lens}}_{i}, z_{i},  \vartheta\right)$ is the function described by Eq.~\ref{eq:srhse} or \ref{eq:timeevol} depending on the SR model, with the parameters $\vartheta = [\alpha^{\mathrm{HSE}}, \beta^{\mathrm{HSE}}, \gamma^{\mathrm{HSE}}]$ defined accordingly. The factors $\ln M^{\mathrm{HSE}}_{i}$, $ \ln M^{\mathrm{lens}}_{i}$, $\delta_{\mathrm{HSE}, i}$, and $\delta_{\mathrm{lens}, i}$ are the HSE and lensing mass of each cluster $i$ and their associated uncertainties, and $z_{i}$ is the redshift of each cluster. We compare the results obtained considering always the systematic uncertainties in the HSE and lensing mass uncertainties. We take the posterior distributions of the parameters for $\alpha^{\mathrm{HSE}}$, $\beta^{\mathrm{HSE}}$, $\gamma^{\mathrm{HSE}}$, $\sigma^{\mathrm{HSE}}$, and $\sigma^{\mathrm{lens}}$. For the scaling relations considering an offset in mass (Eq.~\ref{eq:offset1} and \ref{eq:offset2}), we replace the logarithmic masses and uncertainties by the linear values in the $\hat{\chi}^{2}$ definition in Eq.~\ref{eq:goodnessoffit}. Similarly, we take $A^{\mathrm{HSE}}$ and $B^{\mathrm{HSE}}$ instead of $\alpha^{\mathrm{HSE}}$ and $\beta^{\mathrm{HSE}}$. The $\hat{\chi}^2$ distribution for each SR model fit is shown in Fig.~\ref{fig:chi2}. 

According to the Akaike information criterion \citep[AIC,][]{AIC} and the Bayesian information criterion \citep[BIC,][]{BIC}, the scaling relation of reference and the one considering a deviation from linearity are almost equally probable (see Appendix~\ref{sec:aicbic} for more details).  Furthermore, there is statistically no gain in adding a parameter that describes an evolution with redshift. In other words, redshift evolution does not seem to be favoured by the data.

Anyhow, the intrinsic scatters being free parameters in our LIRA fits, we expect all the models to adjust the data points at the expense of increasing the scatters. From the comparison of all the $\sigma^{\mathrm{HSE}}$ and $\sigma^{\mathrm{lens}}$ (see Tables~\ref{tab:fits} and \ref{tab:timeevol-nolinear-offset}), there is not a statistically significant increase, nor decrease in the intrinsic scatters when changing the number of free parameters in the SR model. 

In conclusion, our best scaling law between X-ray HSE and lensing masses is given by the scaling relation of reference:
\begin{equation}
  \ln M^{\mathrm{lens}} =  \ln M^{\mathrm{True}} \pm 0.257^{+0.080}_{-0.092},
  \label{eq:srlensfinal}
\end{equation}
\begin{equation}
  \ln M^{\mathrm{HSE}} = -0.303^{+0.101}_{-0.095} +  \ln M^{\mathrm{True}} \pm 0.166^{+0.086}_{-0.101},
  \label{eq:srhsefinal}
\end{equation}
which corresponds to a HSE-to-lensing mass bias of
\begin{equation}
 M^{\mathrm{HSE}}_{500}/  M^{\mathrm{lens}}_{500} = (1- b) = 0.739^{+0.075}_{- 0.070} \;\mathrm{(stat.)} \pm 0.226 \; \mathrm{(intrin. \; scatter)},
\end{equation}
assuming Gaussian intrinsic scatters for lensing and HSE masses.

\subsection{Caveats}
\label{sec:caveats}

The two main caveats of the analysis presented in this work are the representativity of the used sample and the inhomogeneity in the estimates of the lensing masses. The former is hardly quantifiable, given that the selection criteria of the \textit{reference} sample (Sect.~\ref{sec:homogsample}) are mainly a combination of the selection criteria used for the ESZ, LoCuSS, LPSZ, and \citet{bartalucci2018} clusters. An equivalent study using a clearly defined selection criterium, as for the LPSZ \citep{mayet2020}, would be of great interest. \\

\begin{figure}[h]
	\centering
	 \includegraphics[trim={0pt 0pt 0pt 0pt},scale=0.35]{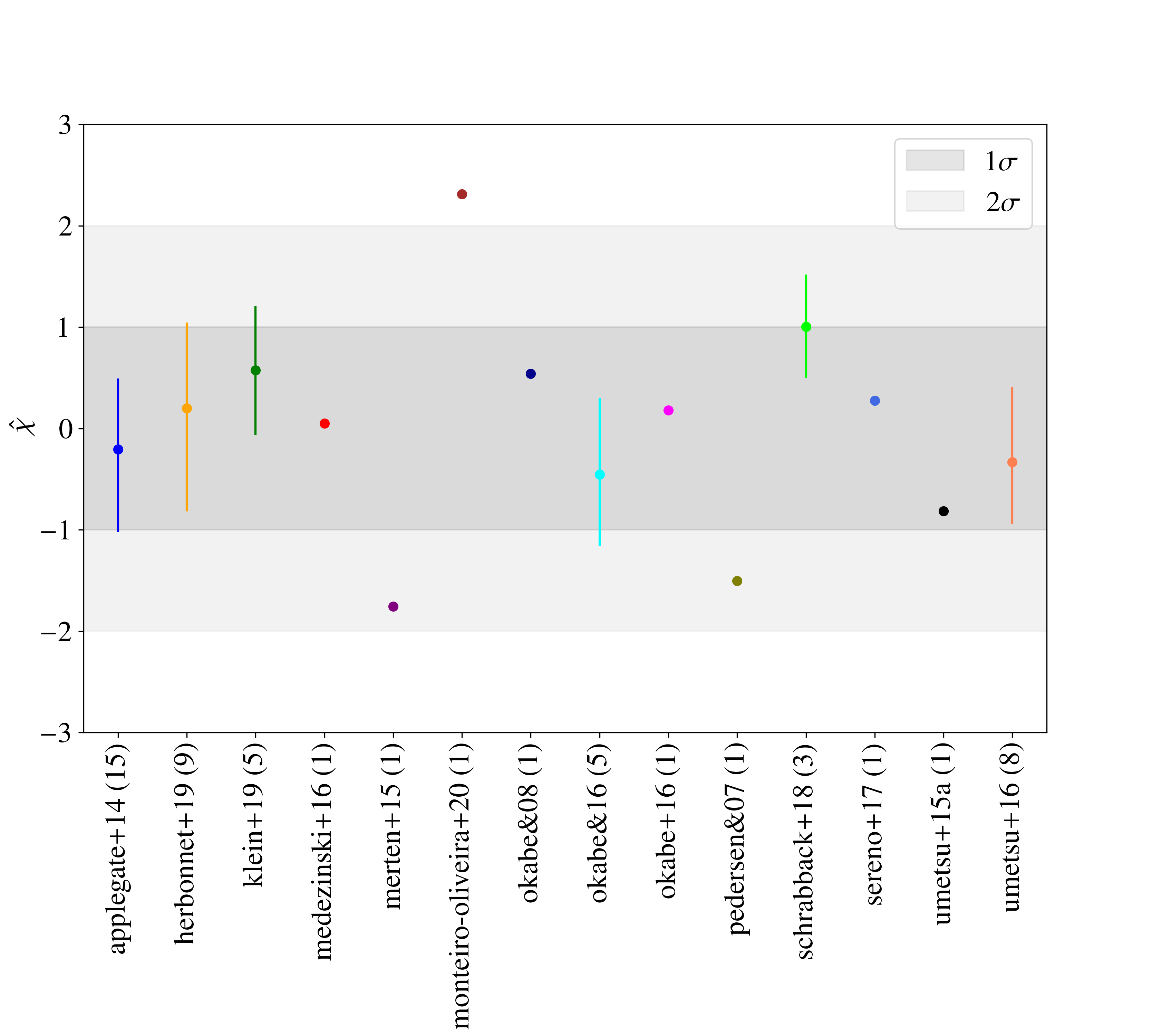}
	\caption{Values of $\hat{\chi}$ for the clusters in the \textit{reference} sample with respect to the reference scaling relation. We distinguish the results from the various lensing works used to build the CoMaLit sample with different colour markers. The number of clusters used in the \textit{reference} sample from each of the works is indicated in brackets. For works with multiple clusters, we give the mean $\hat{\chi}$ and the 16th to 84th percentiles over all those clusters. Grey shaded areas indicate $1\sigma$ and $2\sigma$ regions.}
	\label{fig:impactcomalit}
\end{figure}

Regarding the inhomogeneity of the lensing masses, we have exploited the compilation of mass estimates from different works standardised in the CoMaLit catalogue. We have treated all the CoMaLit masses equally, no matter the work from which the lensing mass has been extracted. But the different quality of the data and/or the methods used in each of the original works make the uncertainties of lensing masses not homogeneous within the CoMaLit sample. By propagating $\sigma_{\mathrm{sys \; lens}}$ we account, to first order, for the overall error of CoMaLit masses with respect to other estimates. A possible improvement would be to measure an independent systematic scatter $\sigma_{\mathrm{sys \; lens}}$ for each of the works used in the CoMaLit sample, but at the expenses of much poorer statistics. Instead, we quantify a posteriori the goodness of our best scaling relation (estimated with all the 53 clusters in the \textit{reference} sample, Eq.~\ref{eq:srlensfinal} and \ref{eq:srhsefinal}) for the cluster masses extracted from each of the different works within the CoMaLit catalogue.  In Fig.~\ref{fig:impactcomalit} we show, for the clusters obtained from each of the lensing works, the corresponding $\hat{\chi}$ defined from Eq.~\ref{eq:goodnessoffit} as
\begin{equation}
    \hat{\chi} =  \frac{ \ln M^{\mathrm{HSE}}_{i}  - \ln M^{\mathrm{HSE}}\left( \ln M^{\mathrm{lens}}_{i},  z_{i}, \vartheta\right)}{\sqrt{ \delta_{\mathrm{HSE}, i}^2 +  \left(\sigma^{\mathrm{HSE}} \right)^2+    \left(\beta^{\mathrm{HSE}} \right)^2 \left[ \delta_{\mathrm{lens}, i}^2 +  \left(\sigma^{\mathrm{lens}} \right)^2 \right]} }.
   \label{eq:chihat}
\end{equation}
For those works with several clusters in our \textit{reference} sample, we give the mean value and the 16th to 84th percentiles over all the used clusters. We observe that only `merten+15' \citep{merten2015}, `monteiro-oliveira+20' \citep{monteiro2020}, and `pedersen\&07' \citep{pedersen2007} cluster masses are at more than $1\sigma$. The cluster from `monteiro-oliveira+20' at more than $2\sigma$ from the scaling relation is Abell1644 (on the top left of all our SRs), which is known for being a cluster in a merger scenario. Thus, we conclude that the scaling relation of reference fits well the large majority of the clusters in the \textit{reference} sample. There is no hint of a too bad or too good fit to any of the subsamples in the  CoMaLit catalogue. 

\section{Comparison to previous results}
\label{sec:comparison}

\begin{figure}[h]
	\centering
	 \includegraphics[trim={0pt 0pt 0pt 0pt},scale=0.35]{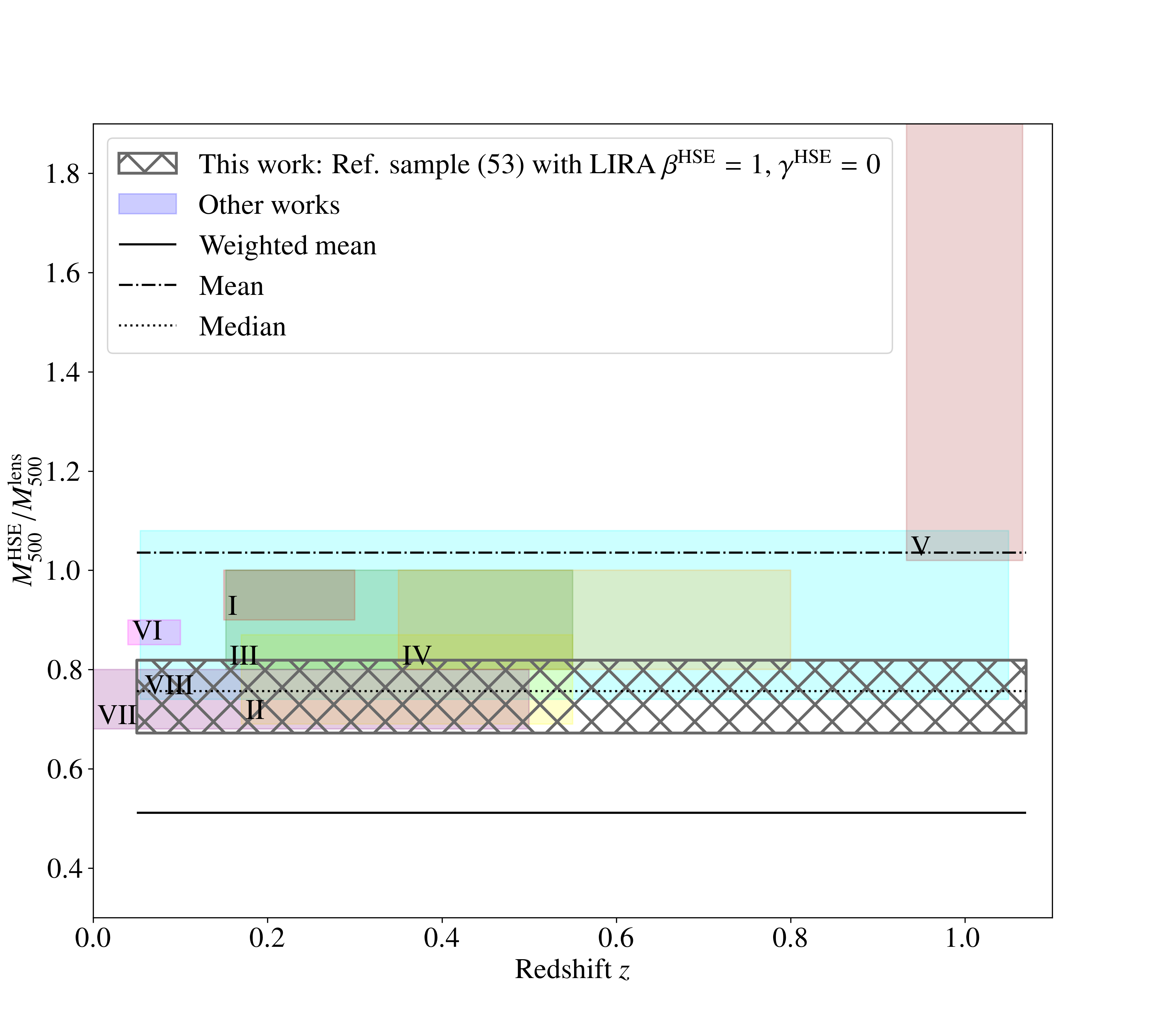}
	\caption{HSE-to-lensing mass ratio with respect to redshift. Shaded areas indicate different results from different works in the literature. See text and Table~\ref{tab:literature} to identify Roman numerals with the works. The horizontal grey hatched area represents the HSE-to-lensing mass ratio measured in this work assuming that HSE and lensing masses scale linearly with the true mass, accounting for the systematic scatter, and considering no evolution with redshift.}
	\label{fig:biasliterature}
\end{figure}

Similar studies to the one presented in this paper were previously done in the literature. However, the methods used to estimate the masses and to compute the HSE-to-lensing bias differ significantly from work to work. Thus, comparisons are delicate. In Fig.~\ref{fig:biasliterature} we present our best bias estimate together with the HSE-to-lensing $M_{500}$ ratios obtained in the works detailed below. We use Roman numerals to refer to each result from the literature. The different results are also summarised in Table~\ref{tab:literature}.

\renewcommand{\arraystretch}{1.4}
    \scriptsize
    \begin{table*}[]
      \scriptsize
      \centering
      \caption{HSE-to-lensing mass bias values from resolved mass profiles.}
        \begin{tabular}{c|c|c|c|c|c}

          \hline
          \hline

          Reference & Sample & \# of clusters &  Redshift & $(1-b) = M^{\mathrm{HSE}}_{500}/M^{\mathrm{lens}}_{500}$ & Notes \\\hline
           
          ~& This work - Reference SR & 53 & $ 0.05< z< 1.07$ & $ 0.739^{+0.075}_{-0.070}$ & Propagating the systematic uncertainties, accounting for intrinsic dispersion \\
          I & \citet{Smith_2015} & 50 &  $0.15<z < 0.3$ & $0.95\pm 0.05$ & Weighted mean \\
          II & \citet{Mahdavi_2008}  & 18 & $0.170 < z < 0.547$ &  $0.78 \pm 0.09$ & $M^{\mathrm{HSE}}(R^{\mathrm{lens}}_{500})$ \\
          III & \citet{Mahdavi_2013} & 50 & $0.152 < z < 0.55$ & $0.8 -1 $ & $M^{\mathrm{HSE}}(R^{\mathrm{lens}}_{500})$ \\
          IV & \citet{israel} & 8 & $0.35 < z < 0.80$ & $0.8 -1 $ & Global temperature profile for the whole sample, $M^{\mathrm{HSE}}(R^{\mathrm{lens}}_{500})$ \\
          V &  \citet{bartalucci2018} & 4 & $0.933 < z< 1.066$ & $1.39^{+0.51}_{-0.37}$ & Weighted mean\\
          VI & \citet{eckert2022} & 12 & $0.047 <z <0.09 $ & $0.85- 0.9$ & ~\\
          VII & \citet{lovisari2020} & 62 &  $z<0.5$ & $0.74 \pm 0.06$ & Accounting for intrinsic dispersion\\
          VIII &  \citet{sereno2020} & 100 &  $0.054 <z <1.050$ &  $0.91 \pm 0.17 $  & Accounting for intrinsic dispersion, temperature measured within 300~kpc\\
           \hline    
        \end{tabular}
        \vspace*{0.2cm}  
        \begin{tablenotes}         
        \small
      \item \textbf{Notes.} We report our reference result and different values from the literature. The last column indicates the singularity of each analysis.
       \end{tablenotes}        
        \vspace*{0.2cm}
        \label{tab:literature}
    \end{table*}
\normalsize

The HSE-to-lensing mass bias was measured in \citet{Smith_2015} with the 50 clusters from the LoCuSS sample ($0.15<z < 0.3$). By using resolved HSE mass estimates, they computed the weighted mean HSE-to-lensing bias: $1-b = 0.95\pm 0.05$ (I in Fig.~\ref{fig:biasliterature}). Uncertainties were calculated from the standard deviation of 1000 bootstrap samples geometric means. Following the equations \citep[Eq.~ 1 and 2 in][]{Smith_2015} used to calculate the weighted mean in \citet{Smith_2015} we obtain for our \textit{reference} sample a mean bias of: $1-b =0.763$ and $0.818$ not including and including, respectively, the systematic error in the uncertainty of each mass estimate. Considering, as in \citet{Smith_2015}, only the clusters in the redshift range $0.15<z < 0.3$, we obtain $1-b =0.769$ and $0.720$ with and without the systematic scatter. The difference between the bias estimated in \citet{Smith_2015} and the results obtained in our work could originate from the larger HSE mass estimates in \citet{Smith_2015}. Bright blue markers in the left panel in Fig.~\ref{fig:initialselection} show that HSE masses used in \citet{Smith_2015} tend to be larger than the \textit{homogeneous} ones.

In \citet{Mahdavi_2008} authors compared the HSE and lensing masses evaluated at the same radius, in particular at the $R_{500}$ measured from the lensing mass profile of each cluster. With a sample of 18 clusters, \citet{Mahdavi_2008} concluded that at $R_{500}^{\mathrm{lens}}$ the ratio of masses is $M^{\mathrm{HSE}}/M^{\mathrm{lens}} = 0.78 \pm 0.09$ (II). Extending the analysis, the HSE-to-lensing mass bias obtained in \citet{Mahdavi_2013} is consistent with no bias for cool-core clusters, while $(1-b) \sim 0.8$ for non-cool core clusters (III). In the same line, authors in \citet{israel} concluded, from the study of 8 clusters with redshifts $0.35 < z < 0.80$, that HSE and lensing masses differ by 0 to 20\% (IV).

By using very high redshift clusters ($0.933 < z< 1.066$), \citet{bartalucci2018} obtained that HSE masses from X-rays are a factor of $1.39^{+0.51}_{-0.37}$ (V) larger than weak lensing estimates, in contradiction with the rest of the results. The clusters in \citet{bartalucci2018} are the highest redshift clusters in our \textit{reference} sample (Sect.~\ref{sec:homogsample}). Using the same HSE masses as in \citet{bartalucci2018}, but with the CoMaLit lensing estimates we obtain an error-weighted mean ratio of $M^{\mathrm{HSE}}_{500}/M^{\mathrm{lens}}_{500} =1.56 (1.58)$ not including (including) the systematic error in the uncertainty of each mass estimate. Instead, the error-weighted mean ratio for our full \textit{reference} sample is $M^{\mathrm{HSE}}_{500}/M^{\mathrm{lens}}_{500} =0.47 (0.51)$. For the clusters in the X-COP sample, \citet{eckert2022} found that HSE masses estimated using XMM-\textit{Newton} data are 10 to 15\% lower than the lensing estimates in \citet{herbonnet2020} (VI). With a different approach and assuming that the gas fraction in clusters is constant, \citet{eckert2019} obtained that HSE masses are biased (with respect to the true total mass) by 7\% at $R_{500}$. These results differ significantly from the bias values obtained in this paper. HSE masses in \citet{eckert2019} were reconstructed making use of excellently well resolved mass profiles, therefore, unless there are unidentified systematic effects, HSE masses in \citet{eckert2019} should be very reliable. The small bias values obtained for the low redshift ($0.047 < z < 0.09$) clusters in \citet{eckert2019} could then indicate that there is indeed a redshift dependence in the HSE bias and that low redshift cluster HSE masses are less biased.

Regarding also the evolution of the bias with redshift, which we have largely discussed in Sect.~\ref{sec:bias} and \ref{sec:SR}, the weak tendency for a larger bias at higher redshift seems to be in line with the results from \citet{wicker} and \citet{Smith_2015}.

Particularly interesting are the comparisons to \citet{sereno2020}, \citet{lovisari2020}, and \citet{serenoettoricomalit} works, where the methods are equivalent to the ones employed in this paper, making use of the LIRA code and accounting for the intrinsic dispersion of HSE and lensing masses to the SR. The analysis in \citet{lovisari2020} compares the HSE masses obtained with XMM-\textit{Newton} data \citep[from][]{lovisari2020apj} to lensing estimates in the CoMaLit LC$^2$ catalogue, for 62 clusters from the \textit{Planck}-ESZ sample with $z<0.5$. With this sample, authors obtain $1-b = 0.74 \pm 0.06$ (VII) and no redshift evolution. This is in excellent agreement with our result. In \citet{lovisari2020} the results found with CoMaLit lensing masses are also compared to those obtained with other lensing masses from other works in the literature: the HSE-to-lensing mass ratio spans from $\sim 0.6$ to $\sim 1$ depending on the used dataset. 

Conclusions are along the same line in \citet{serenoettoricomalit}, where different samples with HSE and lensing mass estimates are used to measure the scaling relation and, consequently, the HSE-to-lensing bias. The effect that intrinsic scatters have on the determination of scaling relations is also studied in \citet{serenoettoricomalit}. They conclude that not taking into account explicitly the scatter of masses makes scaling relations flatter, as we see when using BCES instead of LIRA \citep[also in agreement with][]{lovisari2020apj}.  While the intrinsic scatter for lensing masses obtained in \citet{serenoettoricomalit} is of the order of the expected values from simulations ($\sim 10-15 \%$), the intrinsic dispersion for HSE masses is larger than expected ($\sim 20-30\%$). An underestimation of the statistical uncertainties in HSE masses could be the reason, according to \citet{serenoettoricomalit}, for this large scatter. Accounting for the systematic scatter in the uncertainty of each cluster mass, as described in this paper, could help to have more realistic uncertainties for the HSE mass estimates. The HSE-to-lensing mass ratio in \citet{serenoettoricomalit} depends again on the used sample and data and spans from $\sim 0.5$ to $\sim 1$. We have reproduced the same result by separating the sample in redshift ranges. 

Also \citet{sereno2020} used the Bayesian hierarchical modelling from \citet{serenolira} to fit a scaling relation between HSE masses from XMM-\textit{Newton} data and weak lensing masses of clusters in the Hyper Suprime-Cam Survey \citep{xxlII}. The median redshift of the 100 clusters in the sample is $z = 0.30$, spanning from $z=0.054$ to $z=1.050$. Thus, the analysis in \citet{sereno2020} is probably the closest study to our work. Nevertheless, to get temperature profiles that reach $R_{500}$ with X-ray data (to compute then the HSE mass), in \citet{sereno2020} a model was iteratively fitted to the integrated temperature measured per cluster within 300~kpc, well below $R_{500}$. Assuming $\beta^{\mathrm{HSE}} =1$, $\beta^{\mathrm{lens}} =1$, and $\alpha^{\mathrm{lens}} =0$ they obtained: $\alpha^{\mathrm{HSE}}= -0.04\pm 0.08$, $\sigma^{\mathrm{HSE}}= 0.31 \pm 0.05$, and $\sigma^{\mathrm{lens}}= 0.37 \pm 0.06$. According to \citet{sereno2020}, the difference between HSE and lensing masses is of $ b = 0.09 \pm 0.17 $ (VIII). The $\alpha^{\mathrm{HSE}}$ from \citet{sereno2020} is at $3\sigma$ from our result with the full \textit{reference} sample ($\alpha^{\mathrm{HSE}} = -0.338^{+0.105}_{-0.097}$ without accounting for the systematic uncertainties). Their values for $\sigma^{\mathrm{HSE}}$ and $\sigma^{\mathrm{lens}}$ agree with the intrinsic scatter values that we obtain when we do not account for the systematic uncertainties ($\sigma^{\mathrm{HSE}} = 0.304^{+0.069}_{-0.072}$ and $\sigma^{\mathrm{lens}} = 0.305^{+0.080}_{-0.083}$).

In addition, the behaviour of the HSE-to-lensing mass bias could vary with the overdensity at which masses are measured. By estimating weak lensing masses and HSE masses from X-rays at $R_{200}$, \citet{jee2011} concluded that for a sample of 14 very massive and distant clusters ($0.83 < z < 1.46$), the HSE and lensing masses are compatible. However, the HSE masses were obtained from the extrapolation of a singular isothermal sphere profile to reach $R_{200}$, which likely limits the validity of their HSE mass estimates. Similarly, in \citet{amodeo} authors compared $M_{200}$ masses reconstructed from \textit{Chandra} data (although the radial reach of \textit{Chandra} is way below $R_{200}$) to their lensing estimates, and concluded that both mass estimates are in agreement. No evolution with redshift was detected in \citet{amodeo}. We prefer to avoid extrapolating the mass profiles to reach $R_{200}$. 

\section{Summary and conclusions}
\label{sec:conclusions}

In this work we have investigated the HSE-to-lensing mass bias with masses inferred at $R_{500}$ from resolved profiles. We carefully selected the clusters and obtained a \textit{reference} sample with 53 clusters with redshifts spanning from $z=0.05$ to $1.07$. This is the largest redshift range analysed homogeneously with this type of data, having access to X-ray HSE masses obtained from resolved profiles. HSE masses were estimated with the XMM-\textit{Newton} mass reconstruction reference pipeline and lensing masses were extracted from the LC$^2$ CoMaLit catalogue. 

In order to account for possible systematic effects in the reference analysis, we compared the XMM-\textit{Newton} and CoMaLit masses to other estimates from the literature. The obtained systematic scatters were propagated to our analyses, but all the main conclusions remain unchanged when considering or not these additional systematic dispersions on the HSE and lensing mass uncertainties.

We performed different tests in the measurement of the HSE-to-lensing mass scaling relation and bias, varying the redshift range and the scaling relation model. Our main conclusions are the following:
\begin{enumerate}
\item Assuming that HSE and lensing masses scale linearly with the true mass and considering $\sigma_{\mathrm{sys \; HSE}}^2$ and $\sigma_{\mathrm{sys \; lens}}^2$, we measure for the 53 clusters in the \textit{reference} sample a HSE-to-lensing mass ratio of $M^{\mathrm{HSE}}_{500}/ M^{\mathrm{lens}}_{500} = (1-b) = 0.739^{+0.075}_{- 0.070} \;  \mathrm{(stat.)} \; \pm \; 0.226 \; \mathrm{(intrin. \; scatter)}$.

\item We find that the best scaling relation between HSE and lensing masses is our scaling relation of reference, where we assume that there is no evolution with mass and redshift and that HSE and lensing masses scale linearly. We obtain: $\alpha^{\mathrm{HSE}} = -0.303^{+0.101}_{-0.095}$, $\sigma^{\mathrm{HSE}} = 0.166^{+0.086}_{-0.101}$, and $\sigma^{\mathrm{lens}} =  0.257^{+0.080}_{-0.092}$.

\item When we let the SR evolve with redshift, we observe a trend towards a larger discrepancy between HSE and lensing masses at high redshift, but it is not statistically significant. In conclusion, there is no evidence of evolution with redshift. The dependence of the HSE-to-lensing mass bias on the mass of the clusters is not confirmed either.
  
  \item Given the size of the sample, single clusters can be driving the fits and special care needs to be taken for clusters with very small uncertainties. We have investigated the case of CL~J1226.9+3332 galaxy cluster, whose impact is crucial when determining the bias at high redshift.

  \item Ignoring the intrinsic scatter of HSE and lensing masses with respect to the true mass of clusters introduces a bias in the measurement of the HSE-to-lensing mass bias.

\end{enumerate}

Additional considerations are needed to compare the HSE-to-lensing mass bias obtained in this work to the bias needed to reconcile cluster number counts and CMB power spectrum results for several reasons: 1) the HSE masses used in cluster number count analyses are not direct HSE mass measurements, but masses obtained from a SZ (or X-ray) measurement through a SZ-mass (or X-ray-mass) scaling relation, 2) lensing masses can also be biased with respect to the true mass of clusters \citep{Becker_2011}, and 3) this sample is not representative of the cluster population in any given survey. Instead, this study provide a step forwards in our understanding of the deviation from hydrostatic equilibrium of galaxy clusters and of the impact of systematic and intrinsic errors.


\begin{acknowledgements}
We thank M. Sereno \& S. Ettori for very useful comments. This project was carried out using the Python libraries \texttt{matplotlib} \citep{Hunter2007}, \texttt{numpy} \citep{Harris2020}, \texttt{astropy} \citep{Astropy2013, Astropy2018},  \texttt{scipy} \citep{2020SciPy-NMeth}, and \texttt{ChainConsumer} \citep{Hinton2016}. This work benefited the financial support from CNRS/INSU and CNES. AF also thanks financial support by Universidad de La Laguna (ULL), NextGenerationEU/PRTR and Ministerio de Universidades (MIU) (UNI/511/2021) through grant ``Margarita Salas''. 

\end{acknowledgements}
\normalsize

\bibliographystyle{aa} 
\bibliography{mybibliography} 
%


\begin{appendix}

  
\section{Measuring the systematic scatter}
\begin{figure*}
      
   \begin{minipage}[t]{0.5\textwidth} 
    \includegraphics[trim={10pt 0pt 0pt 0pt},scale=0.32]{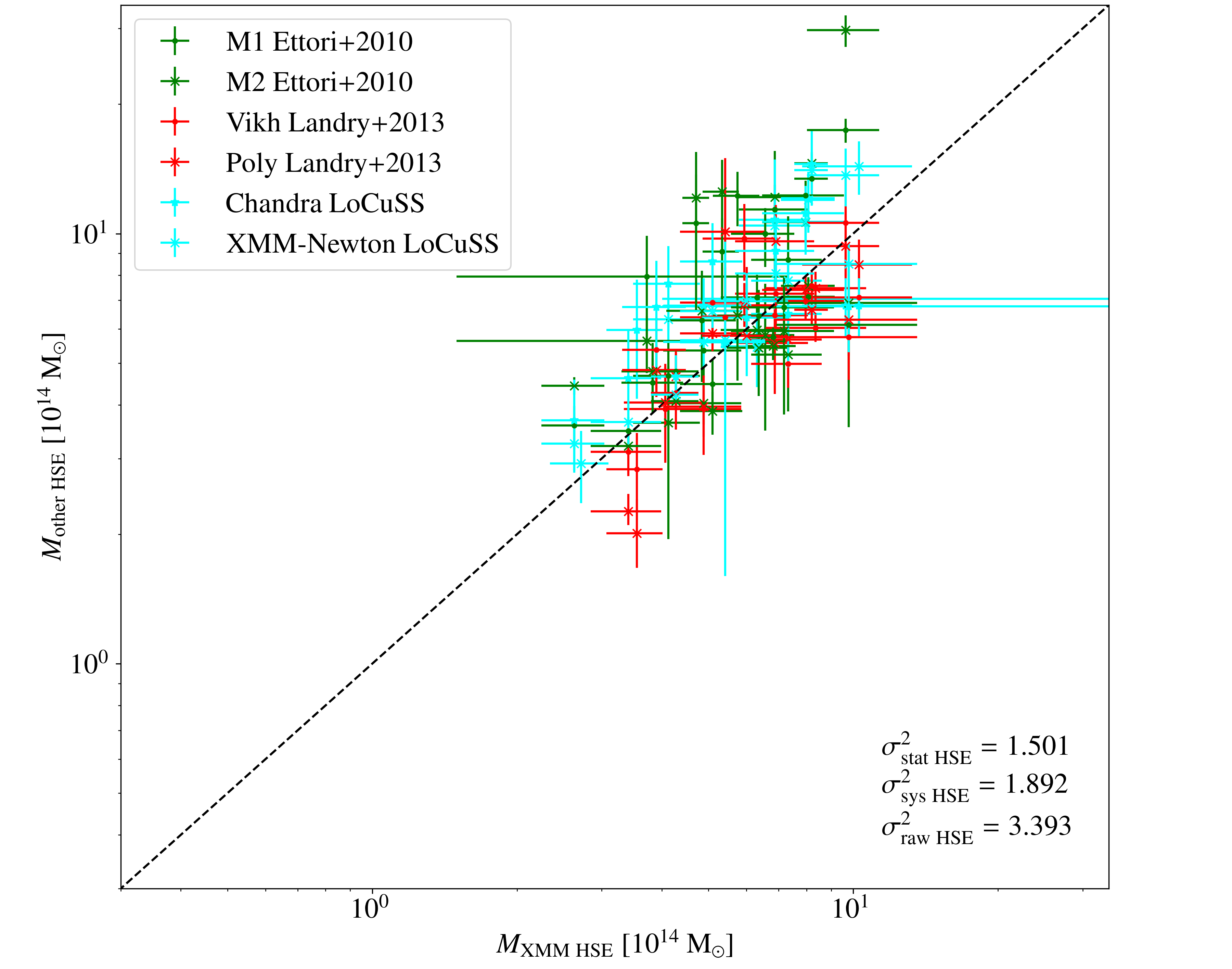}
   \end{minipage}
   \hfill   
   \begin{minipage}[t]{0.5\textwidth}
    \includegraphics[trim={10pt 0pt 0pt 0pt},scale=0.32]{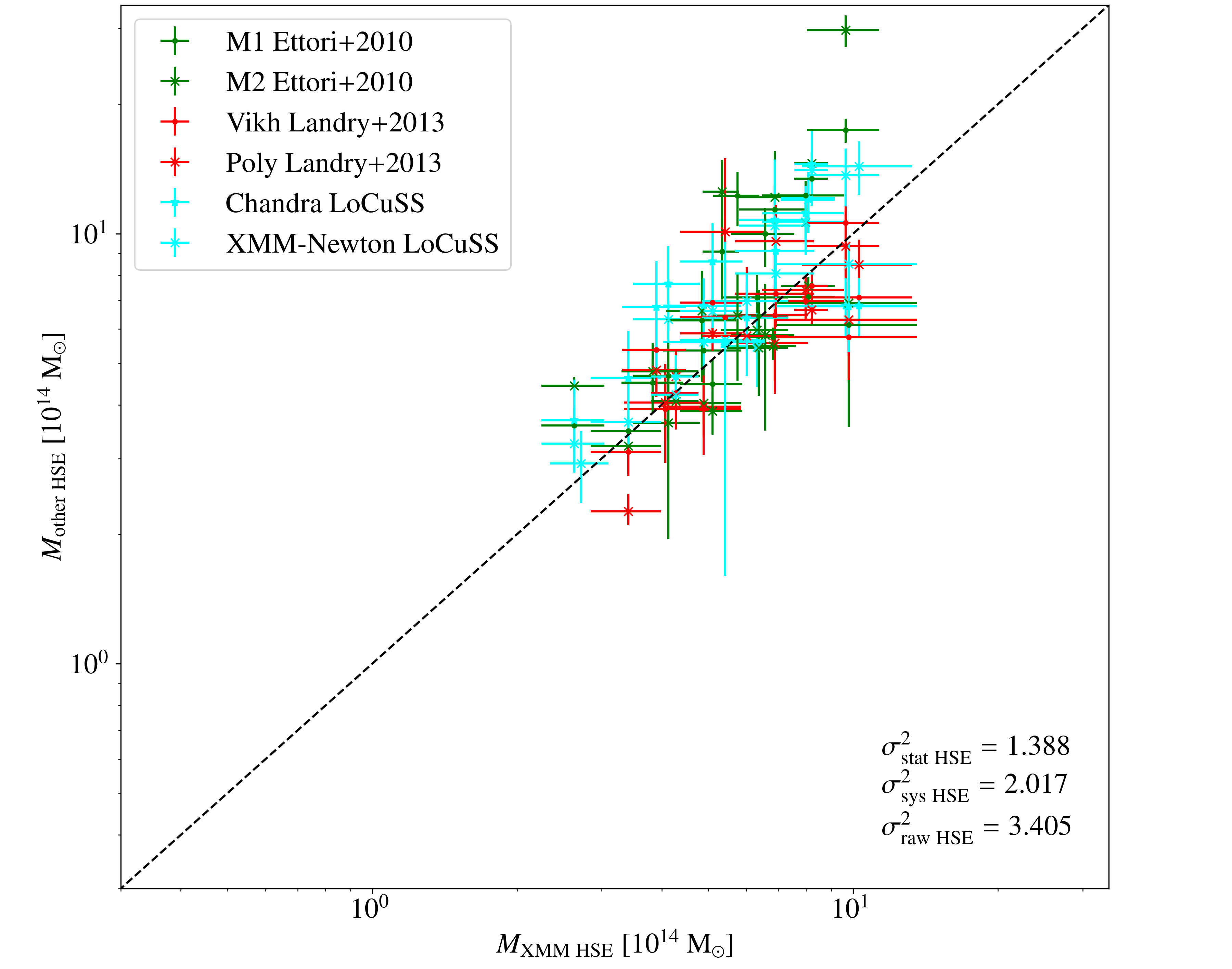} 
   \end{minipage}
   \hfill
    \begin{minipage}[b]{0.5\textwidth}
      \includegraphics[trim={10pt 0pt 0pt 0pt},scale=0.32]{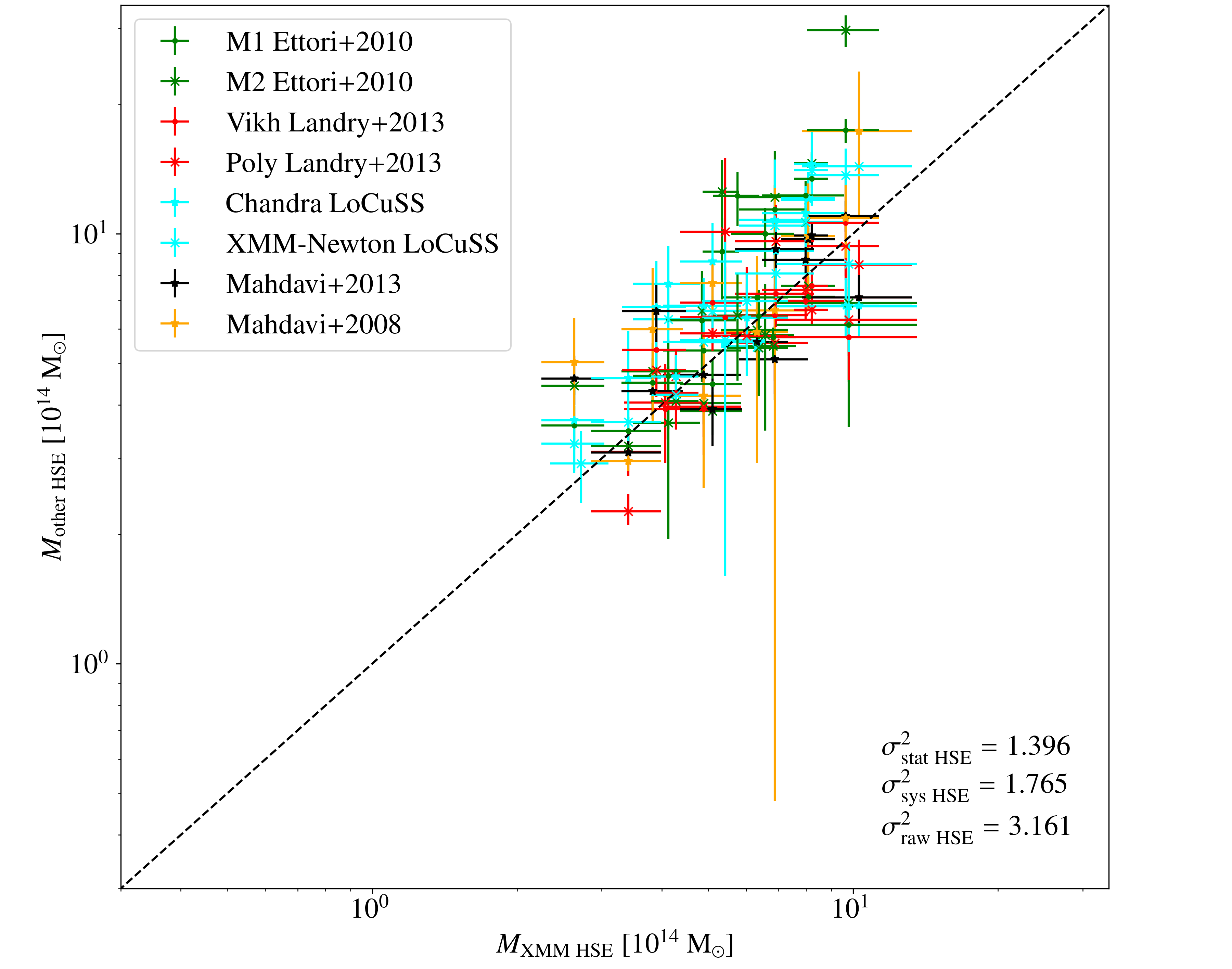}
    \end{minipage}
     \hfill
    \begin{minipage}[b]{0.5\textwidth}
        \includegraphics[trim={10pt 0pt 0pt 0pt},scale=0.32]{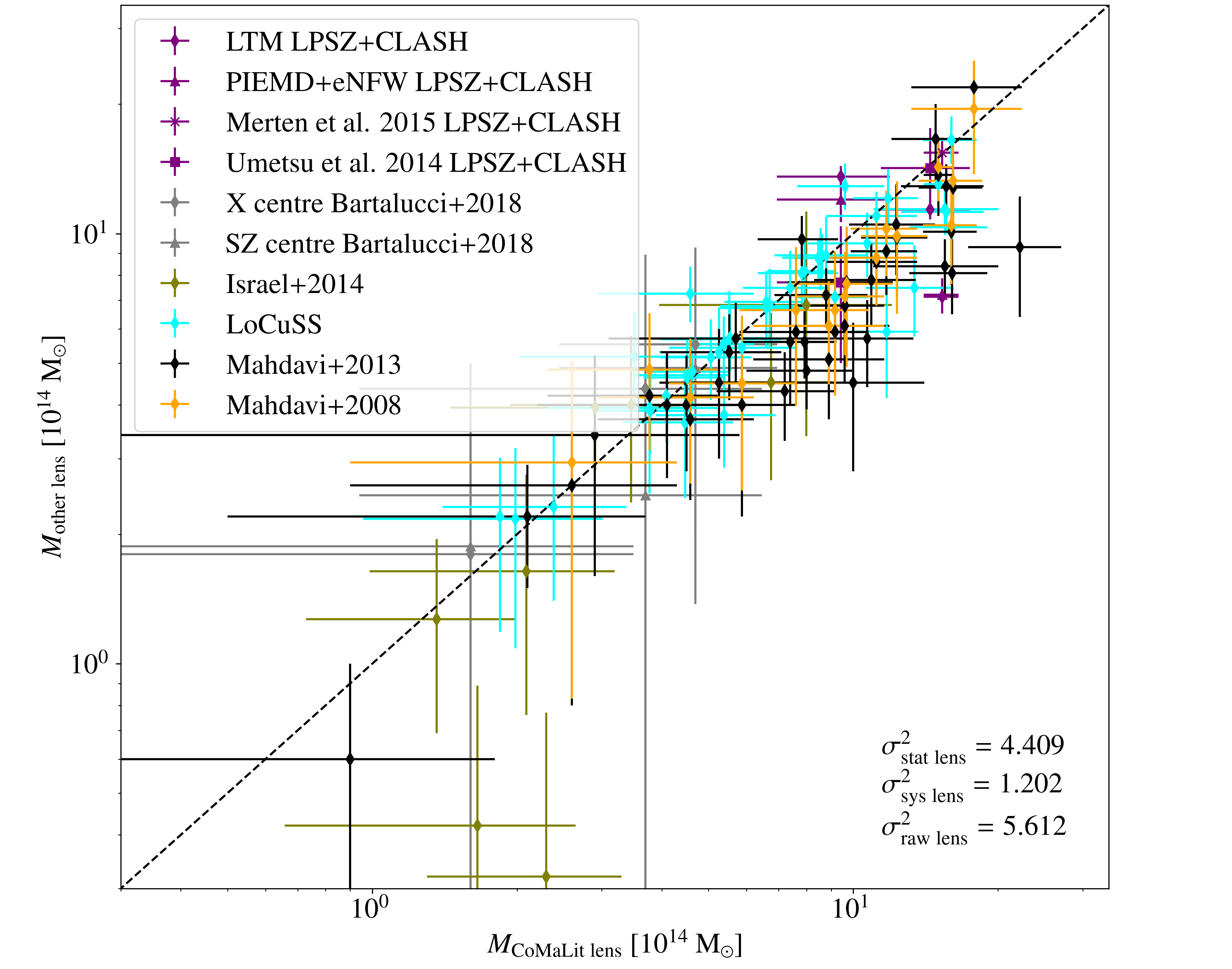}
    \end{minipage}
    
    \caption{Comparison of HSE and lensing mass estimates from the \textit{homogeneous} samples in this work (XMM-\textit{Newton} and CoMaLit) with respect to other estimates from the literature (\textit{comparison} sample). Top: Relation between X-ray masses from literature and from the XMM-\textit{Newton} reference pipeline without accounting for $M^{\mathrm{HSE}}(<R_{500}^{\mathrm{lens}})$. In the left (right) the clusters with very large uncertainties and with very different XMM-\textit{Newton} and CoMaLit centres are considered (not considered). Bottom: Same figure as Fig.~\ref{fig:initialselection}, but not accounting for clusters with very large uncertainties and with very different XMM-\textit{Newton} and CoMaLit centres. The dashed lines show the 1:1 relation. We give the statistical, systematic, and raw variances in units of $10^{28}$ M$^2_{\odot}$ corresponding to the data points in each figure.}

    \label{fig:initialstrangemiscenter}
\end{figure*}

We present in Fig.~\ref{fig:initialstrangemiscenter} different measurements of the systematic scatter for HSE masses and lensing masses obtained from the reference analyses. Top panels do not account for the HSE masses obtained by evaluating the HSE mass profiles at the $R_{500}$ from lensing. In the bottom panels in Fig.~\ref{fig:initialstrangemiscenter} we present the systematic scatters once the cluster selection criteria from Sect.~\ref{sec:refsample} have been applied. We summarise all the scatter values in Table~\ref{tab:scatters}.

\renewcommand{\arraystretch}{1.4}    
    \scriptsize
    \begin{table*}[]
    \scriptsize
    \centering
     \caption{Raw, statistical, and systematic variances of the HSE and lensing mass estimates from the \textit{homogeneous} samples in this work (XMM-\textit{Newton} and CoMaLit) with respect to other estimates from the literature (\textit{comparison} samples). }
        \begin{tabular}{c|c|c|c|c}
          \hline
          \hline
          Sample &     $\sigma^2_{\mathrm{raw \; HSE}}$ &  $\sigma^2_{\mathrm{stat \; HSE}}$ &  $\sigma^2_{\mathrm{sys \; HSE}}$ & Figure \\ 
         ~ &  $[10^{28} \; \mathrm{M}_{\odot}^2]$ &  $[10^{28} \; \mathrm{M}_{\odot}^2]$ &  $[10^{28} \; \mathrm{M}_{\odot}^2]$ \\\hline
  
     All & 3.231 & 1.507 & 1.724 & Fig.~\ref{fig:initialselection} \\
     Without $M^{\mathrm{HSE}}(<R_{500}^{\mathrm{lens}})$ & 3.393 & 1.501 & 1.892 & Fig.~\ref{fig:initialstrangemiscenter}\\
     Without clusters discarded in Sect.~\ref{sec:refsample} & 3.161 & 1.396 & 1.765 & Fig.~\ref{fig:initialselection}\\
     Without $M^{\mathrm{HSE}}(<R_{500}^{\mathrm{lens}})$ and clusters discarded in Sect.~\ref{sec:refsample} & 3.405 & 1.388 & \textbf{2.017} & Fig.~\ref{fig:initialstrangemiscenter}\\
    \hline
    \hline
     ~ & $\sigma^2_{\mathrm{raw \; lens}}$ &  $\sigma^2_{\mathrm{stat \; lens}}$ &  $\sigma^2_{\mathrm{sys \; lens}}$ & Figure\\
    ~ &  $[10^{28} \; \mathrm{M}_{\odot}^2]$ &  $[10^{28} \; \mathrm{M}_{\odot}^2]$ & $[10^{28} \; \mathrm{M}_{\odot}^2]$ & ~ \\ \hline
  
     All &  5.280 & 4.340 & 0.940 & Fig.~\ref{fig:initialselection} \\
     Without clusters discarded in Sect.~\ref{sec:refsample}  & 5.612 & 4.409 & \textbf{1.202} & Fig.~\ref{fig:initialselection}\\\hline
        \end{tabular}
        \vspace*{0.2cm}  
        \begin{tablenotes}         
        \small
      \item \textbf{Notes.} We report the different values depending on the sample selection criteria, showing in bold the systematic scatters considered for the rest of the analysis.
       \end{tablenotes}        
        \vspace*{0.2cm}
        \label{tab:scatters}
    \end{table*}
\normalsize
\FloatBarrier

\section{Additional checks for the selection of the \textit{reference} sample}
\label{sec:additionalchecks}
In this section we detail the additional checks we have performed in Sect.~\ref{sec:refsample} to verify which clusters can be used for the HSE-lensing mass comparison and define the final sample. 

\subsection{Uncertainties of mass estimates}
\label{sec:uncertainties}

Regarding the uncertainties of the mass estimates in the \textit{homogeneous} sample (the 65 clusters in Sect.~\ref{sec:refsample}), we observe that masses and their uncertainties appear correlated, error bars being larger for more massive objects. As already mentioned, lensing mass uncertainties are larger than HSE ones and this is also true for relative uncertainties when calculated with respect to the value of the measured mass. For some clusters, the uncertainties on HSE masses are suspiciously large (larger than the value of the mass) or abnormal (negative error bars) and we decide to exclude for these reasons the clusters Abell119, Abell521 (the LoCuSS and Mahdavi+2013 cluster out of the plot in the left panel in Fig.~\ref{fig:initialselection}), and SPT-CLJ0516-54.

\subsection{Difference of centres for HSE and lensing masses}
\label{sec:miscentring}
For some of the clusters, the HSE and lensing masses have been reconstructed assuming different cluster centre positions. We investigate in the following if this miscentring is correlated to the mass ratio and redshift and how it may affect the mass estimates. 

\subsubsection{Correlation between miscentring and bias or redshift}

\begin{figure*}[h]
        \begin{minipage}[b]{0.5\textwidth}
        \centering
        \includegraphics[trim={0pt 4pt 0pt 4pt},scale=0.38]{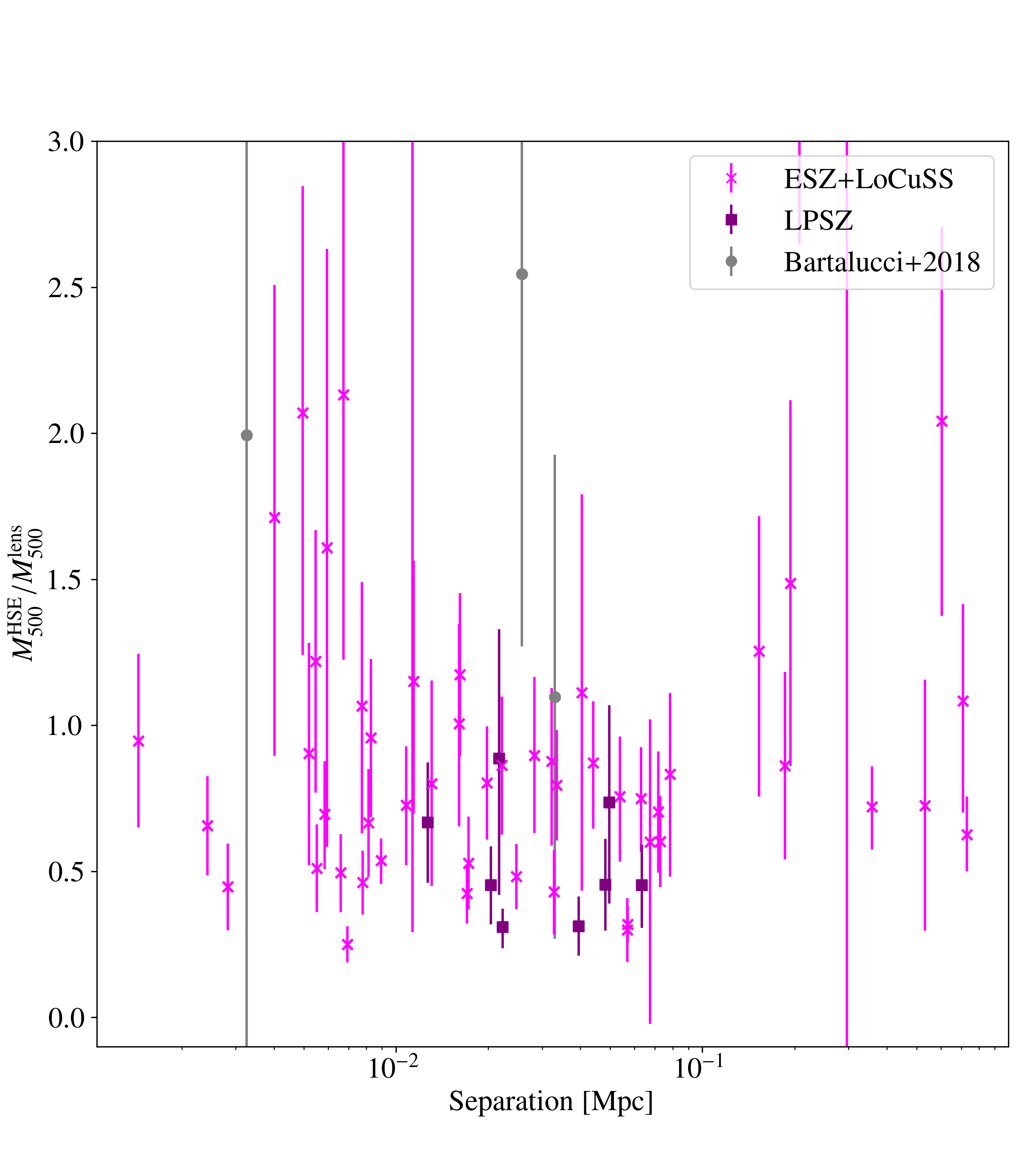}
        \end{minipage}
        \hfill
        \begin{minipage}[b]{0.5\textwidth}
        \centering
        \includegraphics[trim={0pt 4pt 0pt 4pt},scale=0.38]{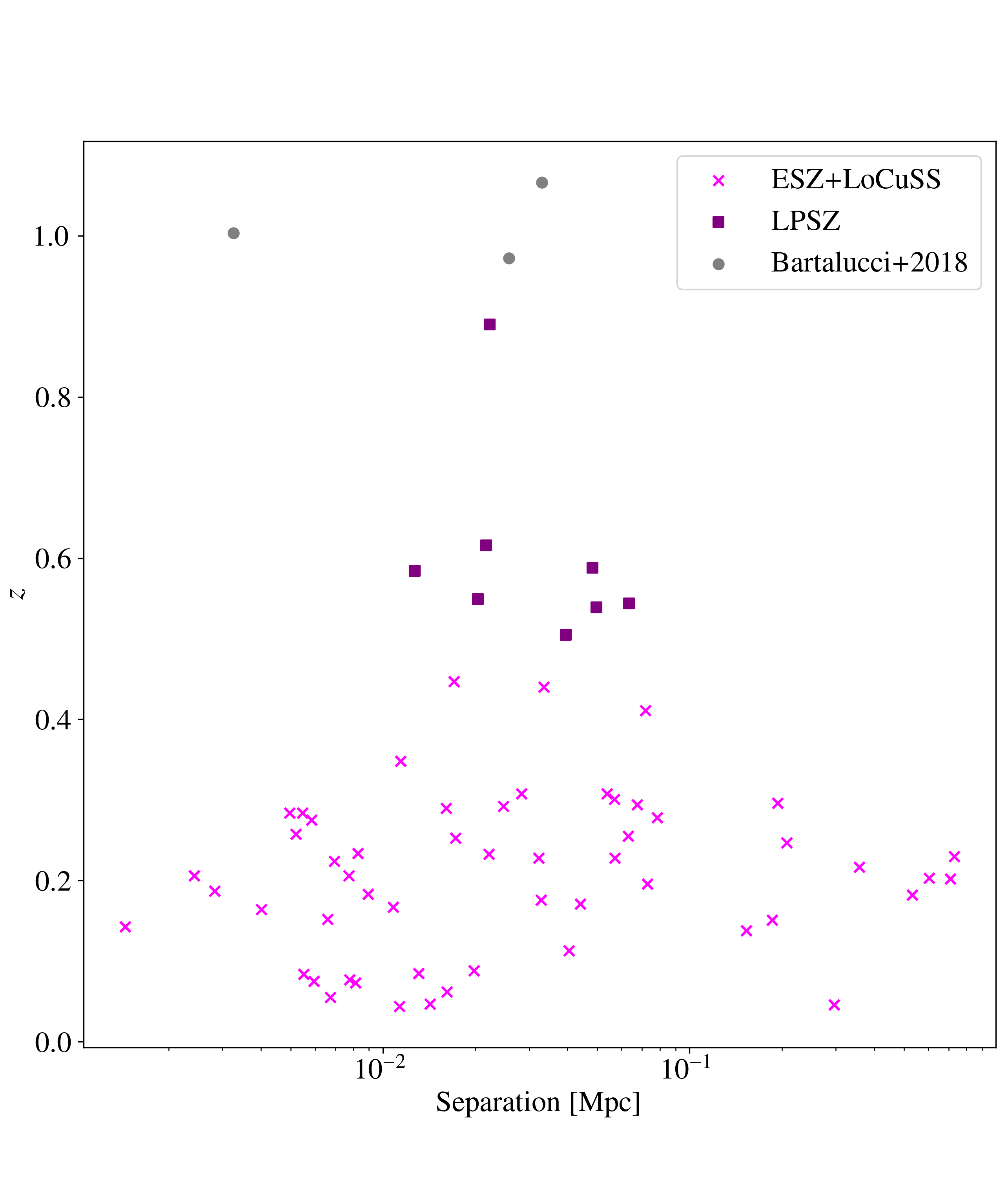}
        \end{minipage}
    \caption{Correlation between the HSE-to-lensing mass ratio (left) and redshift (right) of the 65 clusters in the XMM-\textit{Newton}-CoMaLit \textit{homogeneous} sample with respect to the separation between the centres assumed in the X-ray and lensing analyses. Error bars in the left panel do not account for the systematic scatters.}
    \label{fig:separations}
\end{figure*}

In the left panel in Fig.~\ref{fig:separations} we present the ratios between the HSE and lensing masses of the clusters with respect to the separations between the centres considered in the X-ray and lensing analyses. Each marker corresponds to one of the 65 clusters, showing in magenta crosses, purple squares, and grey circles the clusters from ESZ+LoCuSS, LPSZ, and Bartalucci+2018 samples, respectively. Error bars have been obtained from the propagation of the individual XMM-\textit{Newton} and CoMaLit uncertainties. The separation between X-ray and lensing centres goes from 1.5~kpc to 700~kpc. However, there is no significative correlation between the miscentring and the HSE-to-lensing mass ratio. 

The miscentring could also be related to the redshift of the cluster. In the right panel in Fig.~\ref{fig:separations} we show the redshift with respect to the separation of the centres. There is neither indication of correlation. We do not find any significative correlation between the redshift and the uncertainties of mass ratios either. 

The dynamical state of clusters is an important point to better understand the evolution of the HSE bias. In particular, relaxed clusters tend to have smaller HSE bias as compared to disturbed ones \citep{gianfagna}. The offset between the centres used in lensing and X-ray analyses could be an indicator of the departure from sphericity and equilibrium of clusters, resulting in a difficulty to define the centre. However, the absence of correlation between the separation of centres and the bias or the redshift does not enable any clear dynamical classification of the clusters in our sample.
  
\FloatBarrier

\subsubsection{Simulations to quantify the impact of miscentring}

It is evident that two mass estimates computed assuming very different centres are hardly comparable. Nevertheless, setting a quantitative limit of the acceptable separation between the two centres can be difficult. In this section we make use of simulated mock mass density profiles to check the impact that miscentring has on the mass profiles and, consequently, on $M_{500}$.

We simulate mass density profiles using the \texttt{profiley}\footnote{\url{https://profiley.readthedocs.io/en/latest/index.html}.} Python package, which contains already tested \citep{act2020} functions that describe density profiles, as well as the corresponding miscentred density profiles. We test a variety of density profile shapes following NFW, gNFW, truncated Navarro-Frenk-White \citep[tNFW,][]{tnfw}, and Hernquist \citep{hernquist} models. To initialise the density profiles we consider different $M_{500}$ (in the range $\sim 1$ to $9\times 10^{14}$ M$_{\odot}$) and model parameters. The parameters considered for each model are given in Table~\ref{tab:modelparams}.

\renewcommand{\arraystretch}{1.4}    
    \scriptsize
    \begin{table*}[]
    \scriptsize
        \centering
        \begin{tabular}{c|c|c|c|c|c|c}
          \hline
          \hline
     Model & $c_{500}$ & $\alpha$ & $\beta$ & $\gamma$ & $\tau$ & $\eta$     \\\hline
     NFW & 1.0, 2.25, 3.5, 4.75, 6.0 & ~ & ~ & ~ & ~ & ~ \\
     Hernquist & 1.0, 2.25, 3.5, 4.75, 6.0 & ~ & ~ & ~ & ~ & ~ \\
     gNFW & 6.0 & 0.3, 0.85, 1.4, 1.95, 2.5 & 2.0, 3.0, 4.0, 5.0, 6.0 & 0.3, 0.475, 0.65, 0.825, 1.0 & ~ & ~ \\
     tNFW & 6.0 & ~ & ~ & ~ & 0.2, 1.15, 2.1, 3.05, 4.0 & 0.5, 1.125, 1.75, 2.375, 3.0\\     
    \hline
        \end{tabular}
        \vspace*{0.2cm}  
        \caption{Different mass density models and their corresponding parameters used to study the miscentring effect in Fig.~\ref{fig:miscent}.} 
        \label{tab:modelparams}
    \end{table*}
\normalsize

\begin{figure}[h]
    \centering
    \includegraphics[scale=0.37, left]{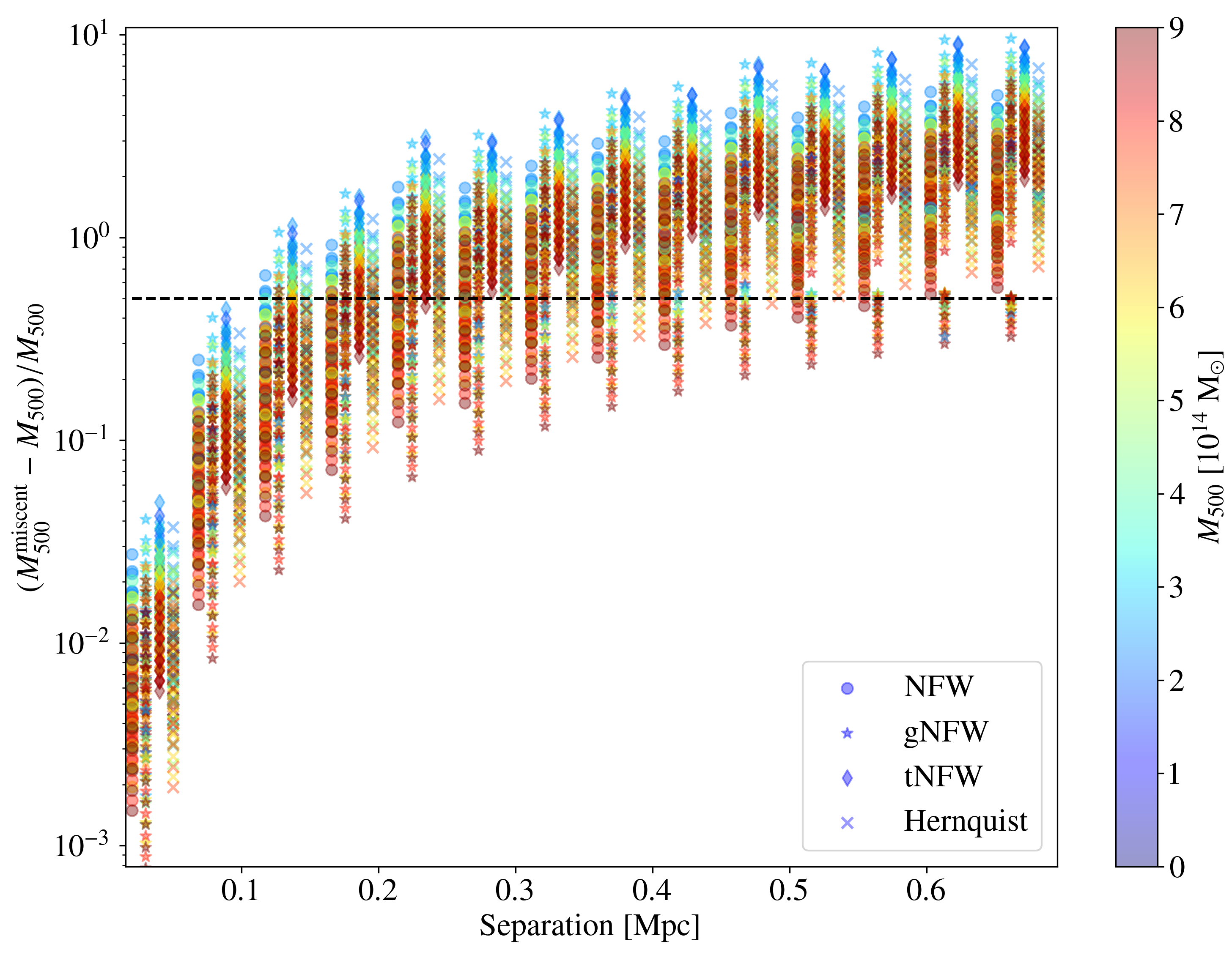}
    \caption{Relative error of $M_{500}$ with respect to the separation between the true and the considered centre for the mass reconstruction. Colours indicate the true $M_{500}$ and markers the mass density model. We consider the same separation for the four models, but gNFW, tNFW, and Hernquist markers have been shifted for visualisation purposes. The black dashed line indicates a $50\%$ error in $M_{500}$.}
    \label{fig:miscent}
\end{figure}

For each density profile we build the corresponding miscentred profiles by displacing the centre by 0.0 to 0.7~Mpc from the original one. We integrate each miscentred density profile to get a mass profile and obtain the miscentred $M_{500}$ estimate at the radius where the overdensity reaches $\Delta=500$, $M_{500}^{\mathrm{miscent}}$. In Fig.~\ref{fig:miscent} we show the relative error of the miscentred masses with respect to the separation to the true centre, for all type of profiles and for different redshifts, $z = [0.1 , 0.45, 0.8 , 1.15, 1.5]$. Each marker indicates a different mass density model and the colours show the true $M_{500}$. For visualisation purposes gNFW, tNFW, and Hernquist markers have been shifted by 0.01, 0.02, and 0.03~Mpc. As expected, the figure shows that computing the density profile from a centre that is separated from the true centre gives biased mass reconstructions. This bias increases with the separation, less massive clusters being more sensitive to this effect. The black dashed line in Fig.~\ref{fig:miscent} indicates $(M_{500}^{\mathrm{miscent}} - M_{500})/M_{500} = 0.5$.\\

\FloatBarrier

Considering the tested density profiles and mass range, the error due to miscentring is smaller than $50\%$ only if the distance to the real centre is smaller than 100~kpc. Therefore, we decide to exclude clusters for which X-ray and lensing centres are separated by more than 100~kpc. These clusters are: Abell3856, Abell3888, Abell773, Abell665, Abell267, 1E0657-56, Abell521, Abell3376E, Abell520, and Abell2163.


\subsection{Final sample}
    \begin{figure*}
        \begin{minipage}[b]{0.5\textwidth}
        \centering
        \includegraphics[trim={0pt 0pt 0pt 0pt},scale=0.35]{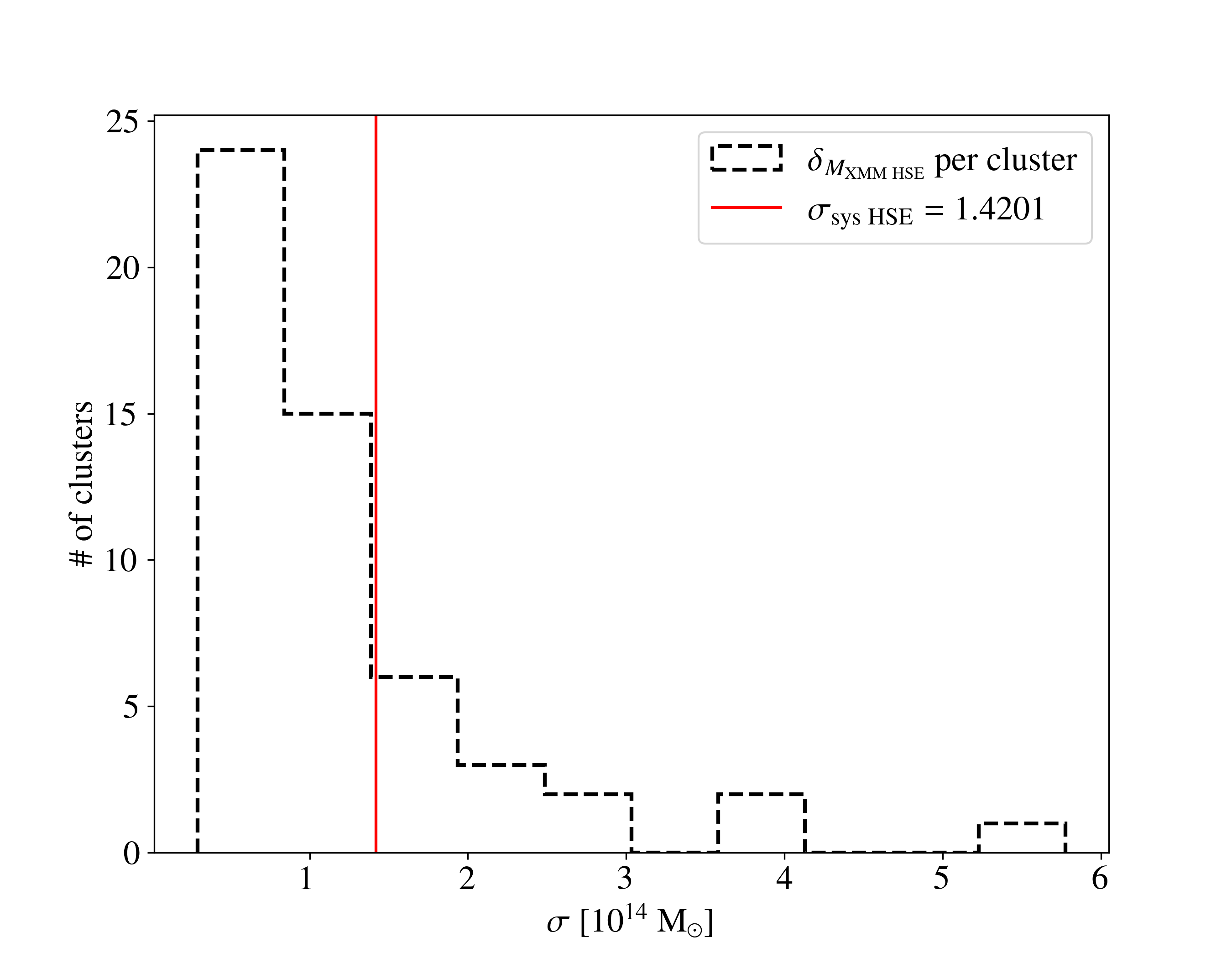}
        \end{minipage}
        \hfill
        \begin{minipage}[b]{0.5\textwidth}
        \centering
        \includegraphics[trim={0pt 0pt 0pt 0pt},scale=0.35]{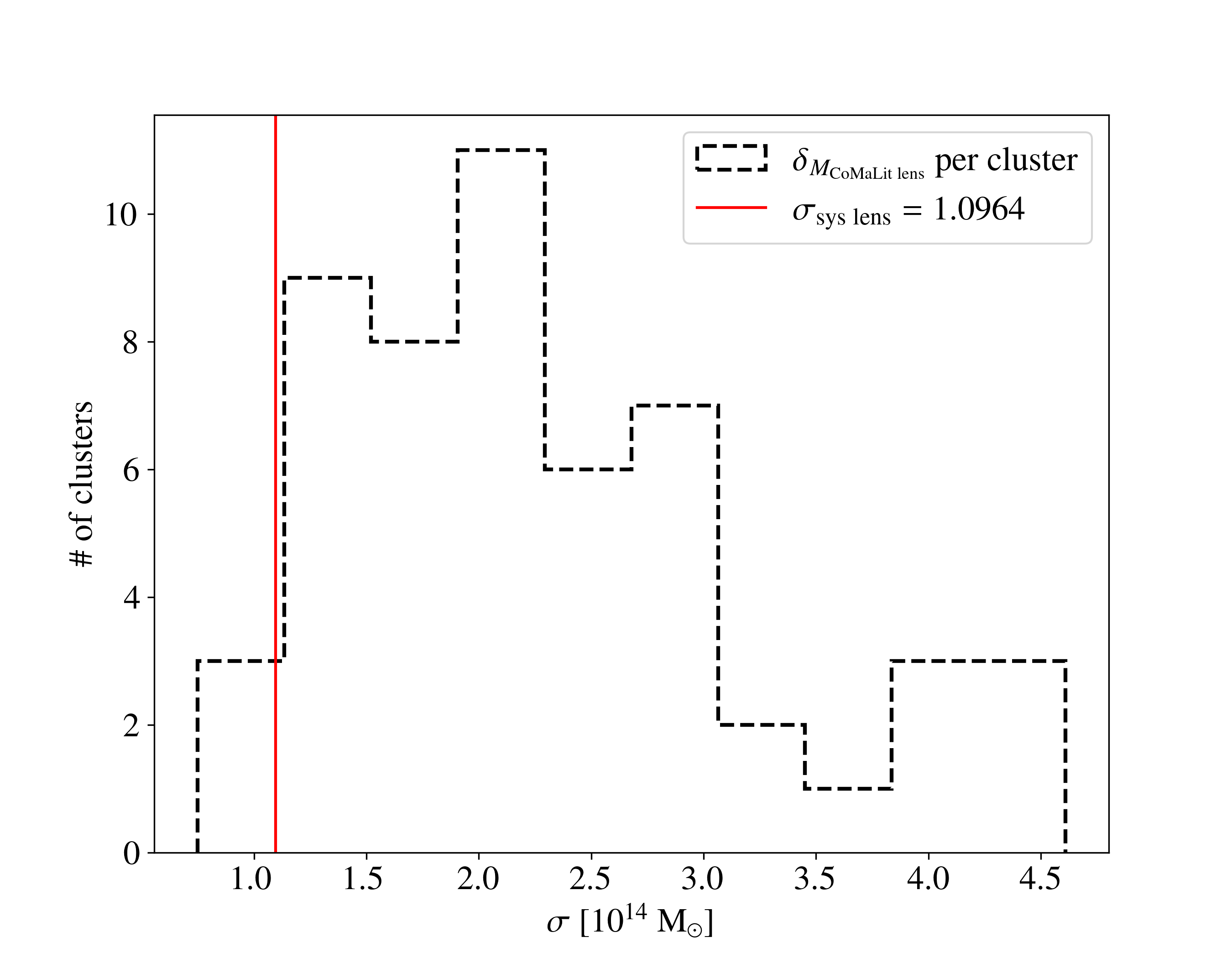}
        \end{minipage}
    \caption{Comparison between statistical uncertainties and systematic scatter. Dashed line histograms show the statistical uncertainties of all HSE mass estimates from XMM-\textit{Newton} reference analysis (left) and CoMaLit lensing mass estimates (right). Vertical red lines give the systematic scatter estimated with respect to other published results in Sect.~\ref{sec:refsample}. }
    \label{fig:uncertcomparison}
    \end{figure*}

    We present in Table~\ref{tab:all53} the 53 clusters in the \textit{reference} sample. We give their names (as named in the CoMaLit LC$^2$ catalogue), redshifts, masses, and mass uncertainties from the analyses of reference, that is, XMM-\textit{Newton} mass reconstruction pipeline masses and CoMaLit estimates.

    The histograms in Fig.~\ref{fig:uncertcomparison} show the distributions of the measurement uncertainties for the XMM-\textit{Newton} and CoMaLit masses of all the clusters in the \textit{reference} sample. The vertical red lines indicate the measured systematic scatter for XMM-\textit{Newton} and CoMaLit masses.

    \renewcommand{\arraystretch}{1.4}    
    \scriptsize
    \begin{table*}[]
    \scriptsize
    \centering
     \caption{Characteristics of the \textit{reference} sample. }
        \begin{tabular}{c|c|c|c|c|c|c|c|c|c}
          \hline
          \hline
     CoMaLit name &      $z$ &   \multicolumn{4}{|c|}{Centre coordinates [deg]} & \multicolumn{4}{|c}{Masses [$10^{14}$ M$_{\odot}$]} \\ 
      ~&     ~ &  CoMaLit $\alpha$ & CoMaLit $\delta$ & XMM $\alpha$ &   XMM $\delta$ &  $M_{500}^{\mathrm{CoMaLit \; lens}}$ &  $\delta_{M_{\mathrm{CoMaLit \; lens}}}$ &  $M_{500}^{\mathrm{XMM \; HSE}}$&  $\delta_{M_{\mathrm{XMM \; HSE}}}$\\\hline

        ABELL2744 & 0.3080 &     3.57875 &    -30.38944 &   3.57752 & -30.38633 & 13.849 &      2.824 & 10.472 &     2.218 \\
        CL0016+16 & 0.5490 &     4.63933 &     16.43694 &   4.63986 &  16.43622 & 17.821 &      4.609 &  8.062 &     1.147 \\
          ABELL68 & 0.2550 &     9.27479 &      9.16000 &   9.27792 &   9.15685 &  9.171 &      1.587 &  6.872 &     1.181 \\
          ABELL85 & 0.0550 &    10.46017 &     -9.30306 &  10.46132 &  -9.30438 &  5.700 &      2.200 & 12.150 &     2.169 \\
        ABELL2813 & 0.2924 &    10.85167 &    -20.62139 &  10.85190 & -20.62294 &  8.557 &      1.450 &  4.131 &     0.665 \\
         Abell209 & 0.2060 &    22.96892 &    -13.61122 &  22.96912 & -13.61121 &  9.614 &      1.965 &  6.309 &     1.010 \\
         ABELL291 & 0.1960 &    30.43417 &     -2.20083 &  30.42942 &  -2.19678 &  4.514 &      0.986 &  2.718 &     0.380 \\
  RXCJ0232.2-4420 & 0.2840 &    38.07750 &    -44.34667 &  38.07721 & -44.34638 &  5.380 &      1.815 &  6.558 &     0.986 \\
         Abell383 & 0.1870 &    42.01417 &     -3.52914 &  42.01418 &  -3.52889 &  5.871 &      1.727 &  2.629 &     0.409 \\
         ABELL478 & 0.0881 &    63.35667 &     10.46694 &  63.35581 &  10.46371 &  8.772 &      2.078 &  7.043 &     0.325 \\
    MS0451.6-0305 & 0.5389 &    73.54767 &     -3.01411 &  73.54705 &  -3.01620 &  9.994 &      4.060 &  7.352 &     1.733 \\
  RXCJ0528.9-3927 & 0.2840 &    82.22083 &    -39.47161 &  82.22109 & -39.47136 &  4.480 &      1.310 &  9.276 &     2.541 \\
  RXCJ0532.9-3701 & 0.2750 &    83.23208 &    -37.02667 &  83.23226 & -37.02703 &  6.960 &      1.430 &  4.847 &     0.852 \\
 SPT-CLJ0546-5345 & 1.0660 &    86.65321 &    -53.76039 &  86.65500 & -53.76000 &  3.700 &      2.760 &  4.060 &     0.500 \\
 SPT-CLJ0615-5746 & 0.9720 &    93.96521 &    -57.77881 &  93.96600 & -57.77960 &  4.700 &      2.252 & 11.960 &     1.750 \\
        ABELL3404 & 0.1670 &   101.37292 &    -54.22697 & 101.37122 & -54.22732 &  8.750 &      2.085 &  6.360 &     0.968 \\
 MACSJ0647.7+7015 & 0.5840 &   101.95946 &     70.24861 & 101.95900 &  70.24810 &  9.427 &      2.493 &  6.296 &     1.014 \\
 MACSJ0911.2+1746 & 0.5050 &   137.79529 &     17.77539 & 137.79704 &  17.77603 & 10.862 &      3.259 &  3.392 &     0.424 \\
         ABELL963 & 0.2060 &   154.26483 &     39.04764 & 154.26530 &  39.04816 &  4.583 &      1.637 &  4.884 &     0.968 \\
        ABELL1300 & 0.3080 &   172.97583 &    -19.92772 & 172.97749 & -19.92848 &  5.950 &      1.695 &  5.341 &     0.476 \\
 MACSJ1149.5+2223 & 0.5440 &   177.39871 &     22.39850 & 177.39763 &  22.40108 & 14.447 &      3.034 &  6.536 &     1.608 \\
        ABELL1413 & 0.1430 &   178.82500 &     23.40503 & 178.82495 &  23.40487 &  7.200 &      2.100 &  6.816 &     0.815 \\
 MACSJ1206.2-0847 & 0.4400 &   181.55062 &     -8.80094 & 181.55208 &  -8.80017 & 12.176 &      2.477 &  9.681 &     1.203 \\
  ZwCl1215.1+0400 & 0.0750 &   184.42137 &      3.65589 & 184.42236 &   3.65650 &  3.500 &      2.200 &  5.629 &     0.566 \\
   CLJ1226.9+3332 & 0.8900 &   186.74271 &     33.54683 & 186.74203 &  33.54627 & 15.298 &      1.275 &  4.732 &     1.042 \\
        ABELL1576 & 0.3010 &   189.24583 &     63.19056 & 189.24408 &  63.18711 & 13.543 &      4.243 &  4.064 &     0.782 \\
       ABELL1644S & 0.0470 &   194.30000 &    -17.41306 & 194.29601 & -17.41110 &  1.309 &      0.748 & 13.664 &     5.775 \\
        ABELL1650 & 0.0840 &   194.67287 &     -1.76139 & 194.67336 &  -1.76223 &  7.100 &      2.000 &  3.629 &     0.315 \\
        ABELL1651 & 0.0850 &   194.84371 &     -4.19603 & 194.84380 &  -4.19831 &  5.600 &      2.400 &  4.486 &     0.429 \\
        ABELL1689 & 0.1830 &   197.87300 &     -1.34100 & 197.87263 &  -1.34172 & 15.033 &      1.025 &  8.071 &     1.078 \\
        ABELL1763 & 0.2279 &   203.82583 &     40.99694 & 203.82979 &  41.00010 & 16.014 &      2.050 &  5.102 &     0.779 \\
        ABELL1795 & 0.0620 &   207.21871 &     26.59300 & 207.22115 &  26.58994 &  9.300 &      2.200 & 10.910 &     0.291 \\
        ABELL1835 & 0.2530 &   210.25804 &      2.87775 & 210.25916 &   2.87824 & 15.510 &      4.503 &  8.200 &     0.660 \\
PSZ2G099.86+58.45 & 0.6160 &   213.69662 &     54.78433 & 213.69522 &  54.78396 &  7.242 &      3.043 &  6.421 &     2.038 \\
        ABELL1914 & 0.1712 &   216.50667 &     37.82722 & 216.51053 &  37.82434 &  7.929 &      1.293 &  6.909 &     1.390 \\
  ZwCl1454.8+2233 & 0.2578 &   224.31292 &     22.34278 & 224.31293 &  22.34242 &  3.771 &      1.457 &  3.407 &     0.584 \\
       Zwicky7215 & 0.2900 &   225.34483 &     42.34750 & 225.34451 &  42.34650 &  5.390 &      1.504 &  5.418 &     1.142 \\
        ABELL2034 & 0.1130 &   227.54875 &     33.51472 & 227.55283 &  33.51044 &  5.169 &      3.100 &  5.750 &     0.643 \\
        ABELL2029 & 0.0770 &   227.73371 &      5.74481 & 227.73502 &   5.74410 & 12.100 &      2.500 &  5.592 &     0.661 \\
        ABELL2065 & 0.0730 &   230.62150 &     27.70769 & 230.62257 &  27.70901 &  8.000 &      2.100 &  5.326 &     0.474 \\
        ABELL2204 & 0.1520 &   248.19650 &      5.57583 & 248.19592 &   5.57544 & 16.051 &      2.963 &  7.966 &     1.605 \\
        ABELL2218 & 0.1760 &   248.95329 &     66.21417 & 248.95972 &  66.21254 &  8.900 &      2.700 &  3.829 &     0.597 \\
        ABELL2219 & 0.2280 &   250.08475 &     46.70833 & 250.08366 &  46.71067 & 11.729 &      1.852 & 10.287 &     2.969 \\
   RXJ1720.1+2638 & 0.1640 &   260.04167 &     26.62464 & 260.04166 &  26.62503 &  3.510 &      1.485 &  6.006 &     1.314 \\
        Abell2261 & 0.2240 &   260.61325 &     32.13258 & 260.61267 &  32.13237 & 15.613 &      3.043 &  3.900 &     0.590 \\
 MACSJ2129.4-0741 & 0.5880 &   322.35717 &     -7.69189 & 322.35913 &  -7.69133 & 13.486 &      3.890 &  6.127 &     1.179 \\
   RXJ2129.7+0005 & 0.2340 &   322.41650 &      0.08922 & 322.41660 &   0.08861 &  4.470 &      1.158 &  4.277 &     0.491 \\
        ABELL2390 & 0.2330 &   328.40446 &     17.69594 & 328.40308 &  17.69493 & 11.183 &      2.396 &  9.652 &     1.668 \\
   RXJ2228.6+2037 & 0.4110 &   337.13658 &     20.62072 & 337.14047 &  20.62040 &  9.728 &      2.626 &  6.844 &     0.819 \\
 MACSJ2243.3-0935 & 0.4470 &   340.83933 &     -9.59522 & 340.83868 &  -9.59470 & 20.294 &      3.865 &  8.625 &     1.317 \\
   RXJ2248.7-4431 & 0.3480 &   342.18317 &    -44.53092 & 342.18243 & -44.53054 & 12.400 &      3.605 & 14.262 &     3.762 \\
        ABELL2631 & 0.2780 &   354.40971 &      0.27069 & 354.40634 &   0.26678 & 11.748 &      1.888 &  9.785 &     3.807 \\
 SPT-CLJ2341-5119 & 1.0030 &   355.30092 &    -51.32850 & 355.30100 & -51.32860 &  1.600 &      1.890 &  3.190 &     0.750 \\

    \hline
        \end{tabular}
        \vspace*{0.2cm}  
        \begin{tablenotes}         
        \small
      \item \textbf{Notes.} Column 1: cluster names from the CoMaLit catalogue (entries Comalit\_Name and Comalit\_Num). Column 2: redshift. Columns 3 to 6: right ascension $\alpha$ and declination $\delta$ of the cluster centres according to CoMaLit or X-rays. Columns 7 to 10: cluster masses and uncertainties from the CoMaLit catalogue and from the XMM-\textit{Newton} analysis.
        \end{tablenotes}        
        \vspace*{0.2cm}
        \label{tab:all53}
    \end{table*}
    \FloatBarrier
\normalsize

\section{Different scaling relation models}

\subsection{Additional figures}

\begin{figure*}
        \begin{minipage}[b]{0.5\textwidth}
        \includegraphics[trim={0pt 0pt 0pt 0pt},scale=0.6]{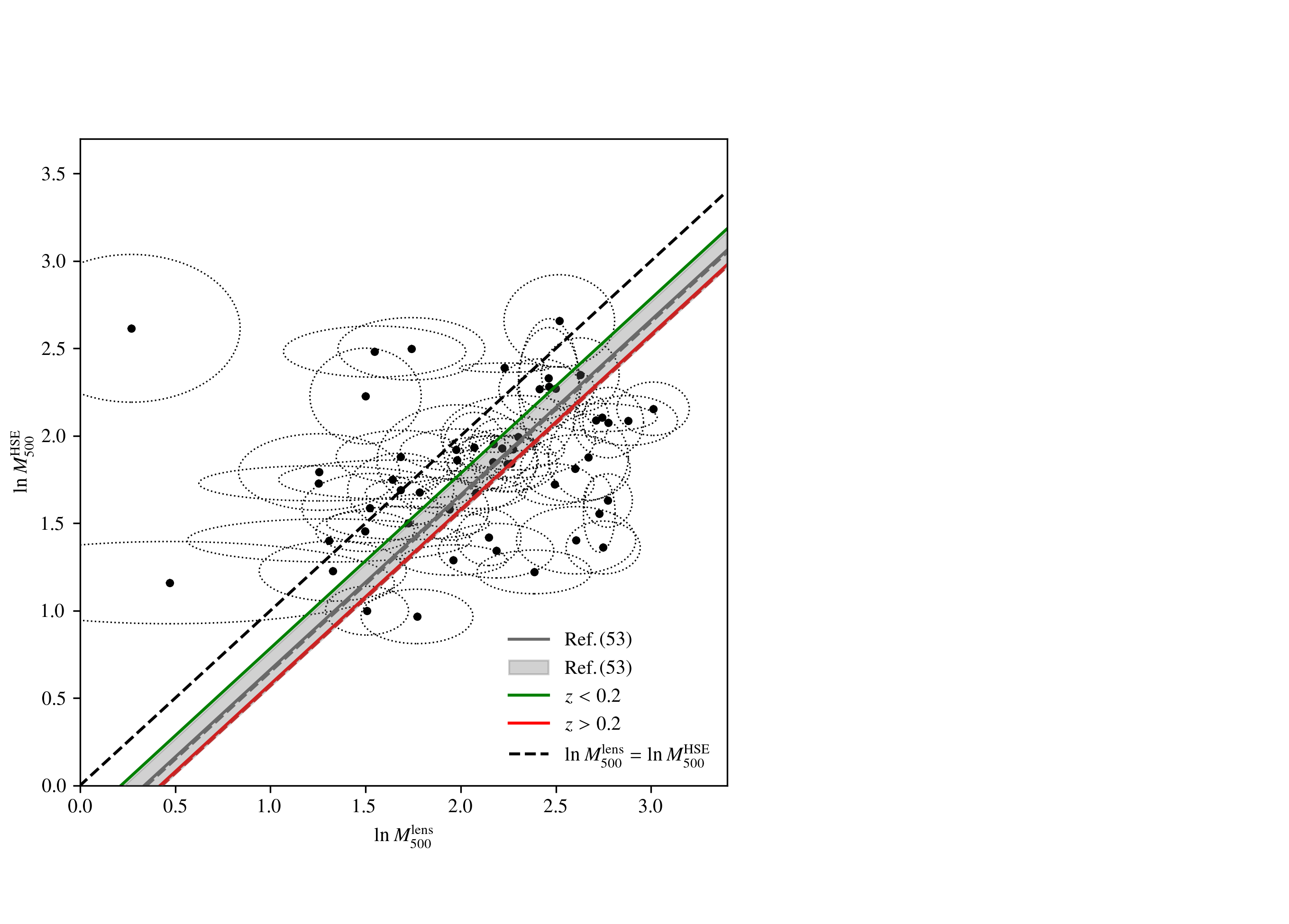}
        \end{minipage}
        \hfill
        \begin{minipage}[b]{0.5\textwidth}
        \includegraphics[trim={0pt 0pt 0pt 0pt},scale=0.3]{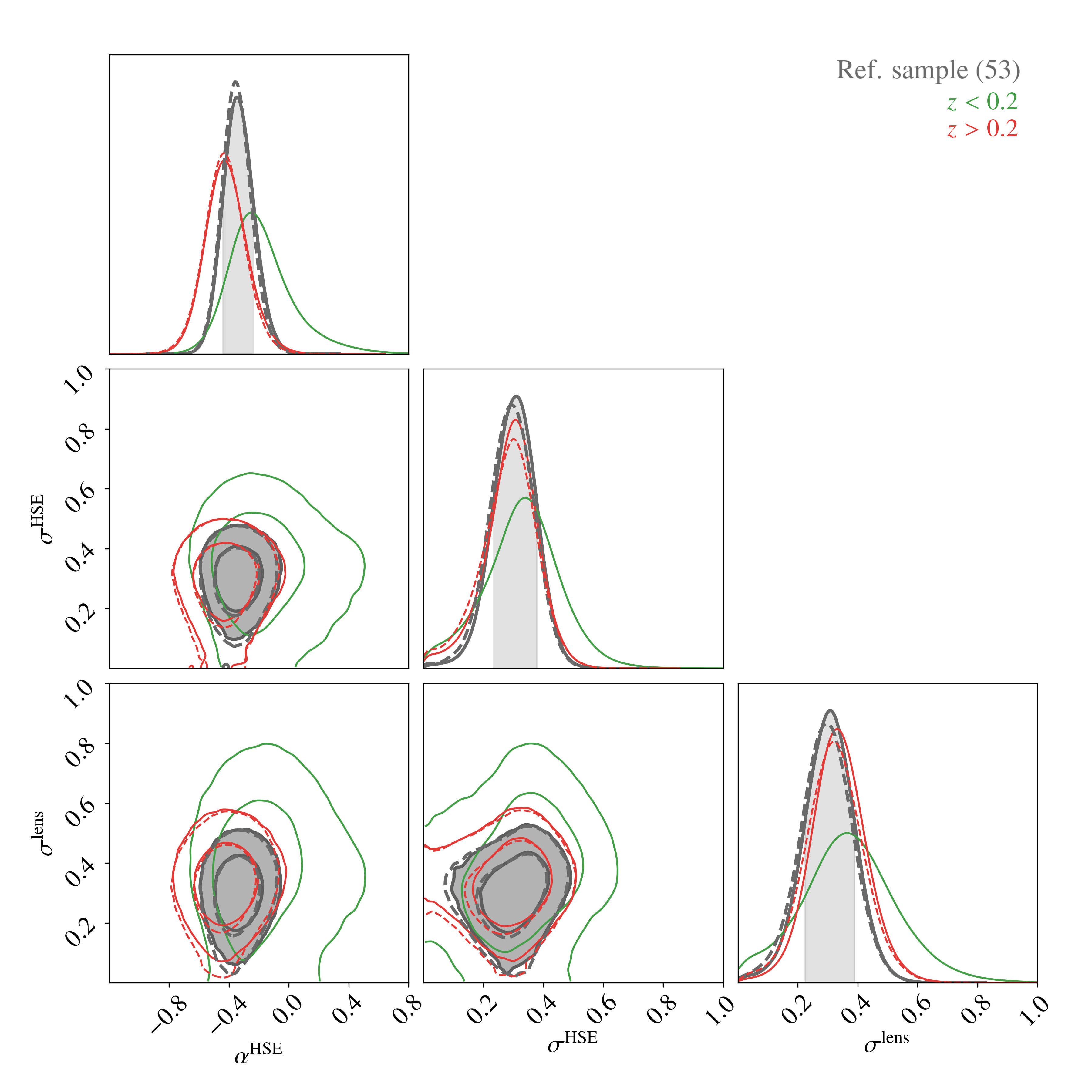}
        \end{minipage}
    \caption{Same as Fig.~\ref{fig:differentzsys}, but without considering the systematic scatter in the fit.}
    \label{fig:differentz}
\end{figure*}

\begin{figure*}
        \begin{minipage}[b]{0.5\textwidth}
        \includegraphics[trim={0pt 0pt 0pt 0pt},scale=0.6]{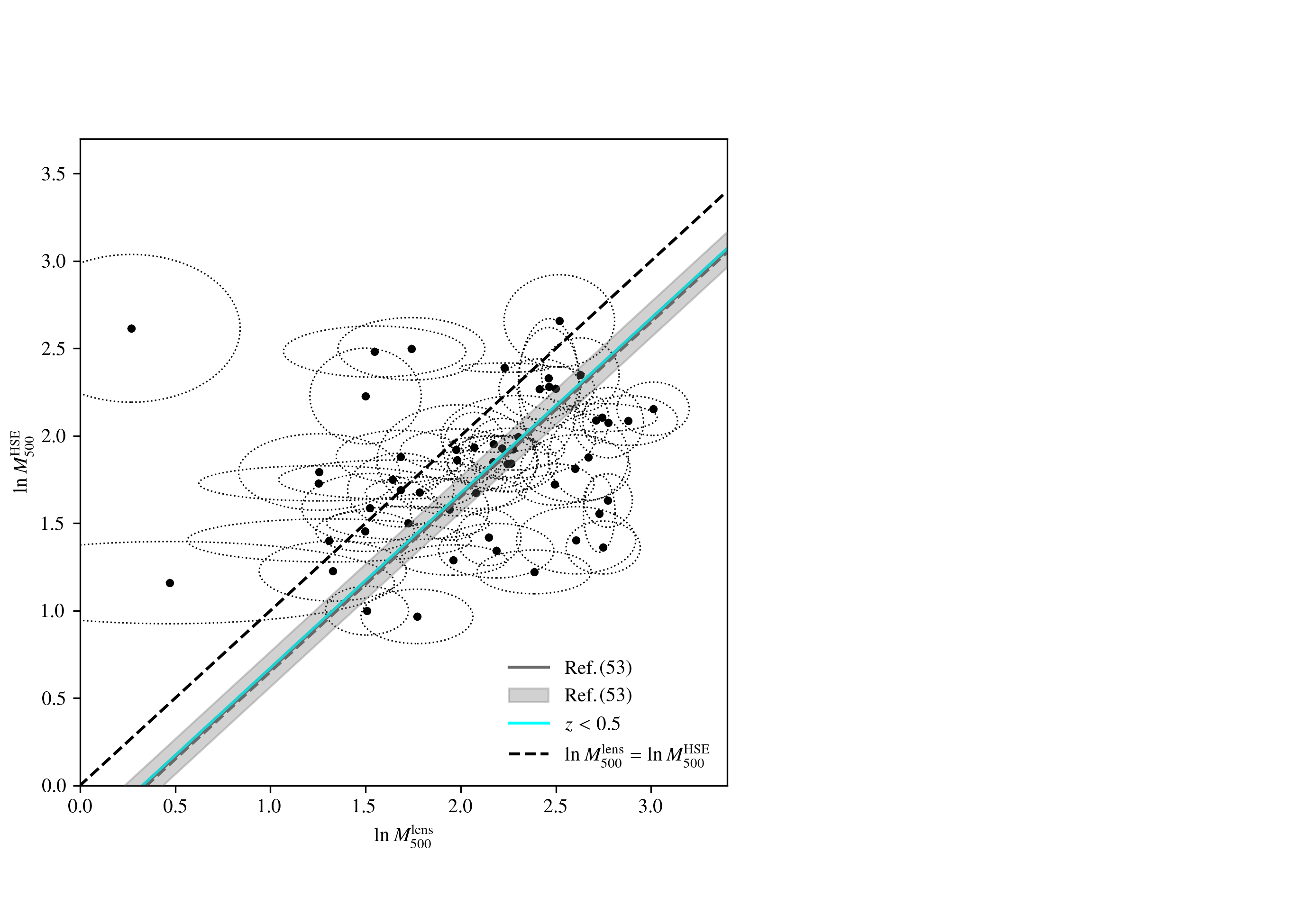}
        \end{minipage}
        \hfill
        \begin{minipage}[b]{0.5\textwidth}
        \includegraphics[trim={0pt 0pt 0pt 0pt},scale=0.3]{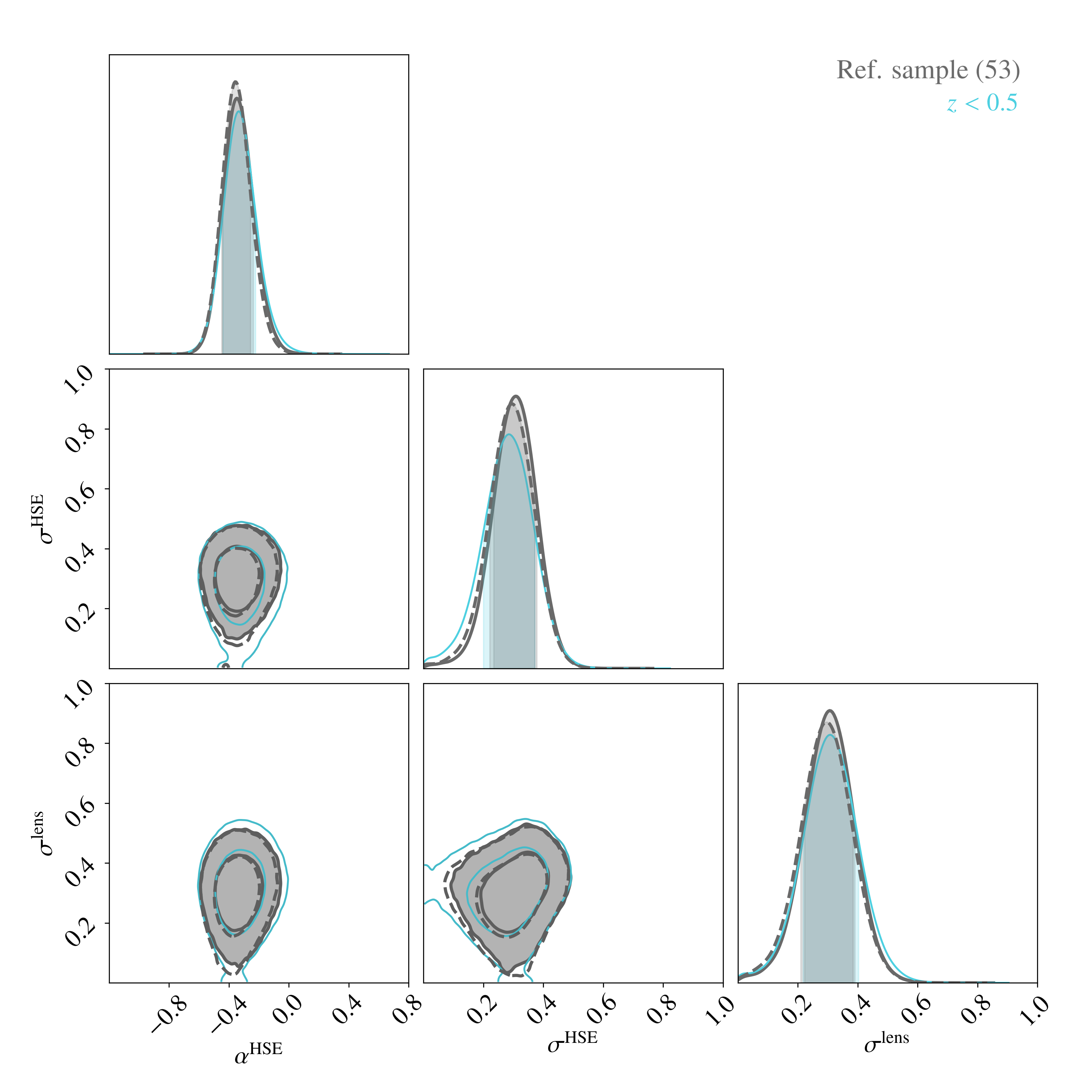}
        \end{minipage}
        \caption{ Scaling relation between HSE and lensing masses for the \textit{reference} sample in grey and a subsample containing only $z<0.5$ clusters in cyan, without considering the systematic scatter in the fit. The black dashed line shows the equality. The corner plots in the right panel are the posterior 1D and 2D distributions of the parameters in the SR. Here $\beta^{\mathrm{HSE}}$ is fixed to 1.}
    \label{fig:different0.5}
\end{figure*}

In this section we present additional figures of the SR model extension fits described in Sect.~\ref{sec:robustness}. We show in Fig.~\ref{fig:differentz} and \ref{fig:different0.5} the scaling relations and posterior parameter distributions for different subsamples defined according to the redshift range, in these cases not including the systematic scatters in the error bars of each cluster. In Fig.~\ref{fig:differentz} we present the scaling relations for the samples with $z<0.2$ and $z>0.2$, while the panels in Fig.~\ref{fig:different0.5} show the results for a $z<0.5$ sample. 

\begin{figure*}
        \begin{minipage}[b]{0.5\textwidth}
        \includegraphics[trim={0pt 0pt 0pt 0pt},scale=0.6]{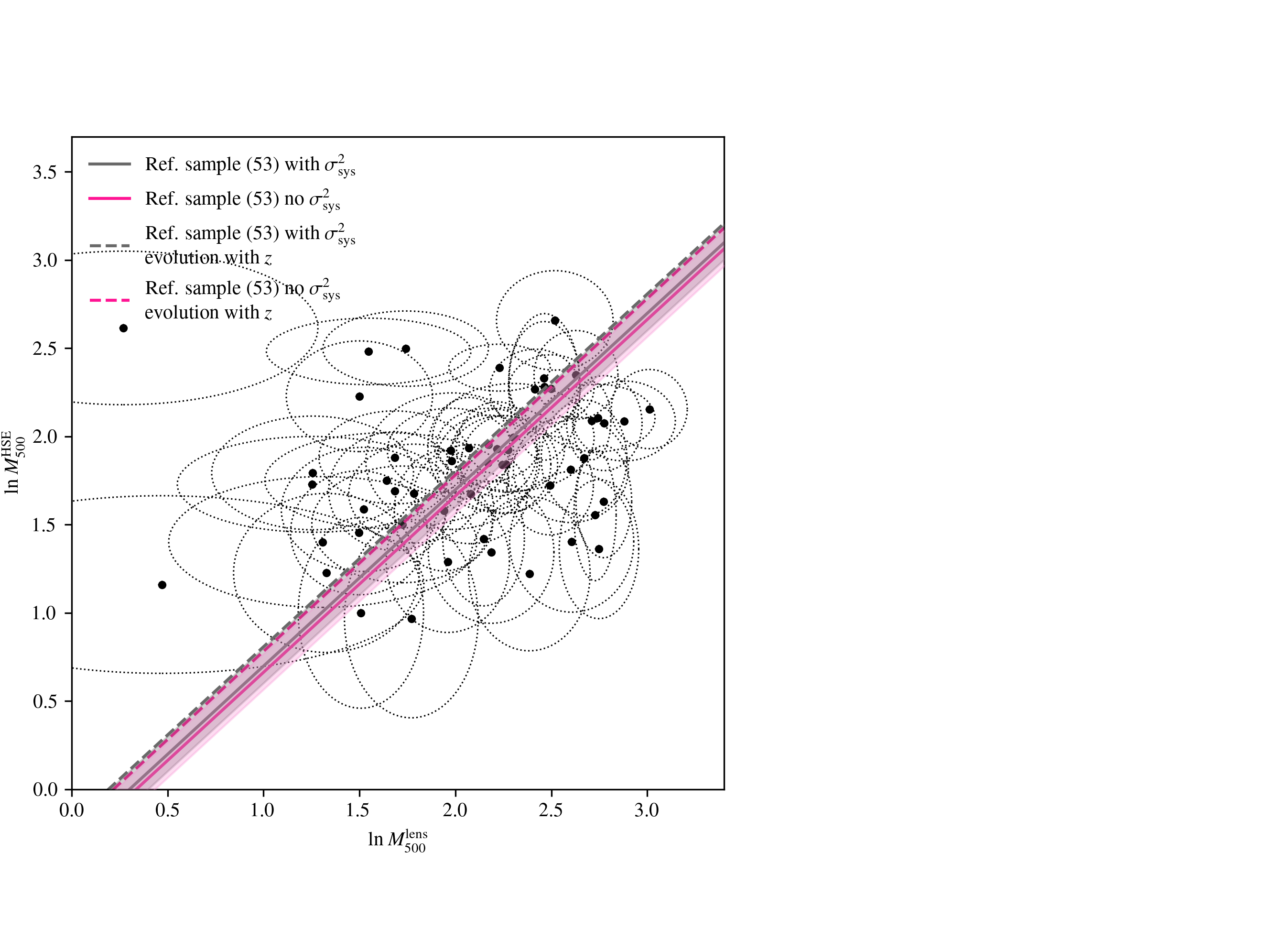}
        \end{minipage}
        \hfill
        \begin{minipage}[b]{0.5\textwidth}
        \includegraphics[trim={0pt 0pt 0pt 0pt},scale=0.3]{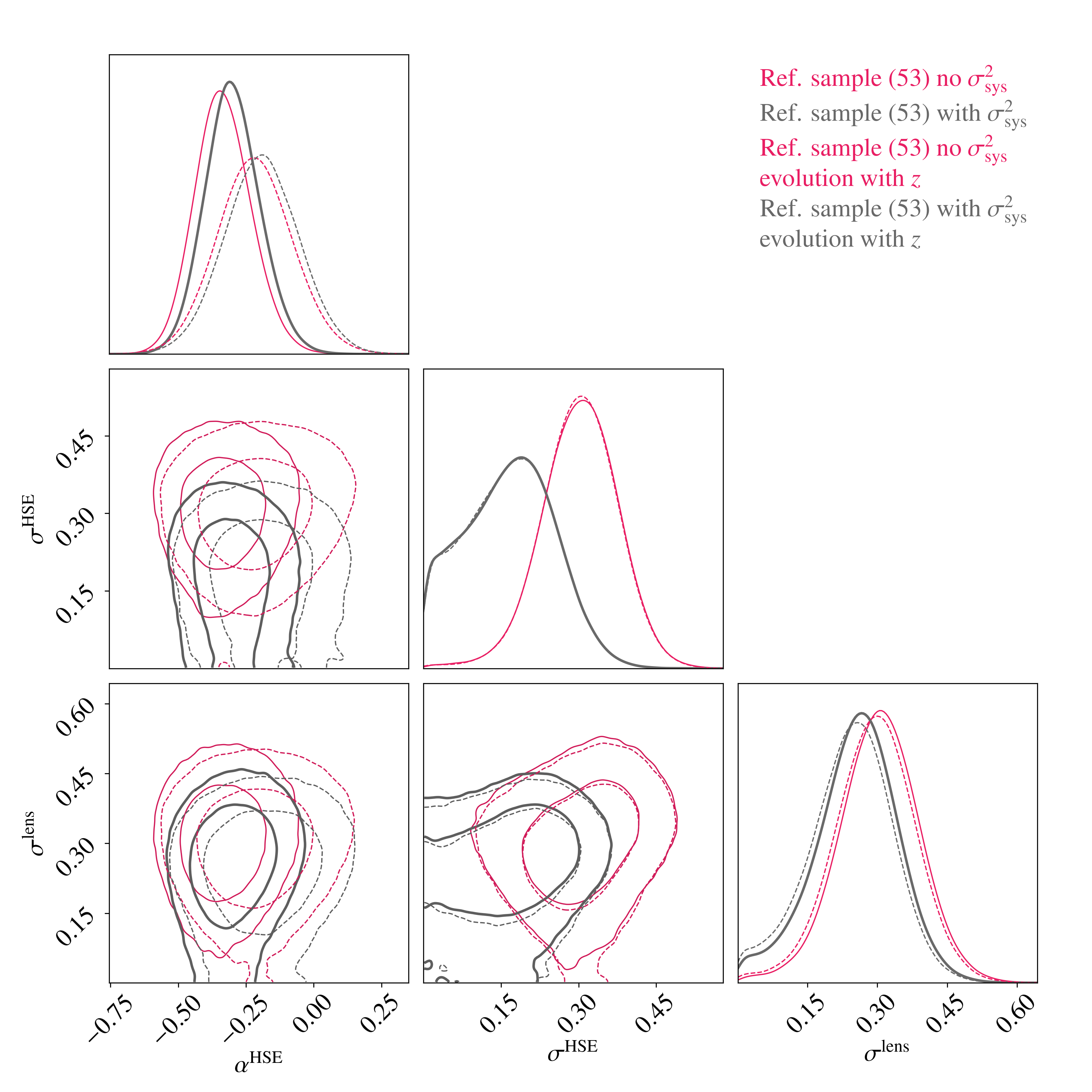}
        \end{minipage}
    \caption{Same as Fig.~\ref{fig:refSRnosysandsys} but with dashed lines showing the results if an evolution with redshift is considered in the scaling relation and solid lines without evolution.}
    \label{nanana}
\end{figure*}

\begin{figure*}
        \begin{minipage}[b]{0.5\textwidth}
        \includegraphics[trim={0pt 0pt 0pt 0pt},scale=0.6]{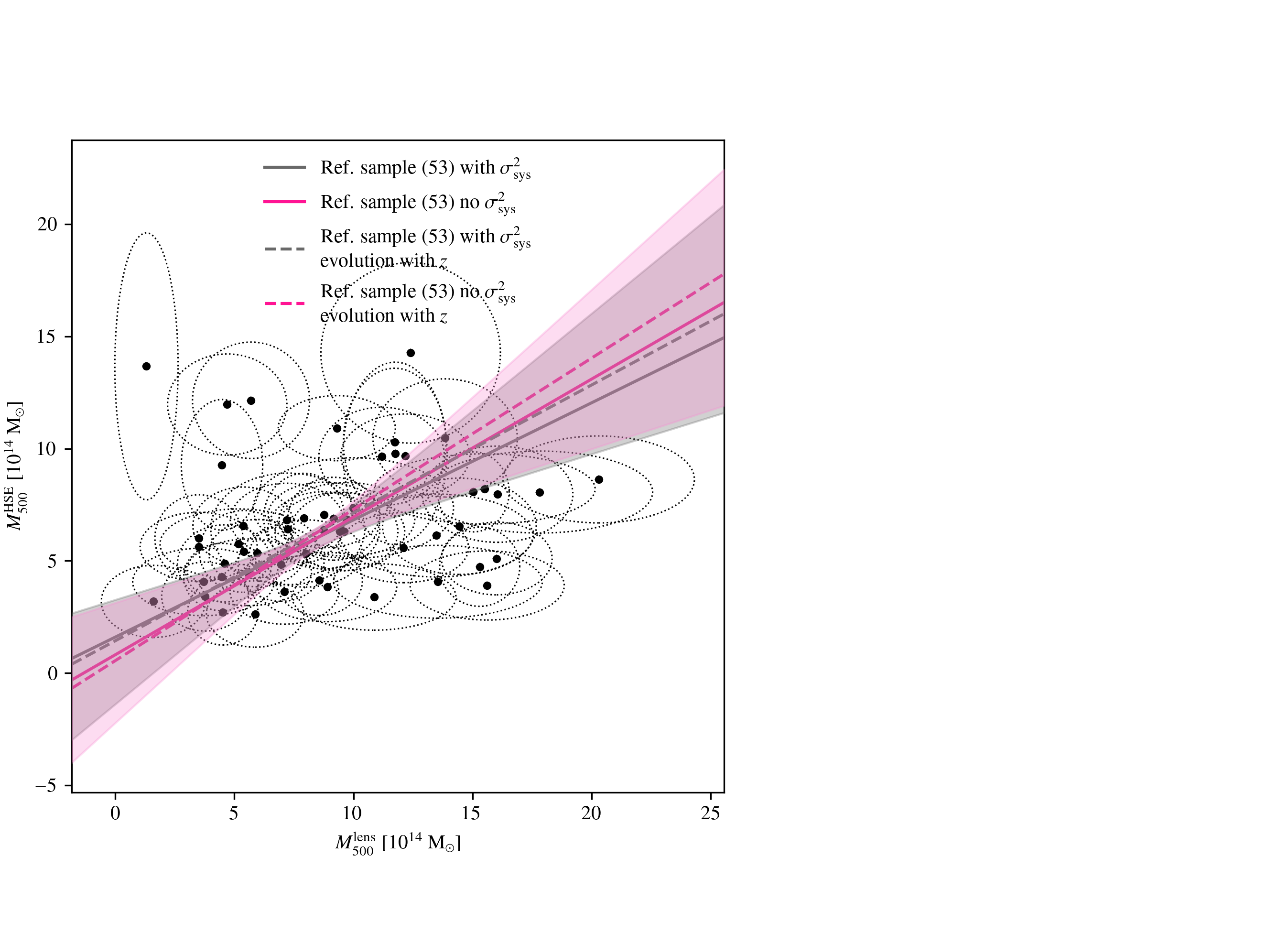}
        \end{minipage}
        \hfill
        \begin{minipage}[b]{0.5\textwidth}
        \includegraphics[trim={0pt 0pt 0pt 0pt},scale=0.3]{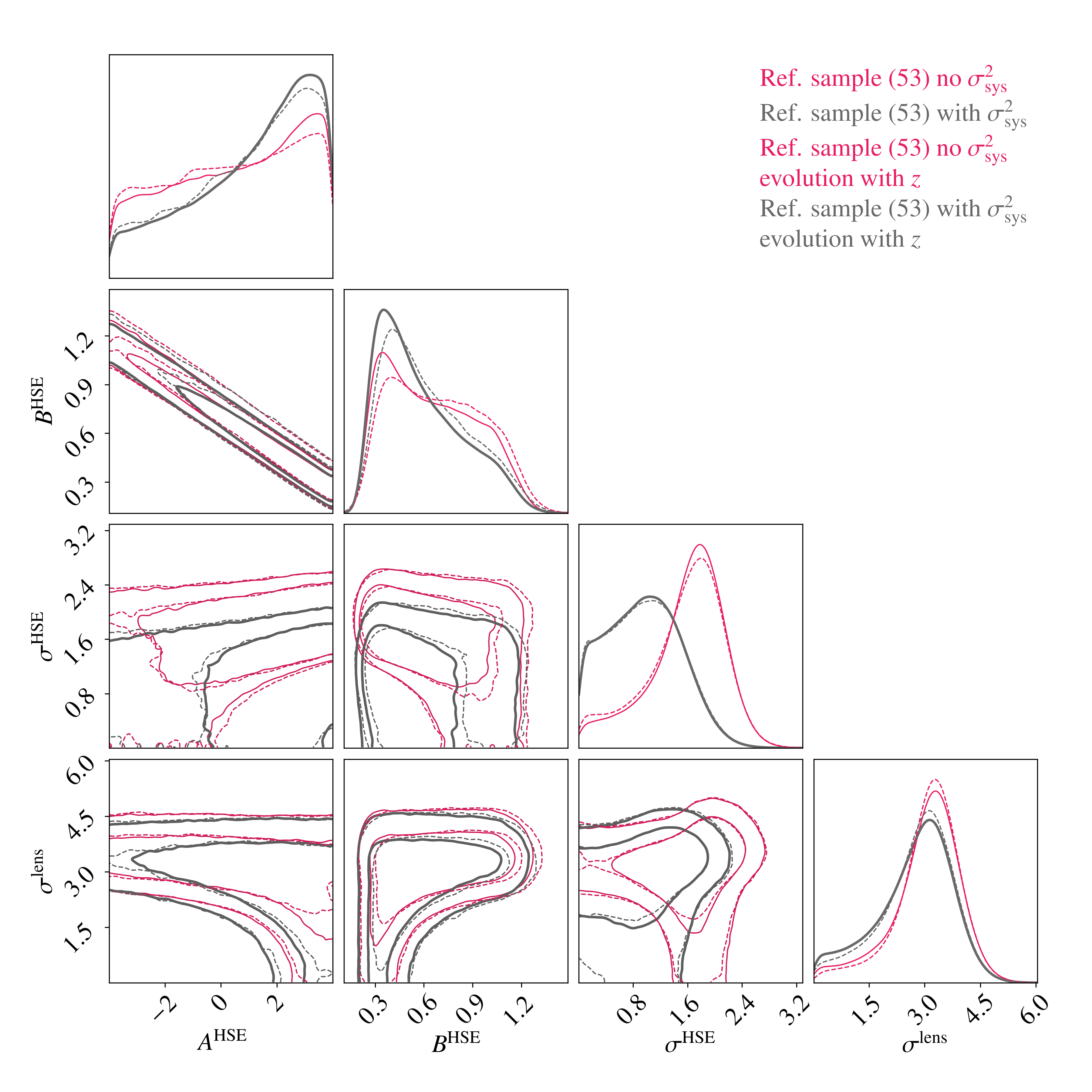}
        \end{minipage}
    \caption{Same as Fig.~\ref{fig:lin} but with dashed lines showing the results if an evolution with redshift is considered in the scaling relation and solid lines without evolution.}
    \label{jiji}
\end{figure*}

Figures \ref{nanana} and \ref{jiji} compare the SR and parameters posterior distributions when considering and not redshift evolution. 


\FloatBarrier

\subsection{Comparison of SR models with AIC and BIC}
\label{sec:aicbic}

\begin{figure*} 
  \begin{minipage}[b]{0.48\textwidth}
    \centering
    \includegraphics[trim={0pt 0pt 0pt 0pt}, scale=0.32]{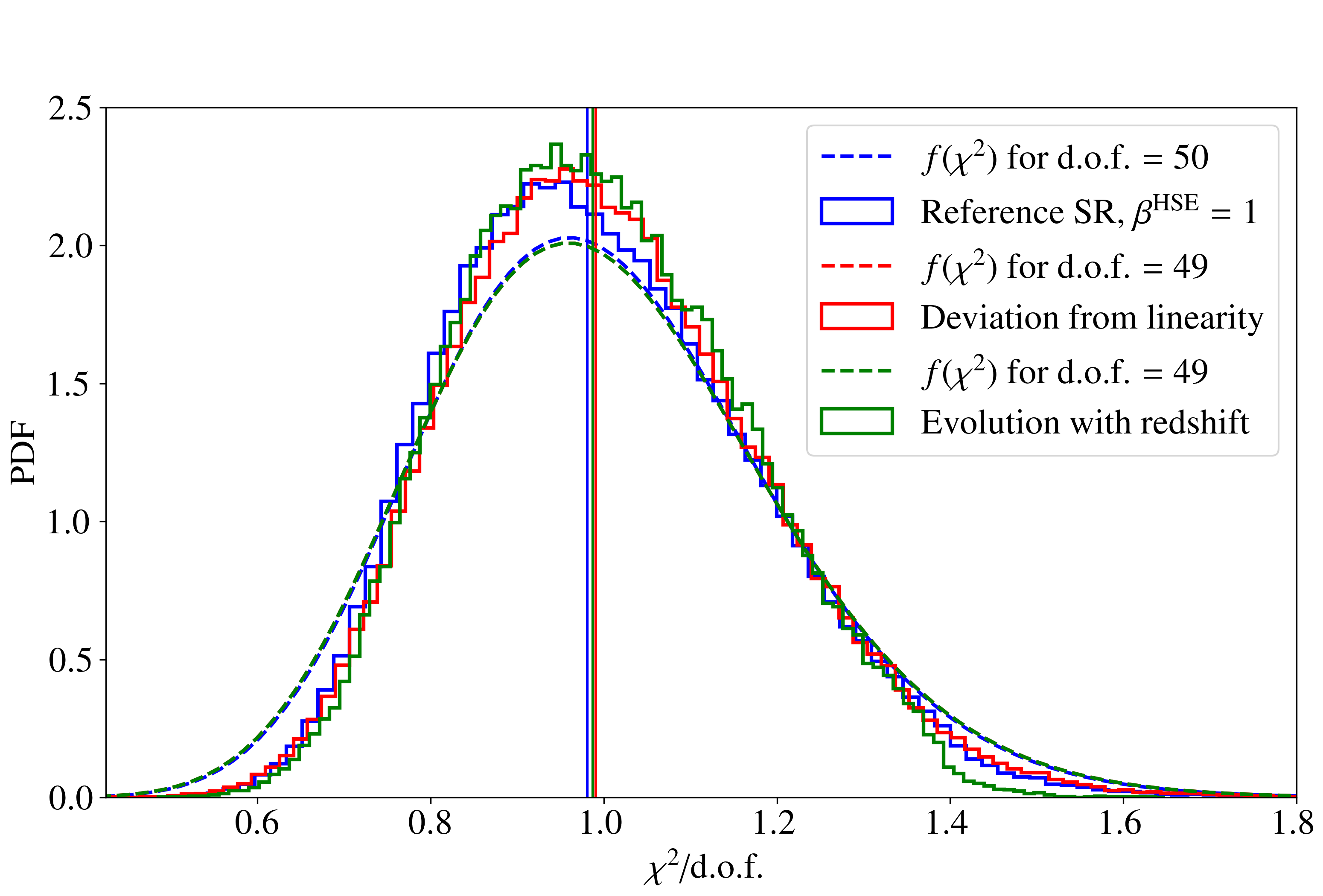}
   \end{minipage}
   \hfill
   \begin{minipage}[b]{0.48\textwidth}
     \centering
     \includegraphics[trim={0pt 0pt 0pt 0pt}, scale=0.32]{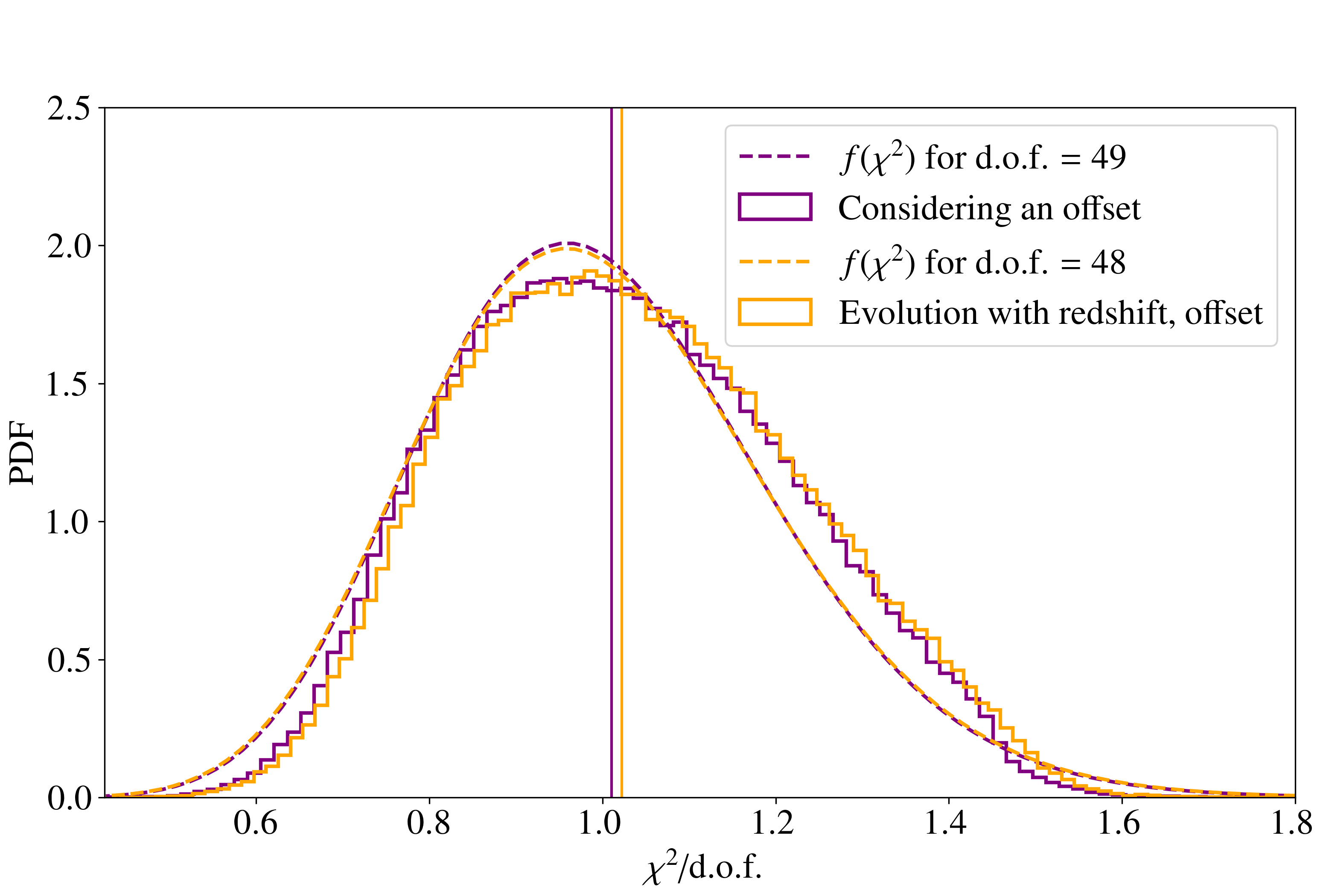}
   \end{minipage}
   \caption{Comparison of the reduced $\hat{\chi}^2$ distributions for different scaling relation fits (solid colour histrograms) and the $\chi^2$-distribution (Eq.~\ref{eq:chi2def}) for the degrees of freedom in each fit (dashed colour lines). Vertical lines show the median value of each histogram.}
   \label{fig:chi2}
\end{figure*}

In Fig.~\ref{fig:chi2} we present the $\hat{\chi}^2$ (defined in Eq.~\ref{eq:goodnessoffit}) distributions normalised by the degrees of freedom (d.o.f.) in each SR model fit as solid line histograms. In the left panel we present the results for the scaling relations fitted in $\ln M^{\mathrm{HSE}} - \ln M^{\mathrm{lens}}$ and in the right for $M^{\mathrm{HSE}} -  M^{\mathrm{lens}}$. The blue histogram in the left panel shows the reduced $\hat{\chi}^{2}$ for the scaling relation of reference in this work and the vertical solid line is the median value of the distribution. For comparison, the dashed lines correspond to the $\chi^{2}$-distributions,
\begin{equation}
  f(\chi^2) = \frac{1}{2^{\nu/2}\Gamma(\nu/2)} e^{-\chi^{2}/2}(\chi^2)^{(\nu/2) -1},
  \label{eq:chi2def}
\end{equation}  
where $\nu$ is the number of the degrees of freedom (d.o.f.). For the scaling relation of reference we have d.o.f. $= 53 - 3 = 50$. The red and green results show the reduced $\hat{\chi}^{2}$ for the fits of the scaling relations when considering a deviation from linearity and an evolution with redshift, respectively. The histograms follow fairly well the $\chi^2$-distributions with $\nu$ degrees of freedom.

We use the Akaike information criterion \citep[AIC,][]{AIC} and the Bayesian information criterion \citep[BIC,][]{BIC} to compare the improvement in the $\hat{\chi}^2$ when adding parameters to the model. We calculate:
\begin{equation}
\mathrm{AIC} = \hat{\chi}^2_{\mathrm{min}} + 2K,
\end{equation}
and
\begin{equation}
  \mathrm{BIC} = \hat{\chi}^2_{\mathrm{min}} + K \ln N,
\end{equation}
where $\hat{\chi}^2_{\mathrm{min}}$ is the minimum of the $\hat{\chi}^2$ values for each model. Here $K$ and $N$ are the number of free parameters in the model and the total number of data points, that is, $N = N_{\mathrm{clusters}} = 53$, respectively.

We report in Table~\ref{tab:chi2criteria} the results for the different scaling relation models and the $\Delta$AIC and $\Delta$BIC differences with respect to the simplest scaling law amongst the nested models.

\renewcommand{\arraystretch}{1.4}
    \scriptsize
    \begin{table*}[]
      \scriptsize
      \centering
       \caption{Statistical results for the scaling relation models presented in this work.}
        \begin{tabular}{c|c|c|c|c|c|c|c}

          \hline
          \hline

           Model & $K$ & $N$ & $\hat{\chi}^2_{\mathrm{min}}$ & AIC & $\Delta$AIC & BIC & $\Delta$BIC \\\hline
           Reference SR & 3 & 53 & 18.126 & 24.126 & 0.000 & 21.422 & 0.000 \\
           Deviation from linearity & 4 & 53  & 16.020 & 24.020 & -0.106 & 21.565 & +0.143 \\
           Evolution with redshift & 4 & 53 & 22.784 & 30.784 & +6.658 & 28.329 & +6.907\\\hline
           Considering an offset & 4 & 53 & 13.929 & 21.993 & 0.000 & 19.475 & 0.000 \\          
           Evolution with redshift and offset & 5 & 53 & 17.992 & 27.992 & +5.999 & 26.039 & +6.564\\
           \hline    
        \end{tabular}
        \vspace*{0.2cm}  
        \begin{tablenotes}         
        \small
      \item \textbf{Notes.} We report the values for the fits accounting for the systematic scatter of HSE and lensing masses.
        \end{tablenotes}        
        \vspace*{0.2cm}
        \label{tab:chi2criteria}
    \end{table*}
\normalsize  
\FloatBarrier

\section{Other HSE mass estimates from XMM-\textit{Newton} data}
\label{sec:otherxmm}
As described in Sect.~\ref{sec:xmm}, all the XMM-\textit{Newton} HSE masses used in this work were obtained from mass profiles, by interpolating the binned profile to get $M_{500}$ and $R_{500}$. For all the clusters in our \textit{reference} sample we also have access to X-ray masses obtained with the $Y_{\mathrm{X}}-M^{\mathrm{HSE}}_{500}$ scaling relation from \citet{arnaud10}. The quantity $Y_{\mathrm{X}}$ is defined as the product of the gas mass ($M_{\mathrm{g}, 500}$) and the spectroscopic temperature ($T_{\mathrm{X}}$) and it is the X-ray analogue of the integrated Compton parameter \citep{kravtsov2006}. We use these masses as $M^{\mathrm{HSE}}$ estimates and fit again the scaling relations in Eq.~\ref{eq:srlens}, \ref{eq:srhse}, \ref{eq:offset1}, and \ref{eq:offset2}.

Median values and uncertainties of the fitted SR parameters are given in Table~\ref{tab:othermasses}. The intrinsic scatters of HSE and lensing masses with respect to the SR are reduced when using $Y_{\mathrm{X}}-M^{\mathrm{HSE}}_{500}$ estimates and the measured HSE bias is also smaller (larger $1-b$).

For ESZ+LoCuSS and LPSZ sample clusters we have a third estimate of the HSE mass, obtained by fitting an NFW mass model to the X-ray profiles. Such masses tend to be in agreement with the interpolated ones (the reference XMM-\textit{Newton} masses used in this work, Sect.~\ref{sec:xmm}), with a mean and median ratio for the 50 cluster estimates of $M_{500} / M_{500}^{\mathrm{NFW}} \sim 1.11$ and $0.95$, indicating that the HSE masses used in the analyses of reference are robust against modelling effects. Masses obtained by fitting an NFW model have in median $30\%$ smaller uncertainties than the interpolated estimates. The scaling relations obtained with NFW fit masses (Table~\ref{tab:othermasses}) and interpolated ones (second row in Table~\ref{tab:fits} for the 50 clusters with $z<0.9$) are almost identical, with $\alpha^{\mathrm{HSE}}$ centred in the exact same value, but with $10 -20\%$ smaller intrinsic scatters when using NFW masses.

\renewcommand{\arraystretch}{1.4}
    \scriptsize
    \begin{table*}[]
      \scriptsize
      \centering
       \caption{Summary of the median values and uncertainties at the 16th and 84th percentiles of the fitted parameters for the HSE-to-lensing mass SR for HSE masses obtained from the $Y_{\mathrm{X}}-M^{\mathrm{HSE}}_{500}$ scaling relation and from the fit of an NFW model to the X-ray data. }
        \begin{tabular}{c|c|c|c|c|c|c}

          \hline
          \hline

             Sample & $\#$ of clusters &  \multicolumn{5}{c}{No $\sigma_{\mathrm{sys}}^2$}  \\
               ~ & ~ &  \multicolumn{5}{c}{~}\\\hline \hline

             $Y_{\mathrm{X}}-M^{\mathrm{HSE}}_{500}$ & ~ & $\alpha^{\mathrm{HSE}}$ &  $e^{\alpha^{\mathrm{HSE}}} = (1-b)$ & $\beta^{\mathrm{HSE}}$ & $\sigma^{\mathrm{HSE}}$ & $\sigma^{\mathrm{lens}}$\\ \hline
             \textit{Reference} sample & 53 & $-0.237_{-0.087}^{+0.095}$ & $0.789_{-0.069}^{+0.075}$  & [1] & $0.105_{-0.067}^{+0.063}$ & $0.247_{-0.056}^{+0.058}$  \\
             \textit{Reference} sample & 53 & $-0.131_{-0.631}^{+0.542}$  & ~& $0.953_{-0.263}^{+0.296}$ & $0.133_{-0.088}^{+0.075}$  & $0.231_{-0.098}^{+0.068}$\\
             \textit{Reference} sample (BCES) & 53 &  $0.504 \pm 0.262$ & ~& $0.662 \pm 0.122$ & $0.250^{*}$ & -  \\\hline

             ~ & ~ & $A^{\mathrm{HSE}}$ [$10^{14}$ M$_{\odot}$] &  $B^{\mathrm{HSE}}$ & ~ & $\sigma^{\mathrm{HSE}}$ [$10^{14}$ M$_{\odot}$] & $\sigma^{\mathrm{lens}}$ [$10^{14}$ M$_{\odot}$]\\ \hline

             \textit{Reference} sample & 53 & $0.990_{-2.050}^{+1.748}$ & $0.672_{-0.205}^{+0.235}$ & ~ & $0.930_{-0.633}^{+0.508}$ & $2.333_{-1.032}^{+0.594}$   \\
             \textit{Reference} sample (BCES) & 53 & $2.401 \pm 0.781$ &  $0.487 \pm 0.095$ &~ & $1.480^{*}$ & -\\\hline
             
             ~ & ~ &  \multicolumn{5}{c}{~}\\\hline \hline
             NFW fit & ~ & $\alpha^{\mathrm{HSE}}$ &  $e^{\alpha^{\mathrm{HSE}}} = (1-b)$ & $\beta^{\mathrm{HSE}}$ & $\sigma^{\mathrm{HSE}}$ & $\sigma^{\mathrm{lens}}$ \\\hline

             $z<0.9$ & 50 & $-0.309_{-0.107}^{+0.119}$ & $0.734_{-0.079}^{+0.087}$ & [1] & $0.249_{-0.063}^{+0.059}$ & $0.212_{-0.086}^{+0.077}$ \\
             $z<0.9$ & 50 & $-0.361_{-0.955}^{+0.692}$  &  ~ & $1.030_{-0.341}^{+0.480}$ & $0.246_{-0.121}^{+0.073}$ & $0.220_{-0.119}^{+0.081}$  \\
             $z<0.9$ (BCES)  & 50 & $0.032 \pm 0.485$ & ~& $0.836  \pm 0.222$ & $0.329^{*}$ & -  \\\hline

             ~ & ~ & $A^{\mathrm{HSE}}$ [$10^{14}$ M$_{\odot}$] & $B^{\mathrm{HSE}}$ & ~ & $\sigma^{\mathrm{HSE}}$ [$10^{14}$ M$_{\odot}$] & $\sigma^{\mathrm{lens}}$ [$10^{14}$ M$_{\odot}$]\\ \hline

             $z<0.9$ & 50 & $1.205_{-2.463}^{+1.490}$ & $0.579_{-0.173}^{+0.287}$ &  ~ & $1.576_{-0.589}^{+0.408}$ & $1.989_{-1.164}^{+0.843}$\\
             $z<0.9$ (BCES) & 50 & $0.530 \pm 1.343$ & $0.646 \pm 0.159$&  ~& $1.944^{*}$ &  -\\\hline
    
        \end{tabular}
        \vspace*{0.2cm}  
         \begin{tablenotes}         
        \small
      \item \textbf{Notes.} We present the results assuming linearity, a deviation from linearity, and an offset between HSE and lensing masses. For the BCES fit, we report the best-fit values and $1\sigma$ uncertainties. $^{(*)}$ We also calculate the scatter with respect to the best BCES scaling relations following Eq.~\ref{eq:sysdef}.
        \end{tablenotes}        
        \vspace*{0.2cm}
        \label{tab:othermasses}
    \end{table*}
    
\normalsize

\end{appendix}

%
%


\end{document}